\documentclass[12pt,a4paper]{article}

\usepackage{vmargin}
\setmarginsrb{1.13in}{1.13in}{1.13in}{1.13in} {0mm}{0mm}{0mm}{10mm}
\usepackage{amsmath,amsfonts,amssymb,bm,amsthm,amscd,bbold,bbm}
\usepackage{graphics}
\usepackage{graphicx}
\usepackage[ruled,vlined,linesnumberedhidden]{algorithm2e}
\SetAlgoSkip{bigskip}
\usepackage{xr}
\usepackage{clipboard}	
\newclipboard{PaperClipboard}	
\usepackage{xr-hyper} 

\usepackage{pifont}
\usepackage{mathtools} 
\usepackage[utf8,latin1]{inputenc}
\usepackage{latexsym}
\usepackage{fancyhdr}
\usepackage{setspace}
\usepackage{enumerate}
\usepackage{comment}

\usepackage[round,longnamesfirst]{natbib}
\usepackage[breaklinks]{hyperref}
\usepackage{cleveref}
\usepackage{pstricks,pstricks-add,pst-plot,pst-node}
\usepackage{color}
\usepackage{float}
\usepackage{subfig}
\usepackage{breakcites}
\usepackage{soul}
\usepackage{multicol}

\usepackage{tikz}
\usetikzlibrary{backgrounds, intersections,patterns,matrix}
\usepackage{pgfplots}

\usepackage[labelfont=bf,textfont=it]{caption}

\usepackage{chngcntr}
\usepackage{apptools}

\DeclareMathAlphabet{\mathpzc}{OT1}{pzc}{m}{it}

\definecolor{MyRed}{rgb}{0.6,0,0}
\definecolor{MyGreen}{rgb}{0,0.5,0.1}
\definecolor{MyBlue}{rgb}{0,0,0.6}
\definecolor{MyGrey}{rgb}{0.3,0.3,0.4}
\newrgbcolor{Myred}{0.6 0 0}

\newrgbcolor{Myred}{0.6 0 0}
\newrgbcolor{Mylightred}{0.74 0.56 0.56}
\newrgbcolor{Mygreen}{0 0.5 0.1}
\newrgbcolor{Myblue}{0 0 0.6}
\newrgbcolor{Mybluee}{0.2 0.2 0.6}
\newrgbcolor{Myyellow}{1 .98 .85}
\newrgbcolor{Mygrey}{.39 .34 .53}
\newrgbcolor{Mylightgreen}{.56 .74 .56}
\newrgbcolor{Mybluegrey}{.7 .71 .9}
\newrgbcolor{Mybeige}{.89 .87 .8}
\newrgbcolor{Mylightgrey}{.86 .86 .86}
\newrgbcolor{Myorange}{1 .4 0}
\newrgbcolor{Mydarkgrey}{.26 .26 .26}
\newrgbcolor{Mywhite}{.99 .99 .99}
\newrgbcolor{Mydarkblue}{0 0 0.5}

\hypersetup{colorlinks,linkcolor=MyBlue,citecolor=MyGrey,filecolor=MyBlue, pdfborder={0 0 0},anchorcolor=MyBlue, urlcolor=MyBlue,pdfstartview=FitH}

\newtheorem{Def}{\color{black}Definition}
\newtheorem{Prop}{\color{black}Proposition}

\newtheorem{Thm}{\color{black}Theorem}
\newtheorem{Lem}{\color{black}Lemma}
\newtheorem{Cor}{\color{black}Corollary}
\newtheorem{Ass}{\color{black}Assumption}
\newtheorem{Proc}{\color{black}Procedure}
\makeatletter  
\def\@endtheorem{\hfill $\diamond$\endtrivlist\@endpefalse } 
\makeatother
\newtheorem{Ex}{\color{black}Example}

\theoremstyle{definition}

\newcommand{\theoremref}[1]{\hyperref[#1]{Theorem~\ref*{#1}}}
\newcommand{\obsref}[1]{\hyperref[#1]{Observation~\ref*{#1}}}
\newcommand{\lemmaref}[1]{\hyperref[#1]{Lemma~\ref*{#1}}}
\newcommand{\defref}[1]{\hyperref[#1]{Definition~\ref*{#1}}}
\newcommand{\corref}[1]{\hyperref[#1]{Corollary~\ref*{#1}}}
\newcommand{\propref}[1]{\hyperref[#1]{Proposition~\ref*{#1}}}
\newcommand{\assref}[1]{\hyperref[#1]{Assumption~\ref*{#1}}}
\newcommand{\remref}[1]{\hyperref[#1]{Remark~\ref*{#1}}}
\newcommand{\exref}[1]{\hyperref[#1]{Example~\ref*{#1}}}
\newcommand{\exsref}[1]{\hyperref[#1]{Examples~\ref*{#1}}}
\newcommand{\appref}[1]{\hyperref[#1]{Appendix~\ref*{#1}}}
\newcommand{\oappref}[1]{\hyperref[#1]{Online Appendix~\ref*{#1}}}
\newcommand{\secref}[1]{\hyperref[#1]{Section~\ref*{#1}}}
\newcommand{\subsecref}[1]{\hyperref[#1]{Subsection~\ref*{#1}}}
\newcommand{\figref}[1]{\hyperref[#1]{Figure~\ref*{#1}}}
\newcommand{\tableref}[1]{\hyperref[#1]{Table~\ref*{#1}}}
\newcommand{\procref}[1]{\hyperref[#1]{Procedure~\ref*{#1}}}
\newcommand{\algoref}[1]{\hyperref[#1]{Algorithm~\ref*{#1}}}
\newcommand{\remarkref}[1]{\hyperref[#1]{Remark~\ref*{#1}}}

\makeatletter
\renewcommand\paragraph{\@startsection{paragraph}{4}{\z@}%
                                    {3.25ex \@plus1ex \@minus.2ex}%
                                    {-1em}%
                                    {\normalfont\normalsize\bfseries\color{MyBlue}}}
\makeatother

\newcommand{\step}[1]{\medskip \noindent\textcolor{MyBlue}{\textbf{#1}}}
\newcommand{\stepnoskip}[1]{\noindent\textcolor{MyBlue}{\textbf{#1}}}

\DeclareMathOperator{\supp}{supp}

\DeclareMathOperator{\Exp}{\mathbbm{E}}

\DeclareMathOperator{\Ind}{\mathbbm{1}}

\AtAppendix{\counterwithin{Lem}{section}}
\AtAppendix{\counterwithin{Prop}{section}}
\AtAppendix{\counterwithin{Cor}{section}}
\AtAppendix{\counterwithin{Def}{section}}
\AtAppendix{\counterwithin{Thm}{section}}

\usepackage{xr}

\pgfplotsset{compat=1.18}

\makeatletter
\newcommand{\labeltext}[2]{%
  \@bsphack
  \MakeLinkTarget*{#1}%
  \def\@currentlabel{#1}{\label{#2}}%
  \@esphack
}
\makeatother

\begin{document}


\title{\textcolor{MyBlue}{Falsification-proof non-market allocation mechanisms }
\thanks{
We thank Ricardo Alonso, Julien Combe, Michael Ostrovsky, James Schummer, Philipp Strack, Andrzej Skrzypacz, Utku \"Unver and Alexander Wolitzky for thought-provoking discussions and suggestions. We also thank Xiao Lin for an outstanding discussion at the ``Theory at Penn State'' conference. Eduardo Perez-Richet acknowledges funding by the European Research Council (ERC) consolidator grant 101001694, and thanks the Fernand Braudel visiting program at the European University Institute for its hospitality. Vasiliki Skreta acknowledges funding by the European Research Council (ERC) consolidator grant  682417 ``Frontiers In Design" and by the NSF grant 2242521 ``Fraud-proof Mechanism Design.'' Francesco Conti, Nathan Hancart and Ignacio N\'u\~nez provided excellent research assistance at various stages. Earlier drafts of this paper were circulated under the title ``Fraud-proof non-market allocation mechanisms''.
}
\\
\textcolor{MyBlue}{(working paper version)}
}

\author{
 Eduardo \textsc{Perez-Richet}\footnote{Sciences Po, CEPR -- e-mail: {\tt eduardo.perez@sciencespo.fr} }
\and 
Vasiliki \textsc{Skreta}\footnote{UT Austin, UCL, CEPR -- e-mail: {\tt vskreta@gmail.com}}
}

\date{\today}

\maketitle \setcounter{page}{1}


\begin{abstract}
Relying on manipulable characteristics to target the allocation of resources is fraught with difficulties. Naively optimal rules lead to falsification and allocative inefficiency. Optimizing the rule for allocative efficiency while accounting for falsification typically induces 
falsification, at a heavy cost to the agents. Moreover, falsification generates negative externalities. We characterize optimal falsification-proof rules under a broad range of falsification technologies. We also provide an honesty condition under which they solve the unconstrained problem of a planner who maximizes a weighted sum of allocative surplus and agents' welfare.

\end{abstract}


\clearpage



\onehalfspacing
\newpage 

\section{Introduction}\label{sec_intro} 
Goods, services, and rewards\footnote{Goods include public housing, seats in schools and vaccines; services include credit, training, education and financial assistance programs; rewards include promotions, labels and certificates.} are often allocated via non-market mechanisms, either due to institutional constraints or 
because monetary transfers are ineffective at targeting deserving recipients.\footnote{See \cite{condorelli2013market} and \cite{akbarpour2020redistributive} for a theory of when non-market mechanisms are optimal.} To target eligible agents, non-market mechanisms must rely on data about their characteristics. For example, seats in schools are assigned using priorities that combine multiple criteria, green labels are awarded based on measured emissions, and public housing is allocated based on criteria such as household income. In many cases, eligibility is assessed through a \emph{score} that measures an agent's characteristics or performance and acts as a proxy for their \emph{worth}. 

However, reliance on the score creates strong incentives to game it. Consequently, practices such as falsification, forgery, greenwashing, teaching to the test, and manipulating statistics are commonplace. For 
example, parents fake addresses to gain admission to desirable public schools \citep{bjerrenielsen2023playing}, firms underreport their workforce size to avoid legal obligations \citep{IPP2022}, and doctors manipulate their patients' priority in organ transplant waiting lists 
\citep{lee2018manipulation, mcmichael2022stealing}.\footnote{\citet{schummer2021influencing} explores the impact of waiting list manipulations in a theoretical model.} Throughout the paper, we use the term \emph{falsification} as a broad category that encompasses all individually costly and socially wasteful activities that agents undertake to produce an altered score.

Falsification is not only wasteful; it can also be socially harmful. First, it may distort achievable allocations, unfairly penalizing agents for whom it is more costly to falsify.\footnote{For example, \cite{bjerrenielsen2023playing} shows that priority gaming in school choice mechanisms resulted in better assignments for those who engaged in such practices and adversely affected others.} Second, it can deteriorate the informational content of the score. This is an instance of Goodhart's law: ``when a measure becomes a target, it ceases to be a good measure.'' For example, greenwashing can blur the assessment of emissions levels. Third, it may make a mechanism politically unsustainable when evidence of falsification becomes 
public. Fourth, it can erode trust and deplete the supply of objects to allocate. For example, a scandal involving the manipulation of the German liver allocation system by transplant providers led to a 20\%-40\% erosion in organ donation \citep{lee2018manipulation}. Fifth, there is evidence that dishonest behavior spreads in society \citep[see, e.g.,][]{galbiati2012tax,ajzenman2018power}.

The main contribution of this paper is twofold. First, we characterize allocatively optimal falsification-proof allocation rules under a broad range of falsification technologies, encompassing \emph{decreasing} (\theoremref{ThOptUDD}) and \emph{increasing differences} (\theoremref{ThOptUID}). Falsification-proof allocation rules ensure \emph{honesty} by giving agents no incentive to falsify, thereby eliminating the private\footnote{\cite{milli2019social} show that optimal binary classifiers under falsification can generate a significant cost in falsification for the agents and argue that it bears disproportionately on disadvantaged groups.} and the social costs of falsification. However, requiring honesty has a price, since it constrains possible designs. Our characterization allows us to conduct a cost-benefit analysis of honesty, balancing its allocative price with its value to agents. Second, we show that under meaningful welfare concerns and increasing differences, falsification-proof rules are not merely constrained-optimal but fully optimal. We consider a designer whose objective is a weighted sum of allocative surplus and agents' welfare. We provide a sufficient condition for honesty to be optimal (\theoremref{thm:fp-optimal}). This honesty condition allows us to isolate the three factors that contribute to making honesty optimal in our model. The first is the weight on agents' welfare, which makes honesty optimal when it is sufficiently high. The remaining factors are high falsification costs and low dependency of an agent's expected worth on their score.

Specifically, we study the problem of allocating a mass of homogeneous prizes to a heterogeneous population of agents using non-market mechanisms based on \emph{scores}. The score is a publicly available but \emph{falsifiable} measure of an agent's private characteristics. If agents do not falsify, they produce their \emph{natural score}, which reflects their true characteristics.\footnote{\cite{frankel2019muddled} introduced the term \emph{natural} action to refer to the unmanipulated action of a 
given type.} Agents who \emph{falsify} produce an altered score at a cost. The falsification cost can exhibit either increasing or decreasing differences in the natural and altered score. We assume that an agent's natural score is positively correlated with their \emph{worth}: the designer's value of assigning a prize to that agent. We define \emph{allocative surplus} as the aggregate worth of rewarded agents. 



Our characterization results make a technical contribution. The constrained optimization problem can be solved with a standard first-order approach under decreasing differences since falsification-proofness constraints bind locally. Under increasing differences, however, the falsification-proofness constraints bind for distant scores. To address this difficulty, we transform the designer's problem into a program that is equivalent to the dual of the classical Monge-Kantorovich optimal transport problem, which we leverage to characterize the optimal allocation rule in \theoremref{ThOptUID}.

What is the welfare impact of inducing honesty? In standard mechanism design, misreporting is costless by assumption, and honesty is inconsequential due to the revelation principle. In contrast, we show in \cite{perez2022test} that under costly falsification with increasing differences, allocatively optimal mechanisms harness falsification to increase allocative efficiency.\footnote{\cite{perez2022test} derive optimal \emph{tests}, which correspond to optimal allocation mechanisms in the  current framework.} Therefore, honesty has a price: requiring falsification-proofness results in a loss of allocative surplus for the designer. In return, we show that honesty also has value for agents. Regardless of their natural score, agents are better off under the optimal falsification-proof mechanism than under the unconstrained optimal mechanism (\propref{PropWelfFP}). Furthermore, we show that this (allocative) price of honesty can be arbitrarily small and the value of honesty (for agents) arbitrarily large when the falsification cost has increasing differences (\propref{PropWelfTradeOff}). These results highlight that, even when abstracting from externalities, falsification-proof mechanisms can enhance agents' welfare at a small allocative price. 

We also study how changes in \emph{gaming ability} affect optimal falsification-proof rules. Gaming ability is defined as the inverse of a multiplicative scaling factor on the cost of falsification. We find that its effect on the allocation rule is nuanced: it benefits low-score agents and harms high-score agents when gaming ability is sufficiently low, whereas the effect becomes uniform at higher levels, with its direction depending on score priority. 

This is a working paper version of the paper with additional results and extensions of the model.  In \secref{sec:FixedCost}, we extend the analysis to falsification cost functions with a fixed cost component under increasing differences. In \secref{sec:MGsetup}, we develop an extended model with multiple identifiable groups, additional sources of heterogeneity across agents, and exogenous quotas. Finally, in \appref{app-sec-comp-stat}, we provide additional comparative statics results with respect to the score distribution and returns to scale in the falsification technology.

\section{The allocation problem}\label{sec:model}
\paragraph{Framework.} The designer allocates a mass $\bar{\rho}\leq 1$ of indivisible and homogeneous prizes to a unit mass of heterogeneous agents without transfers. In many applications, a resource constraint is irrelevant because prizes are immaterial, such as services, certificates, labels, or awards. In this paper, we set $\bar{\rho}=1$. We analyze the problem with a resource constraint $\bar{\rho}<1$ in \secref{sec:MGsetup}.

An exogenous one-dimensional metric, the \emph{score}, measures certain private characteristics of the agent. Each agent has a private type $s\in S \subseteq \mathbbm{R}$, corresponding to their \emph{natural score}---the score they obtain when they do not interfere with the measuring technology. Agents are also characterized by their worth $w$, a scalar capturing the designer's value of allocating a prize to them.  For simplicity, we assume that all agents have a homogeneous value normalized to 1 for a prize,\footnote{In \secref{sec:MGsetup}, we extend our analysis to heterogeneous private values.} and that they do not observe their worth.\footnote{In fact, this is irrelevant since we argue in \secref{sec:MGsetup} that the designer could not elicit this information through the mechanism.}
 

\paragraph{Distributional assumptions.} The joint population distribution of natural scores and realized worth is a Borel probability measure $P$ on $S\times\mathbb{R}$, where $S=[\underline{s},\overline{s}]$ and $-\infty<\underline{s}<\overline{s}<\infty$. For a representative agent, write $(\tilde{s},\tilde{w})\sim P$. The score marginal has an atomless distribution with full support on $S$ and its cumulative distribution function is denoted by $F$. We assume $\Exp[|\tilde{w}|]<\infty$ and that the conditional expectation of worth given the natural score admits a continuous, strictly increasing version $w:S\to\mathbb{R}$, so that $w(\tilde{s})=\Exp[\tilde{w}\mid\tilde{s}]$ almost surely. We denote average worth by $\bar{w}=\Exp[\tilde{w}]=\int_S w(s)\,dF(s)$. No density is assumed for $F$, and no further restrictions are imposed on the conditional distribution of realized worth given the natural score. 


\paragraph{Designer and agents' payoffs.} 
The designer's expected payoff from allocating a prize to an agent with natural score $s$ is $w(s)$. The payoff from the outside option of not allocating is normalized to zero. The monotonicity assumption means that higher natural scores predict higher expected worth. We restrict attention to the nontrivial case $w(\underline{s})<0<w(\overline{s})$. Thus, there is a unique interior \emph{eligibility threshold} $\hat{s}$ satisfying $w(\hat{s})=0$. A score is \emph{eligible} if it yields a nonnegative worth, $s \geq \hat{s}$, and \emph{ineligible} otherwise, $s < \hat{s}$. 

Agents can \emph{falsify} their score at a cost $c(t|s)\geq 0$ to produce a \emph{falsified score} $t$ instead of their natural score $s$. The expected payoff of such an agent receiving an object with probability $\alpha$ is $\alpha-c(t|s)$. We say that agents producing their natural score are being \emph{honest}. Honesty is costless, so $c(s|s)=0$. The falsification cost may reflect technical costs, psychological lying costs, as well as expected penalties.  


\paragraph{Priority.} 
The setup exhibits \emph{high-score priority} if the expected worth exceeds the outside option, $\bar{w}>0$; \emph{low-score priority} if $\bar{w} < 0$; and \emph{neutral priority} if $\bar{w} = 0$. In the absence of information on scores, the designer would allocate prizes to all agents under high-score priority but would not allocate any prizes under low-score priority. We show that priority determines whether it is more important for the designer to ensure that the top score receives a prize with certainty (high-score priority) or that the lowest score does not receive a prize (low-score priority). 


\paragraph{Falsification cost.} We assume that the cost function is \emph{monotonic} for \emph{upward} falsification: if $t\geq s$, then $c(t|s)$ is strictly increasing in $t$ and decreasing in $s$. We also assume that the cost function satisfies a \emph{regularity} assumption.

\begin{Ass}[Regularity]\label{def:reg}
The cost function $c(t|s)$ is continuously differentiable in $t$ on $[s,\overline{s}]$, and in $s$ on $[\underline{s},t]$.
\end{Ass}

We denote the partial derivatives of a regular cost function $c(t|s)$ by $c_t$ and $c_s$. Depending on the context, the cost function may take different forms, so it is useful to rely on flexible assumptions. We characterize optimal allocation rules for the following two salient classes of cost functions.

\begin{Def}[Upward Differences]\label{def:id}
A cost function $c(t|s)$ has upward increasing differences if, for all  $s<s'\leq t<t'$,
\begin{equation}\label{eq:uid}
c(t'|s')-c(t|s')\geq ~c(t'|s)-c(t|s),\tag{UID}
\end{equation}
and upward decreasing differences if, for all  $s<s'\leq t<t'$,
\begin{equation}\label{eq:udd}
c(t'|s')-c(t|s')\leq c(t'|s)-c(t|s).\tag{UDD}
\end{equation}
\end{Def}
These conditions concern only upward falsification because we show that downward falsification is never beneficial under optimal allocation rules. To gain intuition about their interpretation, it is useful to consider different families of cost functions.

\begin{Ex}[The Euclidean family]\label{ExEucl} 
A cost function is \emph{Euclidean} if $c(t|s) = \mathcal{C}(t - s)$ for $t\geq s$, where $\mathcal{C}: \mathbb{R}_+ \to \mathbb{R}_+$ is a continuously differentiable increasing function with $\mathcal{C}(0) = 0$. Euclidean costs are a convenient modeling choice often used in the literature. They satisfy \eqref{eq:uid} if $\mathcal{C}$ is concave and \eqref{eq:udd} if $\mathcal{C}$ is convex. The monotonicity of upward differences captures economies of scale in falsification $t - s$. \eqref{eq:uid} implies increasing returns to scale, while \eqref{eq:udd} implies decreasing returns to scale. Both cases may arise in different contexts. For instance, emission testing software that becomes cheaper to deploy once
the initial investment is made yields increasing returns whereas memorizing increasingly difficult test problems yields decreasing returns.
Another example is when falsification costs arise from a linear detection probability $\pi x$ and a fine $\varphi x + \varphi_0$, both of which increase with the falsification level $x$. In this case, the Euclidean cost $\mathcal{C}(x) = \pi x (\varphi x + \varphi_0)$ satisfies \eqref{eq:udd}. 
\end{Ex}

\begin{Ex}[The shifted Euclidean family]\label{ExShftLin}  We generalize the Euclidean class by considering \emph{shifted Euclidean} cost functions of the form $c(t|s) = \kappa(s)\mathcal{C}(t - s)$ for $t\geq s$. This family captures correlations between the natural score and \emph{gaming ability} $1/\kappa(s)$. For instance, consider the \emph{shifted linear} cost function where $\mathcal{C}(x) = x$. In this case, $c$ satisfies \eqref{eq:uid} if $\kappa(s)$ increases with $s$, and \eqref{eq:udd} if $\kappa(s)$ decreases with $s$. If the natural score reflects financial need or hardship, it is natural to assume that gaming becomes more challenging at higher natural scores. Conversely, if the natural score reflects skill or aptitude, gaming may become easier at higher natural scores.
\end{Ex}

The \eqref{eq:uid} class is particularly suited for studying situations in which falsification involves substantial setup costs.\label{page-fixed} Although our assumptions rule out a fixed cost, we extend our analysis to fixed costs in \secref{sec:FixedCost}. Furthermore, we derive the unconstrained optimal allocation rule under \eqref{eq:uid} in \cite{perez2022test}, allowing us to analyze the welfare consequences of imposing falsification-proofness  in \secref{sec-econ-impl}. 


\paragraph{Falsification-proof allocative surplus optimization.} We are interested in mechanisms that incentivize agents to produce their natural score: \emph{falsification-proof} mechanisms. We show in \cite{perez2023score} that it is without loss of generality to restrict attention to \emph{score-based allocation rules} $\alpha: S \to [0,1]$, where $\alpha(t)$ is the probability that a prize is allocated to an agent who produces score $t$. Then, a mechanism is falsification-proof if and only if it satisfies the constraint\footnote{We can also interpret \eqref{eqFPC} as being motivated by inequality concerns. The cost then acts as a bound on inequality between score pairs. We thank Ricardo Alonso for suggesting this interpretation.} 
\begin{equation}\label{eqFPC}
\forall (s,t),~~ \alpha(t)-\alpha(s)\leq c(t|s).\tag{FPC}
\end{equation}  
Selecting the \emph{honest} best response of the agents, we can then express the designer's problem as
\begin{equation}\label{eqBP}
W(F,c)=\max_{\alpha}\int_S\alpha(s)w(s)dF(s)~~\text{s.t.}~~\eqref{eqFPC},\eqref{eqPC}\tag{P}
\end{equation}
with the \emph{probability constraint} (PC)\labeltext{PC}{eqPC}: $0\leq \alpha(s)\leq 1$ for all $s\in S$.

The objective of the designer in this problem is to maximize \emph{aggregate allocative surplus} (or simply allocative surplus), $\int_S \alpha(s)w(s)dF(s)$.


\paragraph{The weighted-welfare design problem.} We also consider the design problem of a social planner weighing both allocative surplus and agents' welfare:

\[
\underbrace{\iint\alpha(t)w(s)\,d\phi(t|s)dF(s)}_{\text{allocative surplus}}
+\zeta \underbrace{\iint[\alpha(t)-c(t|s)]\,d\phi(t|s)dF(s)}_{\text{agents' welfare}}
\]
where $\phi$ is the falsification strategy of the agents, and $\zeta\geq 0$ is the weight on agents' welfare. Collecting terms yields the objective function
\[\Omega(\alpha,\phi,\zeta)=\iint\alpha(t)\bigl\{w(s)+\zeta\bigr\}\,d\phi(t|s)dF(s)-\zeta\iint c(t|s)\,d\phi(t|s)dF(s).\]
Hence, the weighted welfare objective is equivalent to an allocative 
surplus problem with an upward shifted worth function $w(s)+\zeta$ and a weight $\zeta$ on falsification costs. The weighted welfare design problem is: \footnote{We show in \cite{perez2023score} that restricting attention to score-based rules is without loss of generality among all mechanisms that make deterministic falsification recommendations. Falsification-proof mechanisms make deterministic recommendations by definition; however, this is not necessarily the case for general unconstrained mechanisms.\label{footref-mech-class}}

\begin{align}\label{pgm:wwp}
    \max_{\alpha,\phi}~&\Omega(\alpha,\phi,\zeta)\tag{WWP}\\
    \text{s.t.}~&\eqref{eqPC}\notag\\
    &\phi\bigl(T(\alpha,s)|s\bigr)=1,~~\forall s.\label{eq:IC}\tag{IC}
\end{align}
The incentive constraint \eqref{eq:IC} requires the agents' falsification strategy $\phi:S\to\Delta(S)$ to be concentrated on the set of \emph{optimal falsification targets}
\[T(\alpha,s)=\arg\max_{t}\,\alpha(t)-c(t|s).\]

Clearly, if $\zeta\geq \bar{\zeta}=-w(\underline{s})$, the solution to the weighted welfare problem is to allocate prizes unconditionally, and the agents have no incentive to falsify. In \secref{sec:general fals}, we show that in the \eqref{eq:uid} case, there is a range of weights $\zeta<\bar{\zeta}$ such that the optimal rule for \eqref{pgm:wwp} is falsification proof if the falsification cost is sufficiently high, $c(\hat{s}|\underline{s})>1$. In this case, the solution to \eqref{pgm:wwp} is given by the solution to \eqref{eqBP} with the upward-shifted worth function $w(s)+\zeta$.

\section{Allocatively optimal falsification-proof rules}\label{sec-baseline}

The first-best rule is a step function that allocates to all eligible scores $s\geq\hat{s}$ and to none below. It is not falsification-proof, since an agent just below $\hat{s}$ can gain by falsifying upward. More generally, \eqref{eqFPC} bounds the growth of $\alpha$ between any two scores.  The analysis proceeds as follows. Monotonicity of optimal rules (\lemmaref{lem:mon}) allows us to rewrite allocative surplus via integration by parts in terms of \emph{cumulative surplus}---the total surplus generated by scores above or below any given threshold. The probability constraint caps total growth at $1$, and its Lagrange multiplier $\nu$ pins down the \emph{growth interval} $[s_*,s^*]$, over which $\alpha$ rises from its minimum to its maximum value, as a \emph{Zero Average Surplus} interval \eqref{eqZAS}. We can solve \eqref{eqBP} by first solving a \emph{simplified program} that takes this interval as given, then identifying the correct interval via a \emph{boundary condition} encoding the interplay between falsification costs and the probability constraint. 



\subsection{Cumulative surplus}\label{sec:cumsurpl}

We introduce \emph{cumulative surplus} functions that appear when integrating allocative surplus by parts. They play a central role in the analysis, as they determine the nature of growth intervals for optimal falsification-proof rules.

\paragraph{Cumulative surplus functions.} The \emph{upward cumulative surplus} at $s$ is the total amount of allocative surplus generated by scores above $s$. It corresponds to the marginal gain of uniformly increasing the allocation probability of all such scores:
\begin{align*}
    \mathcal{W}^+(s)=\int_s^{\overline{s}}w(x)dF(x).
\end{align*}

\begin{figure}[!h]
\begin{center}
\scalebox{.8}{\begin{tikzpicture}[
declare function={J(\x)=(1/4)*(\x+(1/2))*(1-\x);},
]
\begin{axis}[
    domain=-1:1, xmin=-1,ymax=9/64,ymin=-.05,xmax=1,
    yscale=4,xscale=0.7,
    axis lines=middle,
    axis equal image,
    xtick={-1,0,1}, ytick={0},
    ticklabel style={font={\small}},
    enlargelimits=true,
    clip mode=individual, clip=false,
]

\addplot[mark=none, dotted] coordinates {(0.25, 0) (0.25, 9/64)}; 
\node[below=2pt]   at (axis cs:0.25,0) {$\scriptstyle \hat{s}$};
\addplot[Mygreen,thick,samples=600]{J(x)};
\node[above=2pt] at (axis cs:-0.4,.12) {\small $\textcolor{Mygreen}{\mathcal{W}=\mathcal{W}^+}$};

\addplot[domain=-1:1,purple,thick]{.09};
\node[above=2pt] at (axis cs:-0.8,0.075) {\small $\textcolor{purple}{\nu}$};
\addplot[mark=none, dotted] coordinates {(-0.2, 0) (-0.2, .09)};
\node[below=2pt]   at (axis cs:-0.2,0) {$\scriptstyle s_*(\nu)$};
\addplot[mark=none, dotted] coordinates {(0.7, 0) (0.7, .09)}; 
\node[below=2pt]   at (axis cs:0.7,0) {$\scriptstyle s^*(\nu)$};
\addplot[domain=-0.2:0.7,Myblue,line width=1.5pt]{0};

\node[circle,draw,brown,fill=brown,inner sep=2pt] at (axis cs:-0.2,0) (s) {};
\node[circle,draw,brown,fill=brown,inner sep=2pt] at (axis cs:0.7,0) (m) {};
\draw[-stealth,thick,brown] (s) to [out=10,in=170]  (m) ;
\node[above=1pt] at (axis cs:.25,.03) {$\scriptstyle \textcolor{brown}{m(\cdot)}$};

\node[rectangle] at (axis cs:.25,-.15) {$\gray \scriptstyle F=U([-1,1]),~w(s)=s-1/4$}; 

\node[rectangle] at (axis cs: 0,.25) {\small \underline{Low-score priority}};

\end{axis}
\end{tikzpicture}
\hspace{1cm}
\begin{tikzpicture}[
declare function={J(\x)=-(1/4)*(\x-(1/2))*(\x+1);},
]
\begin{axis}[
    domain=-1:1, xmin=-1,ymax=9/64,ymin=-.05,xmax=1,
    yscale=4,xscale=0.7,
    axis lines=middle,
    axis equal image,
    xtick={-1,0,1}, ytick={0},
    ticklabel style={font={\small}},
    enlargelimits=true,
    clip mode=individual, clip=false,
]

\addplot[mark=none, dotted] coordinates {(-0.25, 0) (-0.25, 9/64)}; 
\node[below=2pt]   at (axis cs:-0.25,0) {$\scriptstyle \hat{s}$};
\addplot[Mygreen,thick,samples=600]{J(x)};
\node[above=2pt] at (axis cs:0.45,.12) {\small $\textcolor{Mygreen}{\mathcal{W}=\mathcal{W}^-}$};

\addplot[domain=-1:1,purple,thick]{.09};
\node[above=2pt] at (axis cs:-0.8,0.075) {\small $\textcolor{purple}{\nu}$};
\addplot[mark=none, dotted] coordinates {(0.2, 0) (0.2, .09)};
\node[below=2pt]   at (axis cs:0.2,0) {$\scriptstyle s^*(\nu)$};
\addplot[mark=none, dotted] coordinates {(-0.7, 0) (-0.7, .09)}; 
\node[below=2pt]   at (axis cs:-0.7,0) {$\scriptstyle s_*(\nu)$};
\addplot[domain=-0.7:0.2,Myblue,line width=1.5pt]{0};

\node[circle,draw,brown,fill=brown,inner sep = 2pt] at (axis cs:-0.7,0) (s){};
\node[circle,draw,brown,fill=brown, inner sep = 2pt] at (axis cs:0.2,0) (m){};
\draw[-stealth,thick,brown] (s) to [out=10,in=170]  (m) ;
\node[above=1pt] at (axis cs:-.25,.03) {$\scriptstyle \textcolor{brown}{m(\cdot)}$};

\node[rectangle] at (axis cs:-.25,-.15) {$\gray \scriptstyle F=U([-1,1]),~w(s)=s+1/4$}; 

\node[rectangle] at (axis cs: 0,.25) {\small \underline{High-score priority}};
\end{axis}
\end{tikzpicture}}
\caption{Cumulative surplus, matching pairs and growth interval.}\label{FigCumSurpl}
\end{center}
\end{figure}
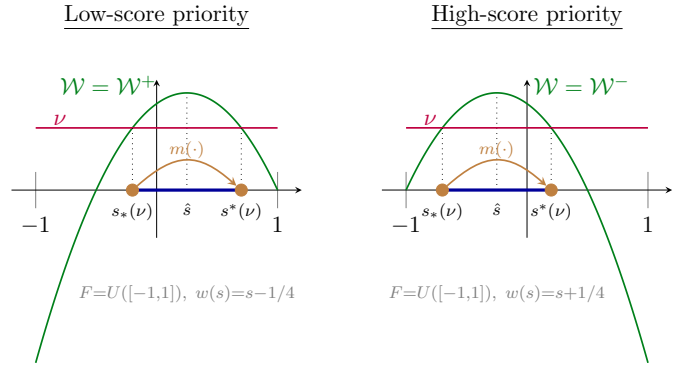

The \emph{downward cumulative surplus} at $s$ is the total amount of negative surplus generated by scores below $s$, and corresponds to the marginal gain of uniformly decreasing the allocation probability of all such scores:
\begin{align*}
   \mathcal{W}^-(s)=-\int_{\underline{s}}^{s}w(x)dF(x)=\mathcal{W}^+(s)-\bar{w}.
\end{align*}
In the analysis, we work with upward cumulative surplus under low-score priority and with downward cumulative surplus under high-score priority. To unify notation, we define a composite \emph{cumulative surplus function}:
\begin{align*}
  \mathcal{W}(z) = \mathcal{W}^+(z)\Ind_{\bar{w} \leq 0} + \mathcal{W}^-(z)\Ind_{\bar{w} > 0}.  
\end{align*}
We next list some key properties of $\mathcal{W}$ illustrated in \figref{FigCumSurpl}.

\begin{Lem}[Properties of cumulative surplus]\label{LemCumSurpl}~
(i) $\mathcal{W}(\cdot)$ is continuous and single-peaked at $\hat{s}$; (ii) For every $\nu\in[0,\mathcal{W}(\hat{s})]$, there exists a unique pair of scores $s_*(\nu)\leq \hat{s}\leq s^*(\nu)$ such that $\mathcal{W}\bigl(s_*(\nu)\bigr)=\mathcal{W}\bigl(s^*(\nu)\bigr)=\nu$; (iii) $\mathcal{W}(s)\geq \nu$ if and only if $s\in\bigl[s_*(\nu),s^*(\nu)\bigr]$; (iv) $s_*(\nu)$ and $-s^*(\nu)$ are continuous and increasing functions; (v) $s^*(0)=\overline{s}$ under low-score priority, and $s_*(0)=\underline{s}$ under high-score priority. Under neutral priority, $[s_*(0),s^*(0)]=[\underline{s},\overline{s}]$.
\end{Lem}

\paragraph{Matching ineligible and eligible scores.} We use cumulative surplus to define a \emph{matching function} between ineligible and eligible scores, as illustrated in \figref{FigCumSurpl}. Specifically, we define the decreasing matching function $m: [s_*(0), \hat{s}] \to [\hat{s}, s^*(0)]$ that maps each ineligible score $s \in [s_*(0), \hat{s}]$ to an eligible score $m(s) \in [\hat{s}, s^*(0)]$ such that $\mathcal{W}(s) = \mathcal{W}\bigl(m(s)\bigr)$. We say that a pair $(s_*, s^*)$ is a \emph{matching pair} if $s^* = m(s_*)$. 

The matching function plays an important role in the analysis for two reasons. First, it describes the (nonlocal) binding constraints in the \eqref{eq:uid} case. Second, together with the Lagrange multiplier $\nu$ on the probability constraint, it determines the \emph{growth interval} of the allocation rule.
 
Point (ii) of \lemmaref{LemCumSurpl} implies that matching pairs characterize a set of intervals $[s_*, s^*]$ around the eligibility threshold $\hat{s}$ that satisfy the following \emph{Zero Average Surplus} condition:
\begin{equation}\label{eqZAS}
\int_{s_*}^{s^*} w(s)dF(s) = 0.\tag{ZAS}
\end{equation}
\eqref{eqZAS} intervals form a nested family of intervals around the eligibility threshold, ranging from $[s_*(0), s^*(0)]$ when $\nu=0$ to $\{\hat{s}\}$ when $\nu=\mathcal{W}(\hat{s})$. 

We show in the next subsection that the growth interval of an optimal rule is necessarily a \eqref{eqZAS} interval. This implies that optimal rules exhibit bunching at the bottom under low-score priority since scores in $[\underline{s}, s_*(0)]$ receive a prize with zero probability. In contrast, there is bunching at the top under high-score priority since scores in $[s^*(0), \overline{s}]$ receive a prize with probability one. 

\subsection{Simplifying the program}\label{sec:simplpgm}

\paragraph{Monotonicity of optimal rules.} We can, without loss of optimality, restrict attention to monotonic allocation rules. Intuitively, replacing a nonmonotonic rule with its lower monotonic envelope below the eligibility threshold and its upper monotonic envelope above the eligibility threshold increases the probability that eligible scores receive a prize while decreasing the probability that ineligible scores do. Furthermore, this transformation preserves falsification-proofness. 

\begin{Lem}[Monotonicity]\label{lem:mon}
If $\alpha$ is feasible for \eqref{eqBP} but not monotonic, there exists a monotonic and feasible rule $\tilde{\alpha}$ that strictly increases allocative surplus.
\end{Lem}

We refer to the monotonicity condition on $\alpha$ as (MON).\labeltext{MON}{eq:mon} It is also useful to note that \eqref{eqFPC} implies that any feasible allocation rule must be continuous. 

\paragraph{Differential program.} Relying on the monotonicity of $\alpha$, we can use integration by parts for Stieltjes integrals to rewrite allocative surplus as a function of either downward or upward cumulative surplus:
\begin{align}
\int_S\alpha(s)w(s)dF(s)&=\bar{w}\underline{\alpha}+\int_S\mathcal{W}^+(s)\,d \alpha(s)\tag{$\underline{\text{DOF}}$}\label{eq:objdn}\\
&=\bar{w}\overline{\alpha} + \int_S \mathcal{W}^-(s) \, d\alpha(s).\tag{$\overline{\text{DOF}}$}\label{eq:objup}  
\end{align}

\eqref{eq:objup} implies that it is optimal to set $\overline{\alpha} = 1$ under high-score priority. Intuitively, an optimal rule must ensure that the top score receives a prize with certainty. \eqref{eq:objdn} implies that it is optimal to set $\underline{\alpha} = 0$ under low-score priority. In this case, an optimal rule must ensure that the lowest score does not receive a prize. The \emph{initial condition} that an optimal allocation rule must satisfy is therefore
\begin{equation}\label{eq:init}
\underline{\alpha}=0 \text{ if }\bar{w}<0;~~~\overline{\alpha}=1 \text{ if }\bar{w}>0;~~~  0\leq \underline{\alpha}\leq\overline{\alpha}\leq 1  \text{ if } \bar{w}=0.\tag{Init}
\end{equation}
If an allocation rule satisfies \eqref{eq:init} and \eqref{eq:mon}, we can rewrite the probability constraint \eqref{eqPC} in the differential form (DPC):\labeltext{DPC}{eq:DPC} $\int_S d\alpha(s)\leq 1$.
We can then write a common program for all priorities using the composite cumulative surplus function $\mathcal{W}$ from \secref{sec:cumsurpl}.

\begin{Lem}[Differential program]\label{lem:diffpgm}
An allocation rule $\alpha$ solves \eqref{eqBP} if and only if $\alpha$ satisfies \eqref{eq:init} and solves the differential program
\begin{equation}\label{pgm:DP}
\max_\alpha~\int_S\mathcal{W}(s)d\alpha(s)~~~\text{\emph{s.t.}}~\eqref{eq:DPC},\eqref{eqFPC}, \eqref{eq:mon}. \tag{DP}
\end{equation}
\end{Lem}

\paragraph{Probability constraint and growth interval.} 
The probability constraint \eqref{eq:DPC} bounds the total growth of the allocation rule and therefore determines the growth interval. Let $\nu \geq 0$ be the Lagrange multiplier on this constraint. The Lagrangian of the differential program is then
\[
\mathcal{L}(\alpha, \nu) = \int_{\underline{s}}^{\overline{s}} \bigl\{\mathcal{W}(z) - \nu\bigr\} d\alpha(z) + \nu.
\]
By \lemmaref{LemCumSurpl}(iii), maximizing this Lagrangian under \eqref{eq:mon} implies that $\alpha$ should put no mass outside $\bigl[s_*(\nu), s^*(\nu)\bigr]$, where the cumulative surplus is below $\nu$ (see \figref{FigCumSurpl}). Growth intervals are therefore \eqref{eqZAS} intervals and are contained within the maximal interval $[s_*(0), s^*(0)]$. 

\paragraph{The simplified program.} We construct a solution to the designer's problem by first solving the \emph{simplified program}
\begin{align*}
\max_{\alpha}~\int_{s_*}^{s^*} w(s)\alpha(s) dF(s) ~~~~
\text{s.t.}~~~\alpha(t) - \alpha(s) \leq c(t|s), \quad \forall s_* \leq s < t \leq s^*,
\end{align*}
which restricts the objective function to a \eqref{eqZAS} interval $[s_*, s^*]$, only imposing \eqref{eqFPC} over this interval and relaxing the probability constraint. We then obtain the optimal rule via the following procedure. 

\begin{Proc}\label{Proc1}~
\begin{itemize}
\item \textbf{Step 1.} Solve the simplified program for all $[s_*, s^*]$. The solution $\alpha$ is determined up to an additive constant.
\item \textbf{Step 2.} Select $(s_*, s^*)$ to saturate the probability constraint: $\alpha(s^*) - \alpha(s_*) = 1$ if possible. Otherwise set $(s_*, s^*) = \bigl(s_*(0), s^*(0)\bigr)$.
\item \textbf{Step 3.} Set the additive constant so that $\alpha(s_*) = 0$ under low-score priority, $\alpha(s^*) = 1$ under high-score priority, or to any value such that $\alpha(s^*)\leq 1$ and $\alpha(s_*) \geq 0$ under neutral priority.
\item \textbf{Step 4.} Set $\alpha(s) = \alpha(s_*)$ for $s < s_*$, and $\alpha(s) = \alpha(s^*)$ for $s > s^*$.
\end{itemize}
\end{Proc}
\lemmaref{LemLagSuff} in \appref{ssecBP} shows formally why \procref{Proc1} yields the solutions of \eqref{eqBP}, adapting standard Lagrangian necessity and sufficiency results to our setup. Step 1 determines the growth rate and shape on the growth interval. Step 2 identifies the growth interval and works like a complementary slackness condition: either the probability constraint is not binding, in which case its Lagrange multiplier is $\nu=0$ and the growth interval is therefore $\bigl[s_*(0), s^*(0)\bigr]$, or it is binding, so the growth interval must satisfy $\alpha(s^*) - \alpha(s_*) = 1$, the Lagrange multiplier satisfies $\nu \geq 0$, and is determined by the binding condition. Step 3 sets the boundary allocation probability based on priority. Step 4 extends the rule beyond the growth interval by assigning minimum probability to scores below $s_*$ and maximum probability to scores above $s^*$.

\subsection{Optimal rules: characterization}\label{sec:uid}

The shape of the optimal rule on the growth interval is determined by the binding falsification-proofness constraints. Under \eqref{eq:udd}, these constraints bind locally, allowing for a first-order approach. In contrast, under \eqref{eq:uid}, the constraints bind non-locally, rendering the first-order approach inadequate. We therefore solve the simplified program by establishing a connection with the dual of the Monge-Kantorovich optimal transport problem. 

\paragraph{Optimal rule under \eqref{eq:udd}.} Under \eqref{eq:udd}, any monotonic allocation rule that satisfies \eqref{eqFPC} must be absolutely continuous and satisfy the necessary first-order condition of the agent's problem: at every $s$ where the derivative $\alpha'(s)$ exists, $\alpha'(s)\leq c_{t^+}(s|s)$, where $c_{t^+}(s|s)$ is the right derivative of $c$ with respect to the target $t$, evaluated at $t=s$. Integrating by parts the objective function of the simplified program and using the absolute continuity of $\alpha$, we can rewrite the simplified program in the following differential form:
\[\max_{\alpha'(s)\leq c_{t^+}(s|s)}~\int_{s_*}^{s^*}\bigl(\mathcal{W}(s)-\nu\bigr)\alpha'(s)ds,\]
where $\nu=\mathcal{W}(s_*)=\mathcal{W}(s^*)$. Since $\mathcal{W}(s)-\nu> 0$ on the interior of $[s_*,s^*]$, the unique solution is to set $\alpha'(s)=c_{t^+}(s|s)$ for almost every $s$. 
 
Following step 2 of \procref{Proc1}, the growth interval $[s_*,s^*]$ of the optimal rule is determined by the \emph{matching condition} $s^*=m(s_*)$ and the \emph{boundary condition}
\begin{equation}
s_*=\min\left\{s\in[s_*(0),\hat{s}]:\int_{s}^{m(s)}c_{t^+}(x|x)dx\leq 1 \right\}. \tag{B}
\end{equation}

The corresponding Lagrange multiplier on the probability constraint is $\nu = \mathcal{W}(s_*)$. In accordance with step 2 of \procref{Proc1}, the boundary condition makes the probability constraint bind whenever possible and otherwise selects the maximal growth interval $[s_*(0), s^*(0)]$, with $\nu = 0$. Whether the constraint binds depends on the magnitude of falsification costs, and its degree of slackness is measured by the \emph{probability gap}
$\Gamma_{udd} = 1 - \int_{s_*}^{s^*} c_{t^+}(x|x) dx$.
The probability constraint is slack, and $\Gamma_{udd}>0$, if and only if the \emph{slackness condition} $\int_{s_*(0)}^{s^*(0)} c_{t^+}(x|x) dx<1$ is satisfied.

The probability gap must be allocated according to priority. In the low-score priority case, it is optimally withheld from agents to ensure $\alpha\bigl(s_*(0)\bigr) = 0$, thereby preventing low-score agents from receiving a prize. In the high-score priority case, it is optimally assigned to agents to ensure $\alpha(s^*(0)) = 1$, so that high-score agents receive a prize with certainty. To capture this choice, we define the \emph{share index}
\[I(r) = \mathbbm{1}_{\bar{w}>0} + r \mathbbm{1}_{\bar{w}=0},\]
which represents the share of the probability gap allocated to agents: it equals 0 under low-score priority and 1 under high-score priority. We refer to $r$ as the \emph{neutral gap share}. 

Under neutral priority, the interval $[s_*(0), s^*(0)]$ must be $[\underline{s}, \overline{s}]$, reducing the gap condition to the slackness condition $\int_{\underline{s}}^{\overline{s}} c_{t^+}(x|x) dx<1$. The designer is then indifferent about the share $r \in [0, 1]$ of the probability gap allocated to agents, and the optimal rule is not unique. Multiplicity therefore arises under neutral priority when the slackness condition holds:
\begin{align}
  \bar{w}=0 \text{ and } \int_{\underline{s}}^{\overline{s}}c_{t^+}(x|x)dx < 1. \tag{Mult}
\end{align}

Although the designer is indifferent, the choice of $r$ plays an important role in solving the problem under a resource constraint (see \secref{sec:MGsetup}). Indeed, choosing $r$ allows the designer to adjust the total mass of prizes accruing to agents in order to satisfy the constraint.

Consequently, the optimal rule under \eqref{eq:udd} is 
\[
\alpha^*_{udd}(s,r)=
\begin{cases}
0&\text{if }s< s_*\\
\Gamma_{udd} I(r)+\int_{s_*}^sc_{t^+}(x|x)dx &\text{if }s\in[s_*,s^*]\\
1&\text{if } s>s^*
\end{cases},
\]
where $(s_*,s^*)$ satisfies the matching and boundary conditions. Note that the conditions $s<s_*$ and $s>s^*$ are vacuous whenever $s_*=\underline{s}$ and $s^*=\overline{s}$ (respectively) so the allocation rule is always continuous.

\begin{Thm}[Optimal rule under \ref{eq:udd}]\label{ThOptUDD}
Suppose the cost function satisfies \eqref{eq:udd}. If \eqref{eq:mult} does not hold, then $\alpha^*_{udd}$ is independent of $r$ and is the unique solution of \eqref{eqBP}. If \eqref{eq:mult} holds, then the set of optimal rules is $\bigl\{\alpha^*_{udd}(\cdot,r)\bigr\}_{r\in[0,1]}$. 
\end{Thm}

Note that $\alpha^*_{udd}$ is flat if $c_{t^+}(x|x)=0$ for almost every $x$, that is, if marginal falsification is uniformly costless. Then the optimal rule is to allocate to all scores under high-score priority and never to allocate under low-score priority. This is, for example, the case with the quadratic cost function $c(t|s)=(t-s)^2$.

\paragraph{Optimal rule under \eqref{eq:uid}.}  Unlike in the \eqref{eq:udd} case, the optimal rule is not pinned down by local slope constraints, but by the matching function $m$ which describes pairs of ineligible-eligible scores for which the \eqref{eqFPC} condition binds:
\begin{equation*}
\alpha^*_{uid}(s,r)=
\begin{cases}
0&\text{if } s<s_*\\
\Gamma_{uid} I(r)-\int_{s_*}^s c_s\bigl(m(x)|x\bigr)dx&\text{if }s\in[s_*,\hat{s}]\\
1-\Gamma_{uid}\bar{I}(r)-\int_{s}^{s^*} c_t\bigl(x|m^{-1}(x)\bigr)dx&\text{if }s\in[\hat{s},s^*]\\
1&\text{if } s>s^*\\
\end{cases},
\end{equation*}
where $\bar{I}(r) = 1 - I( r)$. As in the \eqref{eq:udd} case, the conditions $s<s_*$ and $s>s^*$ are vacuous whenever $s_*=\underline{s}$ and $s^*=\overline{s}$ (respectively); hence, the allocation rule is always continuous. Although they play the same role, the definitions of the boundary condition, the probability gap, and the multiplicity condition differ from the \eqref{eq:udd} case.  

Over the growth interval, the allocation rule grows at rate $-c_s\bigl(m(x) | x\bigr)$ for ineligible scores $x$, and at rate $c_t\bigl(x | m^{-1}(x)\bigr)$ for eligible scores $x$. We prove that \eqref{eqFPC} binds between matching scores and, in particular, for the pair $(s_*, s^*)$. Therefore, the growth interval $[s_*, s^*]$ is uniquely determined by the \emph{boundary condition}
\begin{equation}\label{eq:bounds}
s_* = \min\bigl\{s \in [s_*(0), \hat{s}] : c\bigl(m(s) | s\bigr) \leq 1 \bigr\}.\tag{B}
\end{equation}
Whether the probability constraint binds depends on the magnitude of falsification costs, and its degree of slackness is measured by the \emph{probability gap}, which is now defined as
\begin{equation}\label{eq:gap}
\Gamma_{uid} = 1 - c(s^* | s_*).\tag{Gap}
\end{equation}
The probability constraint is slack, and $\Gamma_{uid}>0$, if and only if the \emph{slackness condition} $c\bigl(s^*(0) |s_*(0)\bigr)<1$ is satisfied.

As in the \eqref{eq:udd} case, the gap is optimally withheld in the low-score priority case and optimally allocated to agents in the high-score priority case. The index $I( r)$ captures this choice. Given the boundary condition \eqref{eq:bounds}, the probability constraint is slack (and hence $\Gamma_{uid}>0$) in the neutral-priority case if and only if the slackness condition $c\bigl(\overline{s}|\underline{s}\bigr)<1$ is satisfied. When it holds, there is a gap $\Gamma_{uid}>0$ between the total growth of the optimal rule and its upper bound (equal to 1). In the \eqref{eq:uid} case, multiplicity therefore arises under the condition
\begin{align}\label{eq:mult}
  \bar{w}=0 \text{ and } c(\overline{s}|\underline{s}) < 1. \tag{Mult}
\end{align}

\begin{Thm}[Optimal rule under \ref{eq:uid}]\label{ThOptUID}
Suppose the cost function satisfies \eqref{eq:uid}. If \eqref{eq:mult} does not hold, then $\alpha^*_{uid}$ is independent of $r$ and is the unique solution of \eqref{eqBP}. If \eqref{eq:mult} holds, then the set of optimal rules is $\bigl\{\alpha^*_{uid}(\cdot,r)\bigr\}_{r\in[0,1]}$. 
\end{Thm}

A key to the \eqref{eq:uid} characterization is identifying the pattern of binding falsification-proofness constraints, which turns out to be pinned down by the matching function $m$. Our proof identifies the simplified problem under \eqref{eq:uid} as the dual of the problem of transporting surplus from ineligible to eligible scores at a minimal (falsification) cost. The \eqref{eq:uid} condition implies that the solution to this problem is deterministic and assortative, translating into the matching function. Allocation probabilities simply correspond to the shadow prices of the optimal transport problem. We discuss this proof intuition in depth in \secref{sec:ot}, while the remaining details are in \appref{AppProofs}.

\subsection{Optimal rules: properties}\label{subsec-gaming} 

\begin{figure}[ht]
\begin{center}
\scalebox{0.48}{\input{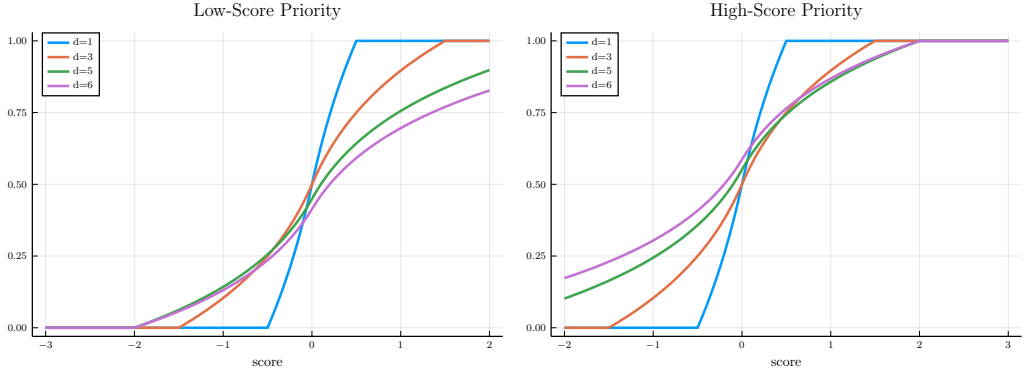}}
\caption{Default setups, Euclidean costs with \eqref{eq:uid}: $\mathcal{C}^d(x)=\frac{\log(1+x)}{\log(1+d)}$ resulting in an S-shape, gaming ability $\gamma=\log(1+d)$ with $d=1,3,5,6$. Binding regime if $d\leq 4$, slack regime with probability gap $\Gamma>0$ otherwise. Under low-score priority, $\Gamma$ is withheld from agents and corresponds to the gap between the $d=5,6$ curves and 1 at $\overline{s}$. Under high-score priority, $\Gamma$ is allocated to agents and corresponds to the gap between the $d=5,6$ curves and 0 at $\underline{s}$. }
\label{fig-gam-comp-statics}
\end{center}
\end{figure}

We discuss different properties of allocatively optimal rules. First, we discuss the consequence of a binding probability constraint. Second, we discuss the shape of optimal allocation rules, providing intuitions for the welfare properties that we study further in the paper. Finally, we consider comparative statics with respect to gaming ability. All results from this section are illustrated in \figref{fig-gam-comp-statics} for the \emph{default setups} with a \eqref{eq:uid} cost function. When the distinction between the \eqref{eq:udd} and \eqref{eq:uid} cases is not necessary, we denote the optimal allocation rule by $\alpha^*$.

\paragraph{Default setups.} We illustrate our results using two \emph{default setups}. In both of them, the worth function is $w(s)=s$, and the eligibility threshold is $\hat{s}=0$. The score distribution is uniform, with support $[-3,2]$ in the low-score priority default setup and $[-2,3]$ in the high-score priority default setup.

\paragraph{The regimes of optimal rules.}
The optimal rule $\alpha^*$ has two regimes depending on whether the probability constraint binds. It is in the \emph{binding regime} if falsification costs are high and the probability constraint binds. Allocative distortions are then concentrated on the growth interval, with bunching and no distortion at the boundaries of the score set: scores in $[s^*,\overline{s}]$ receive a prize with certainty, and scores in $[\underline{s},s_*]$ receive a prize with zero probability. It is in the \emph{slack regime} if falsification costs are low and the probability constraint is slack. In this regime, the optimal rule displays distortions both on the growth interval $[s_*(0),s^*(0)]$ and at the boundaries of the score set. Under low-score priority, there is bunching at the bottom and distortion at the top, $\alpha^*(\overline{s})<1$, while under high-score priority there is bunching at the top and distortion at the bottom, $\alpha^*(\underline{s})>0$. Under neutral priority, bunching and distortions may occur at either or both boundaries, depending on the choice of the neutral gap share $r$.

\paragraph{The shape of optimal rules.} We apply our characterization to the Euclidean\footnote{In \appref{app-sec-comp-stat}, we show that the \emph{shifted linear} family of cost functions (\exref{ExShftLin}) also generates S-shaped rules.} family of cost functions (\exref{ExEucl}). The cost function satisfies \eqref{eq:uid} if $\mathcal{C}$ is concave, and \eqref{eq:udd} if it is convex. We show that optimal rules are S-shaped in the \eqref{eq:uid} case: convex over ineligible scores and concave over eligible scores. Intuitively, this property can boost the probability of allocation for eligible scores and reduces it for ineligible ones. In the \eqref{eq:udd} case, by contrast, the optimal rule is linear since the cost of a marginal falsification is constant and equal to $\mathcal{C}'(0)$.  

\begin{Prop}[Optimal rules under Euclidean cost]\label{PropEucl}
If $\mathcal{C}$ is convex, the cost function satisfies \eqref{eq:udd}, and the optimal rule is linear in $s$ on $[s_*,s^*]$, with $\alpha^*(s)=\alpha^*(s_*)+\mathcal{C}'(0)\bigl(s-s_*\bigr)$. If $\mathcal{C}$ is concave, the cost function satisfies \eqref{eq:uid}, and the optimal rule is convex in $s$ on $[s_*,\hat{s}]$, and concave on $[\hat{s},s^*]$.
\end{Prop}

For a given length of the growth interval, the S-shape concentrates growth close to the eligibility threshold, better approximating the first-best allocation rule, which is a step function at the eligibility threshold, than a linear rule. \propref{PropEucl} suggests that the optimal falsification-proof rule may not be very costly for the designer in the \eqref{eq:uid} case, as its shape reduces allocative distortions. This intuition is confirmed by our welfare analysis in \secref{sec-econ-impl}. 

\paragraph{Gaming ability.} We parameterize the cost function by a scaling factor $\gamma$ that captures \emph{gaming ability}, and perform comparative statics with respect to $\gamma$. Specifically, we consider parameterized cost functions $c^\gamma(t|s)=\frac{1}{\gamma}c(t|s)$, so that higher gaming ability implies lower falsification cost. Let $\hat{\gamma}$ denote the threshold at which the optimal rule switches from the binding regime $\gamma\leq \hat{\gamma}$ to the slack regime $\gamma>\hat{\gamma}$.\footnote{Under \eqref{eq:udd} $\hat{\gamma}=\int_{s_*(0)}^{s^*(0)}c_{t^+}(x|x)dx$ , and under \eqref{eq:uid} $\hat{\gamma}=c\bigl(s^*(0)|s_*(0)\bigr)$. }

\paragraph{Comparative statics.} We show that increasing gaming ability from an initially low level (binding regime) benefits low-score agents and harms high-score agents. If the initial level is sufficiently high (slack regime), an increase affects all agents in the same direction, benefiting them under high-score priority and harming them under low-score priority. The results hold for both the \eqref{eq:uid} and the \eqref{eq:udd} case.

\begin{Prop}[Effect of gaming ability]\label{PropGam}
Consider increasing gaming ability from $\gamma$ to $\gamma'>\gamma$, and denote the corresponding optimal rules by $\alpha^*_{\gamma'}$ and $\alpha^*_{\gamma}$. Then:
\begin{enumerate}[(i)]
\item \textbf{Binding regime:}
If $\gamma\leq \hat{\gamma}$, then $\alpha^*_{\gamma}(s)-\alpha^*_{\gamma'}(s)$ is single-crossing from below.
\item \textbf{Slack regime, 
 low-score priority:} If $\gamma\geq \hat{\gamma}$, and under low-score priority or neutral priority with $r=0$, then $\alpha^*_{\gamma}(s)\geq \alpha^*_{\gamma'}(s)$. Furthermore, the difference in allocation probability increases in $s$ and is equal to 0 at $s_*(0)$.
\item \textbf{Slack regime, 
high-score priority:} If $\gamma\geq \hat{\gamma}$, and under high-score priority or neutral priority with $r=1$, then 
$\alpha^*_{\gamma'}(s)\geq \alpha^*_{\gamma}(s)$. Furthermore, the difference in allocation probability decreases in $s$ and is equal to 0 at $s^*(0)$.
\end{enumerate}
\end{Prop} 

These comparative statics are illustrated in \figref{fig-gam-comp-statics} for the default setup. The left panel illustrates points (i) and (ii); the right panel illustrates points (i) and (iii).

\paragraph{Maximum allocative surplus.} The maximum allocative surplus is nonincreasing in gaming ability because any increase in gaming ability tightens \eqref{eqFPC}.

\begin{Prop}[Properties of the designer's value function]\label{PropWelf}
The value function of \eqref{eqBP}, $W(F,c^\gamma)$, is nonincreasing in $\gamma$. 
\end{Prop}

\subsection{Proof of \theoremref{ThOptUID}: allocation as transport}\label{sec:ot}

We show how to find an optimal rule in the \eqref{eq:uid} case by drawing a connection between the simplified program and optimal transport theory. We consider the following relaxation of the simplified problem:
\label{page-ot1}
\begin{align*}
\max_\alpha~&\int_{s_*}^{\hat{s}}\alpha(s)w(s)dF(s)+\int_{\hat{s}}^{s^*}\alpha(t)w(t)dF(t)\\
\text{s.t.}~&\alpha(t)-\alpha(s)\leq c(t|s),~~\forall s_*\leq s\leq \hat{s} \leq t \leq s^*,
\end{align*}
in which we only require falsification-proofness between ineligible and eligible scores.

In our model, a mass of agents is distributed over the space of scores, which we can think of as locations. Each agent in ``location'' $s$ is endowed with an amount $w(s)$ of surplus. Equivalently, we can describe the problem in terms of masses of negative and positive surplus available at different locations. Each location $s$ then contains a mass $|w(s)| dF(s)$ of negative surplus if $s$ is ineligible, or of positive surplus if $s$ is eligible. We can view falsification as transporting negative surplus from ineligible locations to eligible locations that contain positive surplus.
\label{page-ot2}

To formalize this analogy, we change variables to index locations by their distance to the eligibility threshold, letting $y = \hat{s} - s$ for $s \leq \hat{s}$, and $z = t - \hat{s}$ for $t \geq \hat{s}$. These variables belong, respectively, to the space of negative-surplus locations $Y = [0, \hat{s} - s_*]$ and the space of positive-surplus locations $Z = [0, s^* - \hat{s}]$. By \eqref{eqZAS}, each of these spaces contains the same total mass of surplus. We endow each of them with a probability distribution measuring the fraction of this total mass of surplus, as given by the cumulative distribution functions
\[P(y) = \frac{\mathcal{W}(\hat{s}) - \mathcal{W}(\hat{s} - y)}{\mathcal{W}(\hat{s}) - \mathcal{W}(s_*)},
~~~~~~\text{and}~~~~~~
Q(z) = \frac{\mathcal{W}(\hat{s}) - \mathcal{W}(\hat{s} + z)}{\mathcal{W}(\hat{s}) - \mathcal{W}(s^*)},\]
where the normalizing factor is the total mass. Note that $dP(y) \propto |w(\hat{s} - y)| dF(\hat{s} - y)$ and $dQ(z) \propto |w(\hat{s} + z) | dF(\hat{s} + z)$.
\label{new-OT}

Finally, we rewrite the allocation probabilities as location-specific prices, $\varphi(y) = \alpha(\hat{s} - y)$ and $\psi(z) = \alpha(\hat{s} + z)$, so the program becomes (up to multiplication by the normalizing factor)
\begin{align*}
\max_{\varphi, \psi}~\int_Z \psi(z) dQ(z) - \int_Y \varphi(y) dP(y)
~~~\text{s.t.}~~~ \psi(z) - \varphi(y) \leq  c(\hat{s} + z | \hat{s} - y) \quad \forall y, z.
\end{align*}

To interpret this program in terms of transport, suppose that the designer is a planner who wants to support the production of a locally produced good (surplus) at locations in $Z$, but discourage it at locations in $Y$. As a result, they wish to maximize the profit of producers at eligible locations in $Z$, while minimizing the profit of producers in $Y$. The good costs nothing to produce but can only be produced in quantity $dQ(z)$ at $z \in Z$ and $dP(y)$ at $y \in Y$. Suppose that demand exceeds supply at every location and that the economy is entirely regulated so that the planner can choose the price at which the good is sold at each location. However, producers in $Y$ can be tempted to transport their production to locations in $Z$ at a cost if they can profit from it. The designer is therefore interested in the least costly routes between $Y$ and $Z$. Indeed, the program above is the dual of the optimal transport problem, which seeks to find the least costly way of transporting $P$ to $Q$:
\begin{align*}
\min_{\mu \in \mathcal{M}(P, Q)}~& \int_{Y \times Z} c(\hat{s} + z | \hat{s} - y) d\mu(y, z),
\end{align*}
where $\mathcal{M}(P, Q)$ is the set of joint distributions on $Y \times Z$ with marginals $P$ on $Y$ and $Q$ on $Z$.

By \eqref{eq:uid}, the transportation cost $c(\hat{s} + z | \hat{s} - y)$ is submodular on $Y \times Z$. Under this condition, it is well known\footnote{See, for example, \citet[][chapter 4]{galichon2018optimal}.} that the optimal transportation plan is deterministic and assortative. It is given precisely by the matching function $m$: all surplus at $y$ is transported to location $z$ such that $\hat{s} + z = m(\hat{s} - y)$. 

In terms of our original problem, this implies that \eqref{eqFPC} binds between matched scores $s \in [s_*, \hat{s}]$ and $t = m(s)$. The unique (up to an additive constant) optimal \emph{price functions} $\varphi$ and $\psi$ in closed form are also known, which allows us to solve the simplified problem on any \eqref{eqZAS} interval. The formula of \theoremref{ThOptUID} is then obtained by applying \procref{Proc1}. To complete the proof of the theorem, we show in \appref{AppProofs} that \eqref{eq:uid} implies that $\alpha^*_{uid}$ satisfies the omitted falsification-proofness constraints between scores on the same side of the eligibility threshold $\hat{s}$.


\section{Price of honesty and welfare}\label{sec-econ-impl} 

We assess the welfare impact of honesty in the \eqref{eq:uid} case. To do so, we compare the optimal falsification-proof mechanism to the unconstrained optimal mechanism, which is characterized in \cite{perez2022test}. Honesty comes at a price in allocative surplus since it constrains the mechanism. In return, we show that honesty also has value, as it benefits all agents. We use theoretical results and illustrative examples to compare the magnitudes of these effects. 

In this section, we fix the neutral gap share $r$ and omit the dependence of the optimal rule $\alpha^*$ on $r$ in the notation. As we are throughout in the \eqref{eq:uid} case, we drop the index and denote the optimal falsification-proof rule by $\alpha^*$.


\paragraph{The unconstrained optimal mechanism.} We first describe the unconstrained allocatively optimal mechanism. For clarity, we focus on a low-score priority setting in the discussion, but the results are general. The unconstrained optimal allocation mechanism consists of a \emph{nominal} allocation rule $\alpha^{**}(s)$ paired with an allocatively optimal incentive-compatible falsification strategy. The rule $\alpha^{**}$ ensures that agents with scores within its growth interval are indifferent between not falsifying and falsifying to the top of the interval.\footnote{This is achieved by setting $\alpha^{**}(s)=c(s_+|s_-)-c(s_+|s)$ on the growth interval $[s_-,s_+]$ of the unconstrained rule. See the proof of \propref{PropWelfFP} for a detailed description.} For the mechanism to be allocatively optimal, agents must resolve this indifference by falsifying to the top if eligible and not falsifying otherwise. The optimal mechanism therefore relies on eligible agents incurring falsification costs.

\begin{figure}[ht]
\begin{center}
\scalebox{.34}{\input{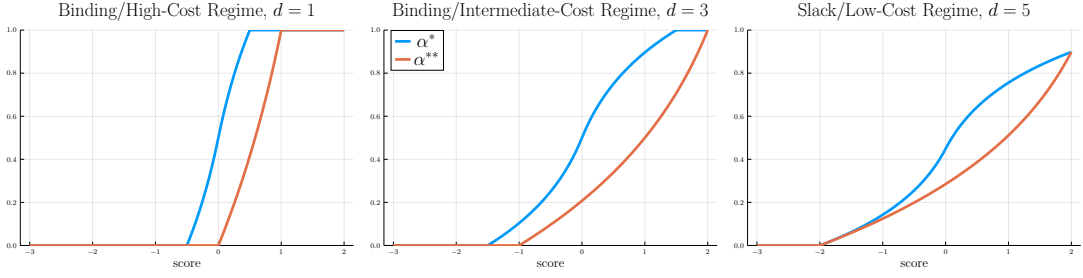}}
\caption{Low-score priority default setup, cost $\mathcal{C}^d(x)=\frac{\log(1+x)}{\log(1+d)}$, gaming ability $\gamma=\log(1+d)$.}\label{FigWelfFPrules}
\end{center}
\end{figure}

The unconstrained optimal rule has three regimes. It is in the \emph{high-cost regime} if $c(\overline{s} | \hat{s}) > 1$. Then the rule achieves the first best by allocating prizes only to eligible agents; however, those with lower scores must falsify. The rule is in the \emph{intermediate-cost regime} if $c(\overline{s} | \hat{s}) \leq 1$ but $c(\overline{s} | s_*(0)) > 1$. Then all eligible agents secure a prize by falsifying to the top score, while some ineligible agents must also receive prizes with positive probability to deter falsification. Finally, the rule is in the \emph{low-cost regime} if $c(\overline{s} | s_*(0)) \leq 1$. The designer then reduces the maximal allocation probability under low-score priority to prevent agents below $s_*(0)$ from obtaining prizes with positive probability. \figref{FigWelfFPrules} illustrates both rules in a low-score priority setting.

\paragraph{The value of honesty for agents.} Unconstrained optimality requires some eligible agents to falsify. However, not falsifying is also a best response to $\alpha^{**}$ for agents who falsify. Therefore, the equilibrium payoff of an agent under $\alpha^{**}$ equals her payoff without falsification, namely $\alpha^{**}(s)$. Thus, comparing $\alpha^{**}(s)$ with $\alpha^*(s)$ suffices to analyze agents' welfare. \figref{FigWelfFPrules} shows an example in which $\alpha^{**}$ lies below $\alpha^*$ everywhere, implying that agents are better off under the falsification-proof rule. This result generalizes: agents benefit from honesty regardless of score.

\begin{Prop}[Value of honesty for agents]\label{PropWelfFP}
In the \eqref{eq:uid} case, all agents are better off under the optimal falsification-proof rule: $\alpha^*(s)\geq \alpha^{**}(s)$, with a strict inequality for a positive mass of agents if the cost function has strict upward increasing differences.
\end{Prop}

Agents with ineligible scores do not falsify under either rule but receive prizes with a higher probability under the falsification-proof rule. When $\alpha^{**}$ is coupled with the allocatively optimal falsification strategy, eligible agents receive prizes with a lower probability under $\alpha^*$ but avoid the falsification costs required by the optimal mechanism. Thus, the loss of allocative surplus due to falsification-proofness may be offset by gains in agents' welfare. We proceed to quantify the relative gains and losses from honesty.

\paragraph{A cost-benefit analysis of honesty.} Consider a concave Euclidean cost function $\mathcal{C}$. Let $A^*(\mathcal{C})$ and $A^{**}(\mathcal{C})$ denote the aggregate payoff of agents under $\alpha^*$ and $\alpha^{**}$, respectively, while $D^*(\mathcal{C})$ and $D^{**}(\mathcal{C})$ denote the designer's payoff. With Euclidean costs, these payoffs take relatively simple forms, whose explicit formulas we provide in the proof of \propref{PropWelfTradeOff}. To measure the welfare impact of honesty, we compare the \emph{gain rate of agents}, $G(\mathcal{C})=\frac{A^*(\mathcal{C})-A^{**}(\mathcal{C})}{A^{**}(\mathcal{C})}$, with the loss rate in allocative efficiency, $L(\mathcal{C})=\frac{D^{**}(\mathcal{C})-D^*(\mathcal{C})}{D^{**}(\mathcal{C})}$. We interpret $G(\mathcal{C})$ as the \emph{value of honesty}, and $L(\mathcal{C})$ as the \emph{price of honesty}.

Controlling for gaming ability, we can make the trade-off arbitrarily favorable to honesty. We control for gaming ability by fixing the maximum amount of falsification an agent would rationally undertake, $d_\mathcal{C}$ (defined by $\mathcal{C}(d_\mathcal{C})=1$) at some value $d$. The nature of the trade-off depends on how $d$ compares to the size of the interval of eligible scores, $\hat{d}=\overline{s}-\hat{s}$, which is the threshold marking the frontier between the intermediate-cost and high-cost regimes of $\alpha^{**}$.

\begin{Prop}\label{PropWelfTradeOff}
Suppose $d\geq \hat{d}$. Then, for every $\varepsilon>0$, there exists a Euclidean cost function $\mathcal{C}$ with $d_{\mathcal{C}}=d$, such that $L(\mathcal{C})\leq \varepsilon$ and $G(\mathcal{C})\geq 1/\varepsilon$. If $d<\hat{d}$, then for every $\varepsilon>0$, there exists a Euclidean cost function $\mathcal{C}$ with $d_{\mathcal{C}}=d$, such that $L(\mathcal{C})\leq \varepsilon$ and $\left|G(\mathcal{C})-\frac{F(\hat{s}+d)-F(\hat{s})}{1-F(\hat{s}+d)}\right|\leq \varepsilon$.
\end{Prop}

To gain intuition, consider the role of returns to scale in falsification. Higher returns to scale correspond to a more concave cost function, which can be achieved while keeping $d_{\mathcal{C}}$ constant. Suppose we are in the intermediate-cost regime with low-score priority for simplicity. A more concave cost makes $\alpha^{**}$ more convex on the same growth interval, reducing agents' payoffs while pushing the designer closer to first-best. This is because making the rule more convex increasingly lowers the probability of receiving a prize for ineligible agents, while eligible agents continue to falsify to obtain a prize. At the same time, $\alpha^*$ approaches first-best because greater concavity increases the convexity of the rule over ineligible scores and its concavity over eligible scores, allowing the S-shaped rule to get closer to the first-best step function.\footnote{See \oappref{subsec-RTS} for comparative statics results on the effect of returns to scale} Hence, \propref{PropWelfTradeOff} can be read as saying that honesty 
yields large gains to agents at a small allocative price when returns to scale are strongly increasing.
  
\begin{figure}[ht]
\begin{center}
\scalebox{.5}{\input{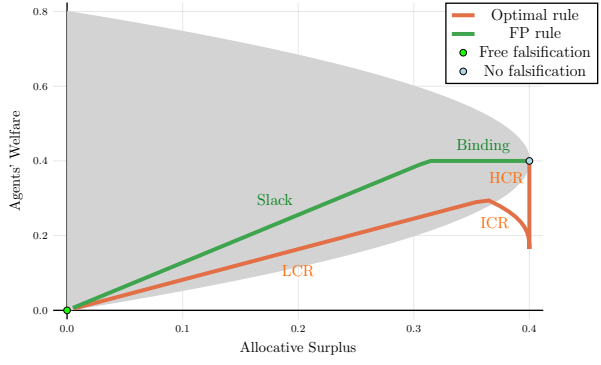}}   
\caption{Payoff trajectories. Low-score priority default setup, cost $\mathcal{C}^d(x)=\frac{\log(1+x)}{\log(1+d)}$, gaming ability $\gamma=\log(1+d):0\to \infty$. The gray zone depicts the set of attainable allocative payoffs (without accounting for falsification costs).}\label{FigWelfFP}
\end{center}
\end{figure}

The condition $d<\hat{d}$ characterizes the high-cost regime for $\alpha^{**}$, in which the designer achieves the first-best allocation. In this case, we cannot make the agents' gain rate arbitrarily large. This is because the agents' payoff does not converge to 0 under the unconstrained optimal mechanism as we make the cost function more concave. Indeed, the unconstrained optimal rule then allocates a prize for sure to all agents with a score above $\hat{s}+d<\overline{s}$ and only requires scores between $\hat{s}$ and $\hat{s}+d$ to falsify. Hence, while the payoff of ineligible agents and falsifying eligible agents still converges to 0, agents with a score above $\hat{s}+d$ receive a prize with certainty without falsifying, so the aggregate payoff of agents under $\alpha^{**}$ converges to $1-F(\hat{s}+d)$.

\paragraph{Quantitative illustration.} We illustrate \propref{PropWelfTradeOff} using the default setups under different Euclidean cost functions. \figref{FigWelfFP} shows the trajectories of allocative surplus and agents' welfare under both rules when varying gaming ability $\gamma=\log(1+d)$ for a family of Euclidean cost functions $\mathcal{C}^d(x)=\frac{\log(1+x)}{\log(1+d)}$. With infinite gaming ability, both rules never allocate (leftmost point in both trajectories). As $\gamma$ approaches 0, both rules converge to the first-best (top-right point in both trajectories). At these extremes, there is no trade-off. For intermediate gaming ability, we see that the optimal rule can be very costly to agents for a moderate gain in allocative surplus. This is especially true at the frontier between the high-cost and intermediate-cost regimes.

\begin{figure}[ht]
\begin{center}
\scalebox{.47}{\input{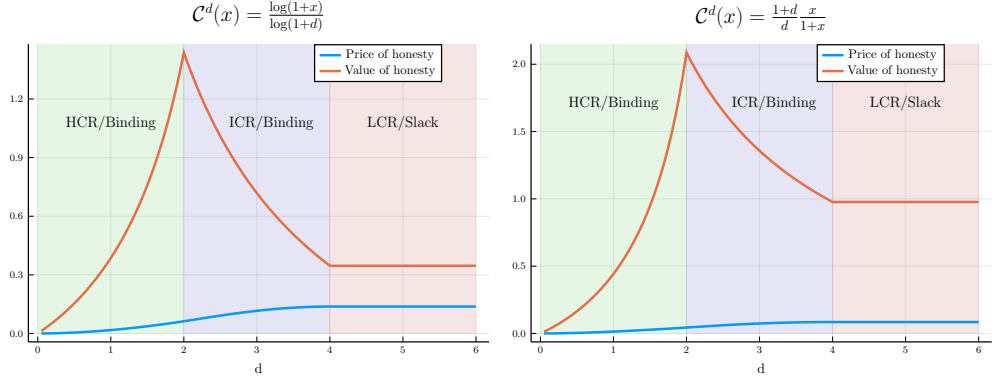}}
\caption{Welfare impact of honesty. Low-score priority default setup, $d:0\to 6$.}\label{FigWelfFP2}
\end{center}
\end{figure}

We quantify the trade-off by plotting the value and the price of honesty, as measured in \propref{PropWelfTradeOff}, for two fixed families of cost functions with varying gaming abilities. This exercise is shown in \figref{FigWelfFP2}. Under both cost functions, the value of honesty dominates its price for all gaming ability levels. The peak in the relative value is reached at the frontier between the high-cost and intermediate-cost regimes. Under the first cost function (left panel), the trade-off is then a 144\% value for a 6.25\% price of honesty, while under the second cost function (right panel), it is a 209\% value for a 4.4\% price. The figure also illustrates the role of returns to scale, as the value is higher and the price lower at every $d$ under the more concave cost function in the right panel.

\propref{PropWelfFP} and \propref{PropWelfTradeOff} suggest a theoretical foundation for falsification-proofness. Beyond the externality motive, requiring honesty can be an effective way to balance allocative surplus and agents' welfare. It can, under some configurations, generate large gains for agents at a moderate cost in allocative surplus.

\section{A welfare foundation for falsification-proofness}\label{sec:general fals}

We show that honesty can emerge as an unconstrained solution to the weighted welfare problem \eqref{pgm:wwp} if the planner places sufficient weight on agents' welfare. \theoremref{thm:fp-optimal} identifies an explicit condition under which honesty is optimal.

\paragraph{Falsification technology.} The analysis in this section assumes \eqref{eq:uid} throughout. However, we also need to strengthen our assumptions on the falsification technology to handle downward falsification. Specifically, we assume that the cost function satisfies the following \emph{downward triangular inequality}:
\begin{equation}\label{eq:dtri}
\forall t<m<s,~~c(t|m)+c(m|s)\geq c(t|s).\tag{DTRI}
\end{equation}
We also assume that $c(t|s)$ is \emph{downward monotonic}; that is, nonincreasing in $t$ and nondecreasing in $s$ for $t<s$. Combined with \eqref{eq:uid}, these assumptions ensure that the cost function satisfies the standard triangular inequality (see \lemmaref{lem-tri}).

\paragraph{Optimality of honesty.} Recall that, in the  weighted welfare problem \eqref{pgm:wwp}, the planner maximizes
\[\Omega(\alpha,\phi,\zeta)=\iint\alpha(t)w_\zeta(s)\,d\phi(t|s)dF(s)-\zeta\iint c(t|s)\,d\phi(t|s)dF(s)\]
subject to \eqref{eqPC} and \eqref{eq:IC}, where $w_\zeta(s)=w(s)+\zeta$. 

The first term of this objective is an allocative surplus term with the upward-shifted worth function $w_\zeta(s)$. This induces a downward shift of the eligibility threshold to $\hat{\sigma}(\zeta)$, where $\hat{\sigma}(\cdot)$ is the decreasing function $\hat{\sigma}(x)=w^{-1}(-x)$, with the convention that $\hat{\sigma}(x)=\overline{s}$ if $-x\geq w(\overline{s})$, and $\hat{\sigma}(x)=\underline{s}$ if $-x\leq w(\underline{s})$. Then $\hat{\sigma}(x)$ is simply the eligibility threshold associated with a shift $x$ of the worth function. In particular, $\hat{\sigma}(0)=\hat{s}$.

A sufficient condition for honesty to be optimal is
\begin{equation}\label{eq:honest}
    c\Bigl(\hat{s}\,|\,\hat{\sigma}(\zeta)\Bigr)\geq 1.\tag{H}
\end{equation}

Honesty is trivially optimal if $\zeta\geq \bar{\zeta}=-w(\underline{s})$, since the optimal rule for the planner is then to allocate prizes unconditionally. The honesty condition \eqref{eq:honest} extends the range of weights $\zeta$ for which honesty is optimal, as long as the cost of falsification is sufficiently high; that is, precisely if $c(\hat{s}|\underline{s})>1$. Then, we show that our solution to \eqref{eqBP} with the shifted worth function $w_\zeta(s)$ solves \eqref{pgm:wwp}. Let $\alpha^*_\zeta(s,r)$ denote a solution to this problem.

\begin{Thm}[Weighted welfare optimality of honesty]\label{thm:fp-optimal}
Under \eqref{eq:uid} and \eqref{eq:dtri}, \eqref{eq:honest} implies that the optimal falsification-proof rule $\alpha^*_\zeta$ solves \eqref{pgm:wwp}.
\end{Thm}

The \emph{honesty condition} \eqref{eq:honest} allows us to isolate the factors that can contribute to making honesty welfare-optimal. First, a \emph{lower gaming ability}. Considering a family $c^\gamma(t|s)=\frac{1}{\gamma}c(t|s)$ of cost functions as in \secref{subsec-gaming}, it is clear that decreasing gaming ability $\gamma$ increases the cost of falsification and makes \eqref{eq:honest} more likely to hold. Second, a \emph{lower dependency} of expected worth on the score. We can think of this dependency as captured by the slope of $w$. If $w(s)=\omega (s-\hat{s})$, for example, where $\omega$ is the coefficient of the score in the linear regression of the worth, then the honesty condition \eqref{eq:honest} becomes $c(\hat{s}\,|\,\hat{s}-\zeta/\omega)\geq 1$, whose left-hand side increases as $\omega$ decreases. Third, a \emph{higher welfare weight} on agents: increasing $\zeta$ lowers $\hat{\sigma}(\zeta)$ making \eqref{eq:honest} more likely to hold. As already noted, above the threshold $\bar{\zeta}=-w(\underline{s})$, the solution to \eqref{pgm:wwp} is to allocate prizes unconditionally, making honesty trivially optimal. The honesty condition \eqref{eq:honest} provides a lower threshold $\hat{\zeta}$ defined by $c\bigl(\hat{s}|\hat{\sigma}(\hat{\zeta})\bigr)=1$ as long as $c(\hat{s}|\underline{s})>1$. For $\zeta\in\bigl[\hat{\zeta},\bar{\zeta}]$, the rule $\alpha^*_\zeta$ paired with the honest strategy for the agents is optimal for the social planner. 

Given \theoremref{thm:fp-optimal}, it is useful to understand how the shift changes the allocation rule. The next proposition shows that increasing $\zeta$ raises the allocation probability smoothly for all scores.  

\begin{Prop}\label{PropShift}
The allocation rule $\alpha^*_\zeta(s,r)$ is weakly increasing in $\zeta$. Furthermore, the correspondence mapping $\zeta$ to $\{\alpha^*_\zeta(s,r)\}_{r\in[0,1]}$ is upper hemicontinuous.
\end{Prop}

By \theoremref{ThOptUID}, $\{\alpha^*_\zeta(s,r)\}_{r\in[0,1]}$ is a singleton unless \eqref{eq:mult} holds for the shifted worth function, which implies, in particular, that $\zeta=-\bar{w}$. In particular, $\alpha^*_\zeta(s,r)$ is independent of $r$ and continuous in $\zeta$ at every $\zeta\neq-\bar{w}$.

\paragraph{Proof overview of \theoremref{thm:fp-optimal}.} The proof has three main steps. First, we show that it is without loss of optimality to restrict attention to allocation rules that satisfy \eqref{eqFPC} and are top-flat in the sense that they allocate with certainty to all scores $s\geq \hat{\sigma}(\zeta)$ such that $c\bigl(s|\hat{\sigma}(\zeta)\bigr)\geq 1$. Note that falsification-proofness does not imply honesty: in particular, optimality for \eqref{pgm:wwp} might require agents to falsify even though honesty is a best response. Second, we introduce a falsification budget $\Phi\geq 0$, and let $Z(\Phi)$ be the value function of the problem of maximizing allocative surplus under $w_\zeta$ with the additional constraint that the aggregate cost of falsification must be equal to the falsification budget $\Phi$. If $\Phi=0$, this program is exactly \eqref{eqBP} with the shifted worth function $w_\zeta$. If $\Phi$ is equal to the aggregate cost of falsification when all eligible scores below the cutoff falsify to the cutoff under $\alpha^{**}$, then this program is the unconstrained allocative surplus optimization problem with the shifted worth function. Then, \eqref{pgm:wwp} reduces to finding the falsification budget $\Phi$ that solves $\max_{\Phi\geq 0}\,\bigl[Z(\Phi)-\zeta\Phi\bigr]$. Third, we show that the top-flatness property implies that, for any $\Phi'>\Phi$, $Z(\Phi')-Z(\Phi)\leq w_\zeta(s^{**}_\zeta)\bigl(\Phi'-\Phi\bigr)$, where $s^{**}_\zeta$ solves $c\bigl(s^{**}_\zeta|\hat{\sigma}(\zeta)\bigr)=1$. Then $\Phi=0$ (honesty) must be optimal whenever $w_\zeta(s^{**}_\zeta)\leq \zeta$, or equivalently $s^{**}_\zeta\leq \hat{s}$. By definition of $s^{**}_\zeta$, this is equivalent to \eqref{eq:honest}.

\paragraph{A falsification-proofness principle (without honesty).} An interesting byproduct of our proof of \theoremref{thm:fp-optimal} is a new \emph{falsification-proofness principle} (\propref{prop:FPprinc}): we prove that it is without loss of generality to restrict attention to allocation rules that satisfy \eqref{eqFPC} in the sense that we can make any allocation rule falsification-proof without altering the original optimal falsification strategy or the final allocation. However, falsification-proofness does not imply honesty, and inducing falsification might actually be optimal in \eqref{pgm:wwp} if \eqref{eq:honest} does not hold. The falsification-proofness principle is a useful technical analog of the revelation principle, reducing the space of mechanisms that must be considered for cost functions that satisfy the triangular inequality. 

\section{Fixed cost}\label{sec:FixedCost}

In this section, we extend our analysis to a falsification cost function with a fixed cost in the \eqref{eq:uid} case. The falsification cost from $s$ to $t$ is then given by
\[\bar{c}(t|s)=k\Ind_{t\neq s}+c(t|s),\]
where $c$ is a cost function that satisfies \eqref{eq:uid} and our standard assumptions. The falsification-proofness condition becomes:
\begin{equation}\label{eq:FPCfix}
\alpha(t)-\alpha(s)\leq k +c(t|s), \forall t\neq s.\tag{FPCfix}
\end{equation}

We show that the optimal falsification-proof rule has the same structure as the one without fixed cost, with identical derivatives on the eligible and ineligible sides of the growth interval but a jump of size $k$ at the eligibility threshold.  

\paragraph{Monotonicity.} We can adapt the proof of \lemmaref{lem:mon} and show that any falsification-proof rule is dominated by a monotonic falsification-proof rule, so there is no loss of generality in restricting attention to monotonic allocation rules. Monotonicity implies that allocation rules have at most countably many jump discontinuities (removable discontinuities can be removed without changing the allocative surplus, so we can assume them away). Furthermore, \eqref{eq:FPCfix} implies that jumps are of size at most $k$. We can also make $\alpha$ right-continuous without affecting allocative surplus, so we can consider $\alpha$ as a right-continuous monotonic function bounded by $[0,1]$.

We can then integrate by parts the objective function of the designer and rewrite it as the following Stieltjes integral
\[\int_S\mathcal{W}(x)d\alpha(x),\]
while optimally  choosing $\alpha(\underline{s})=0$ under low-score priority and $\alpha(\overline{s})=1$ under high-score priority.

The differential program becomes
\begin{align*}
\max_{\alpha }~&\int_S\mathcal{W}(x)d\alpha(x)\\
\text{s.t.}~&\eqref{eq:FPCfix}
\end{align*}

\paragraph{Lagrangian sufficiency.} Since $\alpha$ is a cdf, it satisfies the constraint $\int_S d\alpha(x)\leq 1$. Putting a Lagrange multiplier $\nu$ on this constraint yields the Lagrangian
\[\mathcal{L}(\alpha,\nu)=\int_S\bigl\{\mathcal{W}(x)-\nu\bigr\}d\alpha(x)+\nu.\]
It is clear that any $\alpha$ that puts mass outside of the interval $[s_*(\nu),s^*(\nu)]$ is strictly dominated. We prove the same Lagrangian sufficiency result as in the paper with the cost function $\bar{c}$. There is, however, an important difference in the proof. In the paper, the continuity property (ii) is immediate since any allocation rule satisfying \eqref{eqFPC} must be continuous. With the fixed cost, \eqref{eq:FPCfix} no longer implies (ii). Instead, we show that for a rule $\alpha$ to maximize the Lagrangian $\mathcal{L}(\alpha,\nu)$ it is necessary to satisfy continuity and exhibit no jump at the extremities of the interval $\bigl[s_*(\nu),s^*(\nu)\bigr]$. This proof relies on \eqref{eq:uid}.

\begin{Lem}\label{lem:lagsuff-fix}
Assume \eqref{eq:uid}. An allocation rule $\alpha$ solves \eqref{eqBP} if and only if it satisfies \eqref{eqPC}, \eqref{eq:mon} and \eqref{eq:init}, and there exists a Lagrange multiplier $\nu\geq 0$ such that:
\begin{enumerate}[(i)]
\item \step{Growth interval:} $\alpha(s)=\alpha(\underline{s})$ for every $s< s_*(\nu)$, and $\alpha(s)=\alpha(\overline{s})$ for every $s> s^*(\nu)$;
\item \step{Continuity:} $\alpha$ is continuous at $s_*(\nu)$ and at $s^*(\nu)$;
\item \step{Relaxed \eqref{eq:FPCfix}:} $\alpha(t)-\alpha(s)\leq \bar{c}(t|s)$ for every $s_*(\nu)\leq s<t\leq s^*(\nu)$;
\item \step{Complementary slackness on \eqref{eqPC}:} If $\bar{w}\leq 0$,  $\nu\bigl(1-\alpha(\overline{s})\bigr)=0$; if $\bar{w}\geq 0$, $\nu\alpha(\underline{s})=0$;
\item \step{Optimality for the simplified program:} For every allocation rule $\hat{\alpha}$ that satisfies (iii),
\[\int_{s_*(\nu)}^{s^*(\nu)}w(s)\alpha(s)dF(s)\geq \int_{s_*(\nu)}^{s^*(\nu)}w(s)\hat{\alpha}(s)dF(s).\]
\end{enumerate}
\end{Lem}

\paragraph{Optimal allocation rule under \eqref{eq:uid}.} As in the standard case, we consider the following relaxation of the simplified problem:
\begin{align}\label{pgm:relaxUIDfix}
\max_\alpha~&\int_{s_*}^{\hat{s}}\alpha(s)w(s)dF(s)+\int_{\hat{s}}^{s^*}\alpha(t)w(t)dF(t)\tag{$\overline{\text{Relax BP}}$}\\
\text{s.t.}~&\alpha(t)-\alpha(s)\leq \bar{c}(t|s),~~\forall s_*\leq s\leq \hat{s} \leq t \leq s^*, s<t \notag
\end{align}
in which we only require falsification-proofness to prevent ineligible scores from falsifying to eligible targets and separate the objective function between ineligible and eligible scores.

We perform the same transformations on this program as in the standard case, with one small difference.  We change the variables of this problem to identify scores by their distance to the eligibility threshold, letting $y = \hat{s} - s$ for $s \leq \hat{s}$, and $z = t - \hat{s}$ for $t \geq \hat{s}$. These variables belong, respectively, to the space $Y = [0, \hat{s} - s_*]$ and to the space $Z = [0, s^* - \hat{s}]$. By \eqref{eqZAS}, each of these spaces harbors the same total mass of surplus. We endow each of them with a probability distribution measuring the fraction of this total mass of surplus, as given by the cumulative distribution functions
\[P(y) = \frac{\mathcal{W}(\hat{s}) - \mathcal{W}(\hat{s} - y)}{\mathcal{W}(\hat{s}) - \mathcal{W}(s_*)},
~~~~~~\text{and}~~~~~~
Q(z) = \frac{\mathcal{W}(\hat{s}) - \mathcal{W}(\hat{s} + z)}{\mathcal{W}(\hat{s}) - \mathcal{W}(s^*)},\]
where the normalizing factor is the total mass. Note that $dP(y) \propto |w(\hat{s} - y) | dF(\hat{s} - y)$, and $dQ(z) \propto |w(\hat{s} + z) | dF(\hat{s} + z)$.

Finally, the only difference comes in the definition of prices. We rewrite the allocation probabilities as location-specific prices, $\varphi(y) = \alpha(\hat{s} - y)+k/2$ and $\psi(z) = \alpha(\hat{s} + z)-k/2$, so the program becomes (up to multiplication by the normalizing factor)
\begin{align*}
\max_{\varphi, \psi}~\int_Z (\psi(z)+k/2) dQ(z) - \int_Y (\varphi(y)-k/2) dP(y)\\
~~~\text{s.t.}~~~ \psi(z) - \varphi(y) \leq  c(\hat{s} + z | \hat{s} - y) \quad \forall y, z.
\end{align*}
Since
\[\int_Z (\psi(z)+k/2) dQ(z) - \int_Y (\varphi(y)-k/2) dP(y)=\int_Z \psi(z) dQ(z) - \int_Y \varphi(y) dP(y)+k,\]
solving this program is equivalent to solving the program:
\begin{align*}
\max_{\varphi, \psi}~\int_Z \psi(z)dQ(z) - \int_Y \varphi(y) dP(y)\\
~~~\text{s.t.}~~~ \psi(z) - \varphi(y) \leq  c(\hat{s} + z | \hat{s} - y) \quad \forall y, z,
\end{align*}
which is exactly the same optimal transport problem we solve in the standard case. Hence, the formulas for the optimal prices $\varphi$ and $\psi$ are identical, and only the mapping from prices $\varphi,\psi$ to allocation probability changes. 

Therefore, we obtain the same formula for the optimal solution to \eqref{eqBP}. However, there is an important difference, which is the introduction of a jump of size $k$ at the eligibility threshold $\hat{s}$. The jump is generated by the boundary condition below
\begin{equation*}
\bar{\alpha}^*_{uid}(s,r)=
\begin{cases}
0&\text{if } s<s_*\\
\bar{\Gamma}_{uid} I(r)-\int_{s_*}^s c_s\bigl(m(x)|x\bigr)dx&\text{if }s\in[s_*,\hat{s})\\
1-\bar{\Gamma}_{uid}\bar{I}(r)-\int_{s}^{s^*} c_t\bigl(x|m^{-1}(x)\bigr)dx&\text{if }s\in[\hat{s},s^*]\\
1&\text{if } s>s^*\\
\end{cases}.
\end{equation*}

The value of $\alpha^*(\hat{s},r)$ is irrelevant, and we have chosen it to make $\bar{\alpha}^*_{uid}$ right-continuous. 

The growth interval $[s_*, s^*]$ is uniquely determined by the \emph{boundary condition}
\begin{equation}\label{eq:boundsfix}
s_* = \min\bigl\{s \in [s_*(0), \hat{s}] : \bar{c}\bigl(m(s) | s\bigr) \leq 1 \bigr\}.\tag{$\overline{\text{B}}$}
\end{equation}
We are using $\bar{c}$ instead of $c$ for the new boundary condition, which generates the jump at $\hat{s}$.

Whether the probability constraint binds or not depends on the magnitude of falsification costs, and its degree of slackness is measured by the \emph{probability gap}, now defined as
\begin{equation}\label{eq:gapfix}
\bar{\Gamma}_{uid} = 1 - \bar{c}(s^* | s_*).\tag{$\overline{\text{Gap}}$}
\end{equation}
Then, the probability constraint is slack, and $\bar{\Gamma}_{uid}>0$, if and only if the \emph{slackness condition} $\bar{c}\bigl(s^*(0) |s_*(0)\bigr)<1$ is satisfied.

Given the boundary condition \eqref{eq:boundsfix}, the probability constraint is slack, and the gap positive $\bar{\Gamma}_{uid}>0$ in the neutral probability case, if and only if the \emph{slackness condition} $\bar{c}\bigl(\overline{s},\underline{s}\bigr)<1 $ is satisfied. When it holds, there is a gap $\bar{\Gamma}_{uid}>0$ between the total growth of the optimal rule and its upper bound (equal to 1). Multiplicity, therefore, arises under the condition:
\begin{align}\label{eq:multfix}
  \bar{w}=0 \text{ and } \bar{c}(\overline{s}|\underline{s}) < 1. \tag{$\overline{\text{Mult}}$}
\end{align}

\begin{Thm}[Optimal rule under \ref{eq:uid} with a fixed cost]\label{ThOptUIDfix}
If \eqref{eq:multfix} does not hold, then $\bar{\alpha}^*_{uid}$ is independent of $r$, and it is the unique solution of \eqref{eqBP}. If \eqref{eq:multfix} holds, then the set of optimal rules is $\bigl\{\bar{\alpha}^*_{uid}(\cdot,r)\bigr\}_{r\in[0,1]}$. 
\end{Thm}

Before proving the theorem, some comments are in order. Note that $\bar{\alpha}^*_{uid}$ closely resembles $\alpha^*_{uid}$. The important difference is in the statement of the boundary condition \eqref{eq:boundsfix} and the definition of the probability gap \eqref{eq:gapfix}. The boundary condition changes the nature of the growth interval, generating a jump between the eligible and ineligible sides of the optimal rule, whereas the growth of the allocation rule on each of these sides is exactly as it is without a fixed cost. It is clear from the structure that increasing the fixed cost $k$ reduces the size of the growth interval and increases both allocative surplus and the welfare of the agents. Naturally, when $k\geq 1$, the optimal falsification rule is the first-best allocation rule, which becomes falsification-proof because of the high fixed cost. 

The proof of the theorem closely follows the proof of \theoremref{ThOptUID} in the main paper.

\section{Heterogeneity and multiple groups}\label{sec:MGsetup}

We extend our model to a richer heterogeneity structure with both observable and unobservable dimensions. On the one hand, unobservable dimensions cannot be elicited by any mechanism, and the falsification-proofness constraint requires preventing falsification between any two scores under the worst-case scenario (i.e., for the type of agents for whom it is the least costly). Therefore, the problem with additional unobservable heterogeneity is similar to the one with no additional heterogeneity, but  where agents have the worst-case falsification cost function. On the other hand, observable heterogeneity leads to the creation of groups of agents within which the optimal allocation problem is identical to the setup of the paper.

\subsection{Model} 

\paragraph{Framework.} Each agent is characterized by a private type $\theta=\bigl(i,s,k\bigr)$ and their \emph{worth} $w$. Agents may know their worth (if $\theta$ is a sufficient statistic for $w$), or not, with no effect on the results. 

The first dimension of the type $i\in I$ encompasses all relevant publicly observable and unfalsifiable characteristics of an agent. We refer to $i$ as an agent's \emph{group}, and assume $I$ is a finite set. The mass of group $i$ is $\mu_i>0$, where $\sum_{i\in I} \mu_i=1$. The second dimension of an agent's type is the natural score $s\in S_i \subseteq \mathbbm{R}$. The last dimension of type, $k\in K_{i,s}$, is a vector of privately known characteristics that includes an agent's value for the good $v(k)>0$, information about their individual falsification cost, and possibly other characteristics correlated with their worth. 

\paragraph{Distributional assumptions.} Each agent draws a vector of characteristics $(\theta,w)$ i.i.d. from a joint distribution. Hence, the different dimensions of an agent's vector of characteristics can be, and typically are, correlated; but they are independent of other agents' characteristics. $F_i$ denotes the cumulative distribution function of natural score conditional on $i$, which we assume to have full support on an interval $S_i=[\underline{s}_i,\overline{s}_i]$, and no atoms. Conditional on $(i,s)$, the remainder of the type vector is fully supported on $K_{i,s}$.  

\paragraph{Designer and agent payoffs.} We assume that the worth $w$ is bounded and integrable conditional on $(i,s)$. We denote the corresponding expected worth by $w_i(s)=\Exp(w|i,s)$, and by $\bar{w}_i=\Exp(w|i)$ the expected worth in group $i$. The designer's payoff from assigning an object to a group $i$ agent with score $s$ is $w_i(s)$. We assume that score and worth are positively related in the sense that, for every group $i$, $w_i(s)$ is strictly increasing. The expected payoff of an agent is $\alpha v-C(t,\theta)$, where $\alpha$ is the probability of getting an object, and $C(t,\theta)\geq 0$ defines the cost for type $\theta$ to produce a score $t$. 

Not falsifying is costless, so $C(s,\theta)=0$ for an agent of type $\theta=(i,s,k)$. The cost of producing score $t$ depends not only on the natural score $s$, but also on $k$. The falsification cost may reflect technical costs, psychological lying costs, and expected penalties.  

\paragraph{Falsification-proof mechanisms.} We restrict the designer to \emph{falsification-proof mechanisms}, that is, mechanisms that incentivize agents to produce their natural scores. Under this assumption, we show in \cite{perez2023score} that it is without loss of generality to restrict attention to \emph{score-based allocation rules} $\bm{\alpha}=(\alpha_i)_{i \in I}$, where $\alpha_i: S_i \to [0,1]$ is the probability that an object is allocated to an agent from group $i$ conditional on their produced score. Such rules only condition on the observable dimension $i$, and the score $s$, but not on $k$. The reason is that once we restrict attention to mechanisms that recommend agents to produce their natural score (which by definition involves deterministic 
recommendations), agents with the same score $s$ will pool on reporting the $(k,v)$ pair that maximizes their allocation probability, so the designer cannot elicit this information.\footnote{Lemma 1 in \cite{akbarpour2020redistributive} makes a similar point in a setting where all dimensions are costless to misrepresent, and transfers are allowed: the principal can only elicit information on the dimensions that correlate with agents' willingness to pay.} Among agents from the same group and with the same natural score, falsification is most tempting for those with the highest valuation and lowest falsification costs. Therefore, a mechanism is falsification-proof if and only if it satisfies the following constraint
\begin{equation}\label{eqFPCgroups}
\forall (i,s,t),~~ \alpha_i(t)-\alpha_i(s)\leq c_i(t|s),\tag{FPC}
\end{equation}  
where $c_i(t|s)=\inf_{k\in K_{is}}\frac{1}{v(k)}C(t,i,s,k)$ is the \emph{least cost} for a member of group $i$ with natural score $s$ to falsify to $t$.\footnote{This infimum exists because $C(t,\theta)$ is bounded below by 0. We assume that this bound is tight in the sense that, for every $i,s,t$ and every $\varepsilon>0$, there exists a strictly positive mass of agents from group $i$ with natural score $s$ whose cost of falsifying to $t$ is lower than $c_i(t|s)+\varepsilon$.}
We refer to $c_i$ as the \emph{least-cost} or simply cost function for group $i$. 

\paragraph{Falsification cost.} We assume that each group's least-cost function satisfies the assumptions we make on the single cost function in the paper. In particular, we assume that each least-cost function satisfies either \eqref{eq:uid} or \eqref{eq:udd}.

 \paragraph{Allocative constraints.} In many applications we consider, allocative constraints are irrelevant because prizes are immaterial, such as services, certificates, labels or awards. To cover other applications, we allow for a \emph{resource constraint} and \emph{quota constraints}. The resource constraint requires 
\begin{equation}\label{eqRC}
\sum_i\mu_i \int_{\underline{s}_i}^{\overline{s}_i}\alpha_i(s)dF_i(s)\leq \bar{\rho}. \tag{RC}
\end{equation}
In addition, the designer may have to satisfy a system of exogenous quotas $\phi=(\phi_i)_{i\in I}$, where $\phi_i\in[0,1]$ is a fraction of objects reserved for group $i$, with $\sum_i\phi_i\leq 1$, and $\phi_i\bar{\rho}\leq \mu_i$. The quota constraints are
\begin{equation}\label{eqQC}
\forall i,~~\mu_i \int_{\underline{s}_i}^{\overline{s}_i}\alpha_i(s)dF_i(s)\geq \phi_i \bar{\rho}. \tag{QC}
\end{equation}
A mechanism is feasible if it satisfies these \emph{allocative constraints}. If $\bar{\rho}=1$ and $\phi=0$, the problem has no allocative constraints. Feasible mechanisms also need to satisfy probability constraints:
\begin{equation}\label{eqPCgroups}
\forall (i,s)~~ 0\leq \alpha_i(s)\leq 1.\tag{PC}
\end{equation}

\paragraph{Designer's program.} The designer's program is to choose a score-based allocation rule $\bm{\alpha}$ that solves:
\begin{align*}
\max_{(\alpha_i)_{i\in I}}&~\sum_i\mu_i \int_{S_i} w_i(s)\alpha_i(s)dF_i(s)~~~
\text{s.t.}~~~\eqref{eqPCgroups}, \eqref{eqFPCgroups}, \eqref{eqRC},\eqref{eqQC}. \tag{P}\label{eqP}
\end{align*}

To solve the designer's problem, we decompose it into a collection of group specific \emph{within problems}, that consist in optimally allocating a fixed mass of objects within a group, and an overall \emph{across problem} of optimally choosing the masses of objects accruing to the different groups while satisfying the allocative constraints.

\paragraph{Within problem.} Let $\rho_i$ be the mass of objects allocated to group $i$. Then the corresponding within problem is
\begin{align*}
W_i(\rho_i)=\max_{\alpha_i}&~\int_{S_i}\alpha_i(s)w_i(s)dF_i(s)\tag{$\text{P}_W$}\label{eqPin}\\
\text{s.t.}&~\eqref{eqFPC},\eqref{eqPC},\\
&~\mu_i\int_{S_i}\alpha_i(s)dF_i(s)=\rho_i,\tag{$\text{RC}_W$}\label{eqRCwp}
\end{align*}
where the within resource constraint \eqref{eqRCwp} must hold with equality. Since we require the whole mass $\rho_i$ to be allocated, the within problem is feasible only if $\rho_i\leq \mu_i$, hence its value function is equal to $-\infty$ otherwise.

\paragraph{Across problem.} A \emph{group allocation profile} $\bm{\rho}=(\rho_i)_{i\in I}$ satisfies the allocative constraints if it belongs to the feasible set
$R=\left\{\bm{\rho}\,:\,\sum_i\rho_i\leq \bar{\rho},~\rho_i\geq \phi_i\bar{\rho}~(\forall i)\right\}$. 
The designer's problem is then summarized by the across problem 
\begin{equation}
\overline{W}(\bm{F},\bm{c})=\max_{\bm{\rho}\in R}~\sum_i \mu_i W_i(\rho_i),\tag{$\text{P}_A$}\label{eqPcross}
\end{equation}
where $\bm{F}=(F_i)_{i\in I}$ and $\bm{c}=(c_i)_{i\in I}$ denote profiles.

\subsection{Optimal within group allocation}

We first derive an optimal allocation rule for the within problem. To simplify notation, we drop the group index \(i\). Let \(\hat{w}/\mu\) be the Lagrange multiplier on the resource constraint, where $\mu$ is the size of the group. The Lagrangian for \(\eqref{eqPin}\) is then
\[
\int_S \alpha(s) \{ w(s) - \hat{w} \} dF(s) + \hat{w} \frac{\rho}{\mu}.
\]
Maximizing the Lagrangian for a fixed value of the multiplier $\hat{w}/\mu$ is therefore equivalent to solving our baseline problem in the paper \eqref{eqBP} with a shifted worth function $w(s)-\hat{w}$. Alternatively, we can view $\hat{w}$ as a new value of the outside option. Let $\alpha^*(\cdot,\hat{w}, r)$ denote the allocatively optimal allocation rule for the shifted worth function. To solve the within problem, we then need to identify the value of the Lagrange multiplier, i.e., the shift that makes \eqref{eqRCwp} hold. To do this, we first study how the baseline allocation $\alpha^*(\cdot,\hat{w}, r)$ varies with the shift $\hat{w}$ and the neutral gap share $r$. 

\paragraph{The mechanics of indirect effects.} Since adjusting the shift and the neutral gap share are the tools that allow for achieving a particular group allocation, understanding how the baseline allocation reacts to such changes reveals the mechanics of the indirect effects. The effect of $\hat{w}$ is the same as that of $-\zeta$ in \secref{sec:general fals}, \propref{PropShift}.

\begin{Prop}[Effect of outside option and neutral gap share]\label{PropOutOpt}
The baseline allocation rule $\alpha^*(s,\hat{w},r)$ is decreasing in $\hat{w}$. It is continuous at $\hat{w}$ and independent of $r$ unless \eqref{eq:mult} holds, in which case it is strictly increasing and continuous in $r$. Furthermore, it satisfies $\lim_{\hat{w}\to\bar{w}^-}\alpha^*(s,\hat{w},r)=\alpha^*(s,\bar{w},1)$ and $\lim_{\hat{w}\to\bar{w}^+}\alpha^*(s,\hat{w},r)=\alpha^*(s,\bar{w},0)$.
\end{Prop}


Intuitively, a higher value of the outside option should reduce the mass of allocated objects in the baseline problem. In fact, a stronger result holds since the effect is uniform across all scores: for every $s$, the baseline allocation probability is decreasing in $\hat{w}$. A higher value of the neutral gap share naturally increases the baseline allocation probability for all scores when it is effective, that is, under \eqref{eq:mult}.

Let $A^*(\hat{w}, r) = \int_S \alpha^*(s, \hat{w}, r) dF(s)$ denote the per capita mass of objects allocated to the group under the baseline rule $\alpha^*(s, \hat{w}, r)$. 


\begin{Cor}\label{CorMass}
The per capita mass of allocated objects $A^*(\hat{w}, r)$, is strictly decreasing in \(\hat{w}\). It is continuous in $\hat{w}$ and independent of $r$ unless \eqref{eq:mult} holds, in which case \(A^*(\bar{w}, r)\) is strictly increasing and continuous in $r$, and satisfies $\lim_{\hat{w} \to \bar{w}^-} A^*(\hat{w}, r) = A^*(\bar{w}, 1)$ and $\lim_{\hat{w} \to \bar{w}^+} A^*(\hat{w}, r) = A^*(\bar{w}, 0)$.
\end{Cor}

\paragraph{Optimal within allocation.} Returning to the within problem, the outside option $\hat{w}$ is equal to the Lagrange multiplier on the resource constraint, scaled by group size $\mu$, and can be interpreted as the shadow price of marginally tightening the constraint. To ensure that \eqref{eqRCwp} holds, we must adjust $\hat{w}$ and $r$ so that $A^*(\hat{w}, r) =\rho/\mu$. \theoremref{ThWithinSol} shows that finding such values of $\hat{w}$ and $r$ is always possible:
\begin{Thm}[Optimal within group allocation]\label{ThWithinSol}
For any $0\leq \rho\leq \mu$, there exists a unique outside option value $\hat{w}(\rho)$ and, under \eqref{eq:mult},
a unique neutral gap share $r(\rho)$, such that $\mu A^*\bigl(\hat{w}(\rho), r(\rho)\bigr)=\rho$. Furthermore, $\hat{w}(\rho)$ is continuous, decreasing in $\rho$ unless \eqref{eq:mult} holds, in which case it is constant at $\bar{w}$. The function $r(\rho)$ is continuous and strictly increasing. The baseline allocation rule $\alpha^*(s,\hat{w}(\rho),r(\rho))$ is then the unique solution to the within problem \eqref{eqPin}. The value function of \eqref{eqPin}, $W(\rho)$ is strictly concave in $\rho$ unless \eqref{eq:mult} holds.
\end{Thm} 

To see why \theoremref{ThWithinSol} holds, note that, by assumption, $w(s)$ is bounded. It is easy to see that for any $\hat{w}$ below the lower bound on $w(s)$, the unique baseline allocation rule allocates with certainty to all scores, regardless of the scores. Similarly, for any $\hat{w}$ above the upper bound on $w(s)$, the baseline allocation rule never allocates any object. Hence, by varying $\hat{w}$ between these bounds, we can find an outside option $\hat{w}(\rho)$ such that the baseline allocation rule satisfies the resource constraint and therefore solves \eqref{eqPin}. If $\hat{w}(\rho) = \bar{w}$ and $c(\overline{s}|\underline{s})< 1$, we also need to adjust $r$ to a unique value $r(\rho)$ in order to allocate exactly $\rho$ objects. The allocation rule $\alpha^*(s, \hat{w}(\rho), r(\rho))$ is then the unique solution to the within problem.

\subsection{Optimal across-group allocation}

We characterize the solution to the across problem and provide an algorithm to determine the optimal allocation profile $\bm{\rho} = (\rho_i)_{i \in I}$. 

\begin{Thm}[Optimal across group allocation]\label{ThAcrossSol}
The across problem \eqref{eqPcross} admits a solution $\bm{\rho}$. Furthermore, $\bm{\rho}$ solves the across problem if and only if there exist a scalar $\lambda_R\geq 0$ and, for each $i$, a scalar $\lambda_i\geq 0$, an outside option value $\hat{w}_i(\rho_i)$, and a neutral gap share $r_i(\rho_i)$ such that:
\begin{multicols}{2}
\begin{enumerate}[(i)]
\item $\lambda_i(\phi_i\bar{\rho}-\rho_i)=0$ for all $i$,
\item $\lambda_R\bigl(\sum_i\rho_i-\bar{\rho}\bigr)=0$,
\item $\hat{w}_i(\rho_i)=\lambda_R-\lambda_i$,
\item $\mu_i A_i^*\bigl(\hat{w}_i(\rho_i),r_i(\rho_i)\bigr)=\rho_i$. 
\end{enumerate}
\end{multicols}
The solution $\bm{\rho}$ is unique if, for each $i$, \eqref{eq:mult} does not hold at $\hat{w}_i(\rho_i)$.
\end{Thm}

The characterization in \theoremref{ThAcrossSol} suggests the following algorithm to find a solution to the across problem. First, we compute the solutions to each within problem, setting all outside options to 0. For these solutions, we check which constraints are binding or violated. Next, we adjust outside options to satisfy all the previously violated constraints with equality when recomputing the corresponding solution. This may lead to hitting additional constraints. Indeed, increasing the allocation to one group to satisfy its quota may lead to the violation of a previously slack resource constraint, or another group's quota constraint if the resource constraint was already binding. If so, we adjust outside options to satisfy with equality all constraints that were binding or violated at any previous step. Since, at any step, the set of constraints that have required adjustment at some previous step is bounded and necessarily increases with each step, the process must eventually end. The allocation profile at which it ends is a solution to the across problem. In the next subsection, we state the algorithm formally  and show that it finds a solution to the across problem.


\subsection{Algorithm for the across problem}\label{AppAlgo}


We present an algorithm that finds a solution to the across problem. We extend the definition of $\hat{w}^\phi_i$ as the unique value of $\hat{w}$ such that $\mu_i A^*_i(\hat{w},r)=\phi_i\bar{\rho}$ for some $r$, and let $r^\phi_i$ be the unique value of $r$ that satisfies this equality if $\hat{w}^\phi_i=\bar{w}_i$ (otherwise, let $r^\phi_i$ be any value on $[0,1]$).

\begin{algorithm}[H]
\SetAlgoLined
$\forall i,~~\rho_i^0\leftarrow \mu_iA_i^*(0,1)$\;
$R^0\leftarrow \mathcal{R}(\bm{\rho}^0)$\;
$Q^0\leftarrow \mathcal{Q}(\bm{\rho}^0)$\;
$k\leftarrow  0$\;
\Repeat{$Q^{k}=Q^{k-1}$ and $R^{k}=R^{k-1}$}{
$k\leftarrow k+1$\;
$\forall \ell \in Q^{k-1},~~\hat{w}_\ell^{k}\leftarrow \hat{w}_\ell^\phi$ and $r_\ell^{k}\leftarrow r_\ell^\phi$\;
\eIf{$R^{k-1}=0$}{$\forall \ell \notin Q^{k-1},~~\hat{w}_\ell^{k}\leftarrow 0$ and $r_\ell^{k}\leftarrow 1$\;}
{
$\hat{w},r\leftarrow$ \emph{Solution of:} $\sum_{\ell\in Q^{k-1}}\phi_\ell\bar{\rho}+\sum_{i\notin Q^{k-1}}\mu_iA_i^*(\hat{w},r\mathbbm{1}_{\hat{w}=\bar{w}_i})= \bar{\rho}$\;
$\forall \ell \notin Q^{k-1},~~\hat{w}_\ell^{k}\leftarrow \hat{w}$ and $r_\ell^{k}\leftarrow r\mathbbm{1}_{\hat{w}=\bar{w}_\ell}$\;
}
$\forall i,~~\rho_i^{k}\leftarrow \mu_iA_i^*(\hat{w}_i^k,r_i^k)$\;
$R^k\leftarrow \mathcal{R}(\bm{\rho}^k)$\;
$Q^k\leftarrow \mathcal{Q}(\bm{\rho}^k)$\;
}
\caption{Algorithm to solve the across problem}\label{Algo}
\end{algorithm}

We did not specify how to find a solution $(\hat{w},r)$ to
$$\sum_{\ell\in Q^{k-1}}\phi_\ell\bar{\rho}+\sum_{i\notin Q^{k-1}}\mu_iA_i^*(\hat{w},r\mathbbm{1}_{\hat{w}=\bar{w}_i})= \bar{\rho}$$
within the algorithm. Note, however, that the left-hand side of the equation can be decreased continuously by continuously raising $\hat{w}$ whenever $\hat{w}\neq\bar{w}_i$, for all $i$, and by continuously decreasing $r$ from 1 to 0 and keeping $\hat{w}$ constant whenever $\hat{w}=\bar{w}_i$ for some $i$. Therefore a simple algorithm can solve this equation.

\begin{Prop}\label{prop:algo}
\algoref{Algo} finds a solution of \eqref{eqPcross} in finitely many steps.
\end{Prop}

\subsection{Comparative statics and welfare results} \label{sec-cs-indirect}

\paragraph{Designer welfare.} We show that designer welfare is decreasing in gaming ability, and increasing with first-order stochastic dominance shifts of the score distribution. Let $\bm{c^\gamma}$ denote a profile of parameterized cost functions $c^{\gamma_i}_i=\frac{1}{\gamma_i}c_i(t|s)$.

\begin{Prop}[Properties of the designer's value function]\label{PropAcrossWelf}
The value function of the across problem, $\overline{W}(\bm{F},\bm{c^\gamma})$, is nonincreasing in $\gamma_i$, and nondecreasing in $F_i$ with respect to the first-order stochastic dominance order. 
\end{Prop}

Any increase in gaming ability tightens \eqref{eqFPC}, thereby reducing welfare. The effect of score distributions is more difficult to analyze, as a first-order stochastic dominance shift in the score distribution of one group may have a non-obvious direct effect on the baseline allocation rule for that group (see \propref{PropDist}), as well as intricate indirect effects. However, an envelope theorem argument implies that we can bypass the analysis of these complex effects and instead focus on the effect of the score distribution on welfare while holding the allocation rule fixed. Even then, the effect remains difficult to analyze because the surplus function $w_i(s) - \hat{w}_i$ takes both positive and negative values on $S_i$. Nevertheless, an argument based on the analysis of cumulative surplus functions and the differential form of the objective function shows that improving the score distribution in the first-order stochastic dominance order increases
the designer's payoff. 

\paragraph{An externality interpretation of comparative statics.} This comparative statics result can be interpreted in terms of externalities. We use gaming ability to parameterize the least-cost function. Recall that only the least-cost agents shape \eqref{eqFPC} and thereby the optimal rule. Their presence imposes an externality on other agents. Reducing gaming ability can be interpreted as a removal of least-cost agents, and the resulting impact on remaining agents reflects the externality exerted by these least-cost agents. For example, when gaming ability is low, \propref{PropGam} implies that least-cost agents exert a positive externality on low-score agents but a negative one on high-score agents.

\paragraph{Comparative statics: the consequences of indirect effects.} Comparative statics in the general problem can have rich effects. There can be numerous scenarios depending on initial conditions regarding gaming abilities and score distributions, the number of groups, quotas, and the mass of objects to be allocated. Instead of listing formal results that exhaust all these cases, we find it more effective to focus on an interesting example. We choose an example that illustrates how a quota can be used to shield a group from the indirect effects of transformations occurring in other groups.

 
\begin{figure}[ht]
\begin{center}
\scalebox{0.65}{
\input{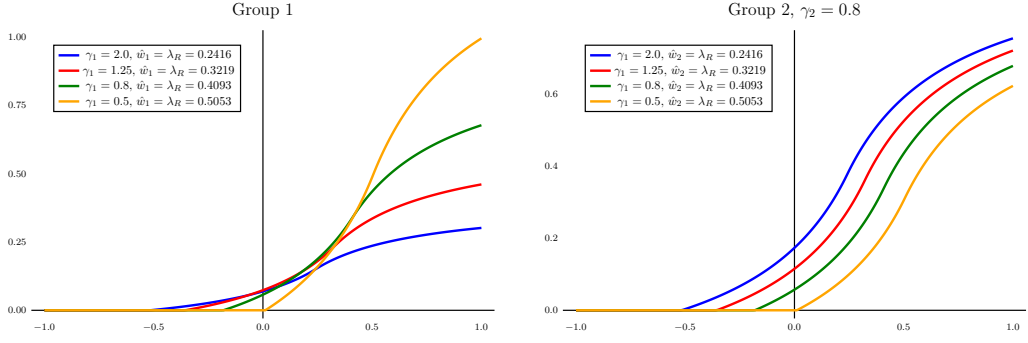}}
\caption{ Group 1's lower gaming ability hurts a low quota group ($\phi_2=0.2$).
Cost $\mathcal{C}^{\gamma_i}(x)=(1/\gamma_i)x/(1+x)$, score distribution $U(-1,1)$, $\phi_2=0.2$, $\gamma_2=0.8$, $\bar{\rho}=0.2$, $\mu_1=\mu_2=0.5$
}\label{FigCSaccross-L}
\end{center}
\end{figure}


\begin{figure}[ht]
\begin{center}
\scalebox{0.45}{
\input{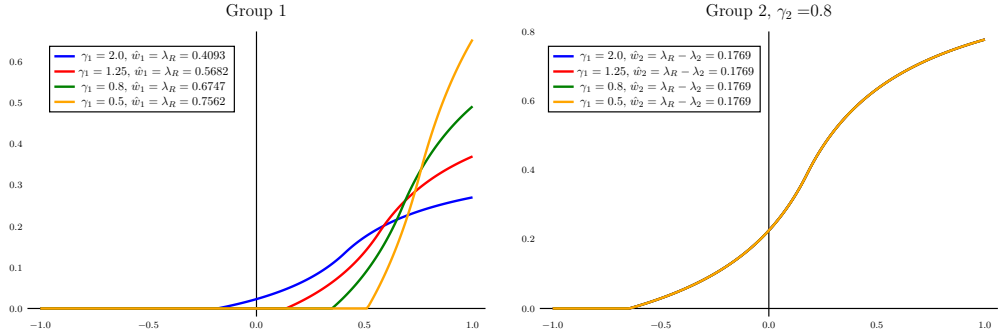}
}
\caption{Group 1's lower gaming ability leaves a high quota group  unaffected}\label{FigCSaccross-H}
\end{center}
\end{figure}

\figref{FigCSaccross-L} and \figref{FigCSaccross-H} illustrate our example. We consider two equally sized groups, $\mu_1=\mu_2=0.5$, and gradually lower the gaming ability of group 1 from $\gamma_1=2$ to $\gamma_1=0.5$ while keeping group 2's gaming ability fixed at $\gamma_2=0.8$, all corresponding to a slack regime. Both groups are otherwise identical, with scores uniformly distributed on $[-1,1]$, and a Euclidean cost function $\mathcal{C}^\gamma(x)=\frac{1}{\gamma}\frac{x}{(1+x)}$. The mass of objects is $\bar{\rho}=0.2$ which causes the resource constraint to bind for all parameters considered. There is a quota $\phi_2$ of objects reserved for group 2. We consider two scenarios: in the low-quota scenario, the quota $\phi_2=0.2$ is never binding, whereas in the high-quota scenario, the quota $\phi_2=0.8$ is always binding.

In each case, the decrease in gaming ability of group 1 generates a direct effect that only affects group 1, and an indirect effect that may affect both groups. For each successive decrease in $\gamma_1$, since we remain under the slack regime (see \propref{PropGam}), the direct effect on group 1 is to increase the allocation probability for all scores, increasing the mass of objects allocated to this group.
  
Consider the low-quota scenario, illustrated in \figref{FigCSaccross-L}. As $\gamma_1$ decreases, the direct effect results in a higher mass of objects being allocated to group 1. However, when keeping the outside option $\hat{w}_1$ at its original value, the direct effect of each decrease in $\gamma_1$ leads to a violation of the resource constraint. To compensate for this direct effect and restore feasibility, the endogenous outside option, which is common to both groups, adjusts upward, generating the indirect effect. This indirect effect shifts group 2's original allocation rule to the right, reducing the probability of receiving an object for all scores. The final allocation rule for group 1 results from the combination of the upward rotation due to the direct effect and the rightward shift due to the indirect effect. It implies that high-score agents gain while low-score agents lose. The overall effect, however, is to increase the mass of objects allocated to group 1 since both groups bear the allocative costs of the indirect effect.

In the high-quota scenario, illustrated in \figref{FigCSaccross-H}, the quota for group 2 is so high that it always binds. To satisfy this quota and ensure that the mass of objects assigned to group 2 remains $0.8\bar{\rho}$, the outside option of group 2 must remain constant. The direct effect of decreasing $\gamma_1$ remains the same as in the first scenario. The direct effect of each decrease in $\gamma_1$ still leads to a violation of the resource constraint. To compensate for this direct effect while maintaining the quota for group 2, group 1 must bear the full allocative cost of the adjustment through an increase in its outside option. This implies that the overall mass of objects allocated to group 1 remains constant, while high-score agents still gain and low-score agents still lose.

\section{Related literature}\label{sec-lit}

We contribute to the literature on optimal allocation mechanisms with privately informed agents. In the seminal contribution of \cite{mye81}, the designer uses monetary transfers to target allocation in order to maximize revenue. However, monetary transfers may lose their effectiveness if the designer has a more general objective, as in \cite{condorelli2013market}, or wishes to maximize a weighted combination of utilitarian and revenue objectives, as in \citet*{akbarpour2020redistributive}. Both studies establish conditions under which the designer optimally refrains from using transfers altogether. We also consider a general objective and exclude transfers. While there may be exogenous reasons to rule out transfers, \cite{condorelli2013market} and \cite{akbarpour2020redistributive} show that doing so is optimal if the designer's utility from allocation is negatively correlated with willingness to pay. In contrast to these works, we study a setting in which agents' private information can be manipulated at a cost. Methodologically, our approach differs from standard techniques based on virtual surplus that work directly with the allocation rule. Instead, we use cumulative surplus and work with the derivative of the allocation rule to identify the growth interval.

We add to a sizable literature on non-market optimal mechanisms that studies the use of alternative targeting tools in lieu of transfers. In our setting, targeting is enabled by the availability of the (possibly falsified) score. \citet{bdl2019} rely on evidence disclosure. \citet{ben2014optimal}, \citet{mylovanov2017optimal}, \citet{erlanson2019note}, \citet{chua2019optimal}, \citet{epitropou2019optimal}, and \citet{li2020mechanism} use ex post (costly) inspection or verification with limited penalties. \citet{hartrough2008} and \citet{dworczak2022equity} consider money (or utility) burning, while \citet{patel2023costly} combine verification and money burning. In \citet{kattwinkel2019allocation}, the designer has access to a private signal correlated with the agent's private information, while in \citet{kattknoe2022}, they can additionally verify the agent's type. In contrast, we consider costly state falsification and impose falsification-proofness. This is similar to money burning in that it is wasteful, but it differs in that the cost of falsification is type-dependent. In a related framework, \citet{augiasperez2023} study the optimal design of allocation mechanisms when agents can improve their scores.

We also contribute to the literature on costly screening. \citet{frankel2020improving} and \citet{ball2020scoring} study the optimal design of linear scores under a gaming technology that amounts to costly falsification. \citet{lan_pla16} characterize optimal tax design under costly income hiding. \citet{kephart2016revelation}, \citet{deneckere2017screening}, and \citet{severinov2019screening} study mechanism design with misreporting costs but focus on settings (mainly with transfers) in which falsification-proofness is without loss. \citet{tan2023price} considers a price-discrimination setting in which consumers can distort their data at a cost to avoid high prices. \citet{li2024screening} study costly screening in a multi-agent setting without transfers and identify conditions under which contests are optimal, as well as situations in which random mechanisms dominate contests. \citet{lac_wei89} investigate the design of risk-sharing contracts with costly state falsification, focusing on optimal falsification-proof contracts. They show that this constraint may lead to a loss, without characterizing the optimal contract. We build on \citet{perez2022test}, where we show that allocatively optimal mechanisms necessarily induce falsification. 

Finally, we add to the growing list of papers using optimal transport in economics, surveyed in \cite{carlier2012optimal} and \cite{galichon2018optimal}. More recently, it has been applied to mechanism design problems with multidimensional private information \citep{daskalakis2017strong,kolesnikov2022beckmann}, information design \citep{arieli2022persuasion,kolotilin2022persuasion,lin2022credible}, and labor-market sorting problems \citep{boerma2021sorting}. Most of these papers rely on duality-based characterizations in optimal transport.\footnote{For example, \cite{lin2022credible} rely on properties characterizing optimal coupling for given marginals to establish their characterization of stable credible signals.} In contrast, our approach relies directly on the dual of the optimal transport problem, which we show to be equivalent to our problem. This requires constructing the marginal distributions, which we obtain by reinterpreting the cumulative surplus generated by allocation rules as probability measures over the spaces of eligible and ineligible scores. Thus, our approach provides an original solution method that differs both from Myersonian techniques in settings with transfers and from Lagrangian techniques in settings without transfers, as in \cite{amador2006commitment} and \cite{amador2020regulating}.

\appendix


\section{Lagrangian for the differential problem}\label{ssecBP}

\begin{Lem}\label{LemLagSuff}
An allocation rule $\alpha$ solves \eqref{eqBP} if and only if it satisfies \eqref{eqPC}, \eqref{eq:mon} and \eqref{eq:init}, and there exists a Lagrange multiplier $\nu\geq 0$ such that:
\begin{enumerate}[(i)]
\item \stepnoskip{Growth interval:} $\alpha(s)=\alpha(\underline{s})$ if $s< s_*(\nu)$, and $\alpha(s)=\alpha(\overline{s})$ if $s> s^*(\nu)$;
\item \stepnoskip{Continuity:} $\alpha$ is continuous at $s_*(\nu)$ and at $s^*(\nu)$;
\item \stepnoskip{Relaxed \eqref{eqFPC}:} $\alpha(t)-\alpha(s)\leq c(t|s)$ for every $s_*(\nu)\leq s<t\leq s^*(\nu)$;
\item \stepnoskip{Complementary slackness on \eqref{eqPC}:} If $\bar{w}\leq 0$,  $\nu\bigl(1-\alpha(\overline{s})\bigr)=0$; if $\bar{w}\geq 0$, $\nu\alpha(\underline{s})=0$;
\item \stepnoskip{Optimality for the simplified program:} For every allocation rule $\hat{\alpha}$ that satisfies (iii),
\[\int_{s_*(\nu)}^{s^*(\nu)}w(s)\alpha(s)dF(s)\geq \int_{s_*(\nu)}^{s^*(\nu)}w(s)\hat{\alpha}(s)dF(s).\]
\end{enumerate}
\end{Lem}
\begin{proof}
We proceed in two steps. The first step is a standard Lagrangian necessity and sufficiency theorem. The second step shows that the conditions of the lemma are equivalent to the Lagrangian conditions. In this proof, we say that an allocation rule $\alpha$ is \emph{feasible} if it satisfies \eqref{eq:mon}, \eqref{eqFPC}, and the boundary normalization: $\alpha(\underline{s})=0$ if $\bar{w}<0$ and $\alpha(\overline{s})=1$ if $\bar{w}>0$. It is immediate to verify that the set of such feasible allocation rules, which we denote by $\mathbbm{A}$, is convex. 

Throughout, we let $[x]_y^z=y\Ind_{x<y}+x\Ind_{y\leq x\leq z}+z\Ind_{x>z}$ for any $x$ and $y<z$.

\step{Step 1:} \emph{An allocation rule $\alpha\in\mathbbm{A}$  solves \eqref{eqBP} if and only if (A) $\alpha$ satisfies \eqref{eqPC}; and there exists $\nu\geq 0$ such that (B) $\nu=0$ or $\overline{\alpha}-\underline{\alpha}=1$, and (C) $\mathcal{L}(\alpha,\nu)\geq\mathcal{L}(\hat{\alpha},\nu)$ for every  allocation rule $\hat{\alpha}\in\mathbbm{A}$.}

\step{\fbox{$\Leftarrow$}~} For any $\hat{\alpha}\in\mathbbm{A}$ that satisfies \eqref{eqPC}, 
\begin{align*}
\int_S \mathcal{W}(z)d\alpha(z)=\mathcal{L}(\alpha,\nu)&\geq \mathcal{L}(\hat{\alpha},\nu)=\int_S\mathcal{W}(z)d\hat{\alpha}(z)+\nu\left(1-\int_S d\hat{\alpha}(z)\right)\\
&\geq \int_S\mathcal{W}(z)d\hat{\alpha}(z),
\end{align*}
where (B) implies the first equality, (C) implies the first inequality, and the last inequality holds because $\hat{\alpha}$ satisfies \eqref{eqPC} and $\nu\geq0$. 

\step{\fbox{$\Rightarrow$}~} For every $b\geq 0$, consider the program where we replace the probability constraint \eqref{eqPC} by the constraint $g(\alpha)\leq b$ where $g(\alpha)=\int_S d\alpha(z)$. Let its value be 
\[h(b)=\sup_{\tilde{\alpha}\in\mathbbm{A}}~\int_S\mathcal{W}(z)d\tilde{\alpha}(z)~~\text{s.t.}~g(\tilde{\alpha})\leq b.\]
Under nonneutral priority, the boundary normalization implies that $g(\tilde{\alpha})\leq 1$ is equivalent to \eqref{eqPC}. Under neutral priority, any $\tilde{\alpha}\in \mathbbm{A}$ with $g(\tilde{\alpha})\leq 1$ can be shifted by $\tilde{\alpha}(\underline{s})$ to satisfy \eqref{eqPC} and \eqref{eq:init} without changing $\int_S\mathcal{W}(z)d\tilde{\alpha}(z)$. Thus, if $\alpha$ solves \eqref{eqBP}, then $h(1)=\int_S\mathcal{W}(z)d\alpha(z)$.

Since the objective and the constraint $g(\cdot)$ are both linear in $\tilde{\alpha}$, and $\mathbbm{A}$ is convex, $h(b)$ is a concave function. It is also obviously nondecreasing. Moreover, \eqref{eqFPC} implies $g(\tilde{\alpha})\leq c(\overline{s}|\underline{s})$, so $h$ is finite. Let $\nu\geq 0$ be the left-derivative of $h$ at $b=1$, which exists by concavity and is nonnegative by monotonicity. 

By assumption, $h(1)=\int_S\mathcal{W}(z)d\alpha(z)$. If $g(\alpha)=1$, we also have $\int_S\mathcal{W}(z)d\alpha(z)=\mathcal{L}(\alpha,\nu)$. Otherwise, we must have $g(\alpha)<1$. But then $\alpha$ must also solve the program for any $b\in[g(\alpha),1]$, implying that $h$ is constant on this interval, and $\nu=0$. Then again, $\int_S\mathcal{W}(z)d\alpha(z)=\mathcal{L}(\alpha,\nu)$. For all $\hat{\alpha}\in \mathbbm{A}$, $\int_S\mathcal{W}(z)d\hat{\alpha}(z)\leq h\bigl(g(\hat{\alpha})\bigr)$, by definition of $h$, and $h\bigl(g(\hat{\alpha})\bigr)\leq h(1)+\nu\bigl(g(\hat{\alpha})-1\bigr)$, by concavity of $h$. Hence, 
\[\mathcal{L}(\hat{\alpha},\nu)=\int_S\mathcal{W}(z)d\hat{\alpha}(z)-\nu\bigl(g(\hat{\alpha})-1\bigr)\leq h(1)=\int_S\mathcal{W}(z)d\alpha(z)=\mathcal{L}(\alpha,\nu).\]

\step{Step 2:} \emph{An allocation rule $\alpha$ and a scalar $\nu\geq 0$ satisfy \eqref{eq:mon}, \eqref{eqPC}, \eqref{eq:init}, and (i)-(v) if and only if $\alpha$ and $\nu$ satisfy  $\alpha\in\mathbbm{A}$, (A), (B) and (C).}

\step{\fbox{$\Rightarrow$}~} (A) is immediate.  (B) must hold for the same scalar $\nu$ by \eqref{eq:init} and (iv). Next, we show that (i)-(iii) imply that $\alpha$ satisfies \eqref{eqFPC}, so that $\alpha\in \mathbbm{A}$ since it already satisfies \eqref{eq:mon} and \eqref{eq:init}. Let $s<t$, $s'=\bigl[s\bigr]_{s_*(\nu)}^{s^*(\nu)}$ and $t'=\bigl[t\bigr]_{s_*(\nu)}^{s^*(\nu)}$. Then
$\alpha(t)-\alpha(s)= \alpha(t')-\alpha(s')\leq c(t'|s')\leq c(t|s)$,
where the first equality is from (i) and (ii), the first inequality is from (iii), and the last inequality is by cost monotonicity. Hence, $\alpha\in \mathbbm{A}$ satisfies (A) and (B).

For every $\hat{\alpha}\in\mathbbm{A}$, its restriction to $I=[s_*(\nu),s^*(\nu)]$ satisfies (iii), and (C) follows from:
\small
\begin{align*}
\mathcal{L}(\alpha,\nu)&= \nu+\int_{I}w(s)\alpha(s)dF(s)\tag{by (i)-(ii) and IBP}\\
&\geq \nu+\int_{I}w(s)\hat{\alpha}(s)dF(s)\tag{by (v)}\\
&=\nu+\int_{I}\bigl[\mathcal{W}(z)-\nu\bigr]d\hat{\alpha}(z)\tag{by \eqref{eqZAS}}\\
&\geq \nu+\int_S\bigl[\mathcal{W}(z)-\nu\bigr]d\hat{\alpha}(z)=\mathcal{L}(\hat{\alpha},\nu),\tag{by \lemmaref{LemCumSurpl} (iii)}
\end{align*}
\normalsize
where we use \lemmaref{LemCumSurpl} (ii), and IBP means integration by parts.

\step{\fbox{$\Leftarrow$}~}  $\alpha\in\mathbbm{A}$ directly implies \eqref{eq:mon}, \eqref{eq:init}, (ii), and (iii). (A) directly implies \eqref{eqPC}. (B) and \eqref{eq:init} imply (iv).

Next, we show that (C) implies (i). Suppose $\alpha$ does not satisfy (i). Then $\alpha$ cannot satisfy (C). To show this, suppose we are in a low-score priority setting, so feasibility requires $\alpha(\underline{s})=0$ (the high-score and neutral priority cases are analogous). Suppose first $\alpha\bigl(s_*(\nu)^-\bigr)\equiv \lim_{s\to s_*(\nu),s<s_*(\nu)}\alpha(s)>\alpha(\underline{s})=0$. Then consider $\hat{\alpha}(s)=\max\bigl\{\alpha(s)-\alpha\big(s_*(\nu)^-\bigr),0\bigr\}$. Then we have
\[\mathcal{L}(\hat{\alpha},\nu)-\nu=\int_{s_*(\nu)}^{\overline{s}} \bigl[\mathcal{W}(z)-\nu\bigr]d\alpha(z)>\int_{\underline{s}}^{\overline{s}} \bigl[\mathcal{W}(z)-\nu\bigr]d\alpha(z)=\mathcal{L}(\alpha,\nu)-\nu,\]
since $\mathcal{W}(z)<\nu$ on $[\underline{s},s_*(\nu))$. Similarly, if $\alpha\bigl(s^*(\nu)^+\bigr)<\alpha(\overline{s})$, a rule such that $\hat{\alpha}(s)=\alpha(s)$ for $s\leq s^*(\nu)$, and $\hat{\alpha}(s)=\alpha\bigl(s^*(\nu)^+\bigr)$ otherwise would yield $\mathcal{L}(\hat{\alpha},\nu)>\mathcal{L(\alpha,\nu)}$, contradicting (C). 

Next, we show (v). By the same construction as in \lemmaref{lem:mon}, any allocation rule that is nonmonotonic is dominated by a monotonic allocation rule for the simplified program. Consider any allocation rule $\hat{\alpha}$ that is monotonic on $[s_*(\nu),s^*(\nu)]$ and satisfies (iii). Since the simplified program only considers scores in $I=[s_*(\nu),s^*(\nu)]$, we can assume that $\hat{\alpha}$ is constant on $[\underline{s},s_*(\nu))$ and $(s^*(\nu),\overline{s})$. Then, let 
$\tilde{\alpha}(s)=\hat{\alpha}(s)+a,$ where $a=-\hat{\alpha}\bigl(s_*(\nu)\bigr)\Ind_{\bar{w}<0}+\bigl(1-\hat{\alpha}\bigl(s^*(\nu)\bigr)\bigr)\Ind_{\bar{w}> 0}$. Then, 
\begin{align*}
\int_{I}\hat{\alpha}(s)w(s)dF(s)=\int_{I}\tilde{\alpha}(s)w(s)dF(s)
&=\int_{I}\bigl[\mathcal{W}(z)-\nu\bigr]d\tilde{\alpha}(z)\tag{by IBP}\\
&=\int_S\bigl[\mathcal{W}(z)-\nu\bigr]d\tilde{\alpha}(z)=\mathcal{L}(\tilde{\alpha},\nu)-\nu\notag\\
&\leq \mathcal{L}(\alpha,\nu)-\nu\tag{by (C)}\\
&=\int_{I}\bigl[\mathcal{W}(z)-\nu\bigr]d\alpha(z)\tag{by (i)}\\
&=\int_{I}\alpha(s)w(s)dF(s).\tag{by IBP}
\end{align*}\end{proof}


\section{Proofs}\label{AppProofs}

 \begin{proof}[Proof of \lemmaref{LemCumSurpl}]
By strict monotonicity of $w(\cdot)$,  $w(s)$ has the same sign as $s-\hat{s}$, implying that both $\mathcal{W}^+$ and $\mathcal{W}^-$ are increasing on $[\underline{s},\hat{s}]$ and decreasing on $[\hat{s},\overline{s}]$, and therefore single-peaked at $\hat{s}$, proving (i). The existence of $s_*(\nu)$ and $s^*(\nu)$ is then ensured by continuity of both $\mathcal{W}^+$ and $\mathcal{W}^-$, and the fact that $\mathcal{W}(s)$ takes weakly negative values at both ends of the score interval. Then (ii), (iii) and (iv) are direct consequences of single-peakedness and continuity. For (v), note that, in the low-score priority, $\mathcal{W}(\overline{s})=\mathcal{W}^+(\overline{s})=0$, while, in the high-score priority case, $\mathcal{W}(\underline{s})=\mathcal{W}^-(\underline{s})=0$. 
\end{proof}


\begin{proof}[Proof of \lemmaref{lem:mon}]
Suppose $\alpha$ is feasible but not monotonic. Consider the nondecreasing function
$\tilde{\alpha}(s)=\alpha^-(s)\Ind_{s\leq \hat{s}}+\alpha^+(s)\Ind_{s> \hat{s}},$
where $\alpha^-:[\underline{s},\hat{s}]\to[0,1]$ is the lower nondecreasing envelope of $\alpha$ on $[\underline{s},\hat{s}]$, and $\alpha^+:[\hat{s},\overline{s}]\to[0,1]$ is the higher nondecreasing envelope of $\alpha$ on $[\hat{s},\overline{s}]$. Then $\tilde{\alpha}$ remains feasible. It obviously satisfies \eqref{eqPC}. Since $\tilde{\alpha}$ is nondecreasing, we only need to check \eqref{eqFPC} for upward falsification. Let $s<t$, and let $s'=\max\bigl\{x\geq s:\tilde{\alpha}(x)=\tilde{\alpha}(s)\bigr\}$, and $t'=\min\bigl\{x\leq t:\tilde{\alpha}(x)=\tilde{\alpha}(t)\bigr\}$. We can assume $s\leq s'<t'\leq t$, for otherwise $\tilde{\alpha}(t)=\tilde{\alpha}(s)$ and the proof is done. Then, 
\[\tilde{\alpha}(t)-\tilde{\alpha}(s)=\tilde{\alpha}(t')-\tilde{\alpha}(s')=\alpha(t')-\alpha(s')\leq c(t'|s')\leq c(t|s),\]
where the first equality obtains by definition of $s'$ and $t'$, and the second equality 
because $\tilde{\alpha}$ must coincide with $\alpha$ wherever it is not flat, and therefore also at the end of every flat interval. The first inequality is due to falsification-proofness of $\alpha$, and the last inequality to cost monotonicity.

Eligible scores are more likely, and ineligible scores less likely to get a prize under $\tilde{\alpha}$ than under $\alpha$. Hence, $\tilde{\alpha}$ dominates $\alpha$ for \eqref{eqBP}. Furthermore, if $\alpha$ is not monotonic, there must exist an interval of scores for which $\alpha$ and $\tilde{\alpha}$ do not coincide. Since $F$ has full support, $\tilde{\alpha}$ is therefore strictly better than $\alpha$.
\end{proof}


\begin{proof}[Proof of \lemmaref{lem:diffpgm}]
By \lemmaref{lem:mon}, imposing \eqref{eq:mon} in \eqref{eqBP} is without loss. We can then integrate the objective by parts as in \eqref{eq:objup} and \eqref{eq:objdn} so optimizing either objective under \eqref{eqPC}, \eqref{eqFPC} and \eqref{eq:mon} is equivalent to solving \eqref{eqBP}. Then using \eqref{eq:objup} under high-score priority and \eqref{eq:objdn} under low-score priority shows that choosing $\alpha$ to satisfy \eqref{eq:init} strictly increases the objective function. It also relaxes the probability constraint \eqref{eqPC} which can be equivalently written as 
\[0\leq \underline{\alpha},\text{ and }\underline{\alpha}+\int_Sd\alpha(z)\leq 1,~~~~~
\text{or}~~~~~~
\overline{\alpha}\leq 1,\text{ and }\overline{\alpha}-\int_S d\alpha(z)\geq 0.\]
Given that $\alpha$ must satisfy \eqref{eq:init} and eliminating the constant $\bar{w}$ from the objective under high-score priority, we obtain the equivalent program \eqref{pgm:DP}. 
\end{proof}

\begin{proof}[Proof of \theoremref{ThOptUDD}]
To keep notations simple, we only indicate the dependence of $\alpha^*_{udd}$ on $r$ when it is useful for the argument. We only need to check that the conditions of \lemmaref{LemLagSuff} are satisfied. Picking $\nu=\mathcal{W}(s_*)$, it is clear that  (i) holds and that $\alpha^*_{udd}$ satisfies \eqref{eq:mon}, \eqref{eq:init}, \eqref{eqPC} and (ii). For (iii), let $s_*\leq s<t\leq s^*$, then
\begin{align*}
\alpha_{udd}^*(t)-\alpha^*_{udd}(s)&=\int_s^t c_{t^+}(x|x)dx\leq \int_s^t c_{t}(x|s)dx=c(t|s),
\end{align*}
where the inequality follows from \eqref{eq:udd}. This also shows that the first-order approach is valid. Point (iv) is immediate to check. Furthermore, the differential program solved by $\alpha^*_{udd}$ is obtained from the program in (v) by using integration by parts on the objective function and absolute continuity of $\alpha$. Therefore, to show that (v) holds, we only need to prove the claim that \eqref{eqFPC} implies that a feasible allocation rule $\alpha$ must be absolutely continuous. For any $s<t$, \eqref{eqFPC} implies $\alpha(t)-\alpha(s)\leq c(t|s)=\int_s^t c_t(x|s)dx$ by the regularity assumption on $c$. \eqref{eq:udd} implies that this is bounded above by $\int_s^t c_t(x|\underline{s})dx\leq (t-s)\max_{x\in S} c_t(x|\underline{s})$. Hence, $\alpha$ must be Lipschitz.

While there may be several values of the Lagrange multiplier $\nu$ that work if $c_{t^+}(x|x)$ is equal to 0 both in the neighborhoods of $s_*$ and $s^*$, the corresponding optimal allocation rules for the program would be identical for all such values, so uniqueness of the solution to the differential program is granted when \eqref{eq:mult} does not hold. If, instead, \eqref{eq:mult} holds, the designer is indifferent between the allocation rules $\alpha^*_{udd}(s,r)$ for any $r\in[0,1]$. The argument is the same as in the \eqref{eq:uid} case.
\end{proof}


\begin{proof}[Proof of \theoremref{ThOptUID}]
To keep notations simple, we only indicate the dependence of $\alpha^*_{uid}$ on $r$ when it is useful for the argument. We check that the conditions of \lemmaref{LemLagSuff} are satisfied. We pick the multiplier $\nu=\mathcal{W}(s_*)\geq 0$. Then $\alpha^*_{uid}$ clearly satisfies (i). It is easy to check that it satisfies \eqref{eq:mon}, \eqref{eq:init} and \eqref{eqPC}, as well as (ii). To see that it satisfies (iv), note that $\alpha^*_{uid}(s^*)-\alpha^*_{uid}(s_*)=c(s^*|s_*)$ since \eqref{eqFPC} is binding for $(s_*,s^*)$. Hence, by \eqref{eq:bounds}, either $\alpha^*_{uid}(s^*)-\alpha^*_{uid}(s_*)$ is equal to 1 and the probability constraint is binding, or it is strictly less than 1, and then $\nu=0$ and $(s_*,s^*)=(s_*(0),s^*(0))$.

By the optimal transport connection established in \secref{sec:ot}, $\alpha^*_{uid}$ solves the relaxed program of that section. To show that it satisfies (v), we need to show that it satisfies (iii) for any pair $s,t$ such that $s,t\in[s_*,\hat{s}]$ or $s,t\in[\hat{s},s^*]$. Take, for example, the first case. Then

\begin{align*}
\alpha^*_{uid}(t)-\alpha^*_{uid}(s)&=-\int_s^tc_s\bigl(m(x)|x)dx\leq -\int_s^tc_s(m(t)|x)dx\tag{by \ref{eq:uid}}\\
&= c\bigl(m(t)|s\bigr)-c\bigl(m(t)|t\bigr)\leq c(t|s)-c(t|t)=c(t|s).\tag{by \ref{eq:uid}}
\end{align*}
The argument is similar in the second case.

For uniqueness, first note that $c(s^*(\nu)|s_*(\nu))$ is decreasing in $\nu$ so there is a single value of $\nu$ that satisfies \eqref{eq:bounds}, and hence satisfies the necessary and sufficient conditions of \lemmaref{LemLagSuff}. Then for this $\nu$ and the corresponding bounds $(s_*,s^*)$, the solution to the optimal transport problem is uniquely determined up to a constant. This constant is uniquely determined either by the probability constraint if it binds, that is, if $c(s^*|s_*)=1$, or by the requirement that $\alpha^*_{uid}(s_*)=0$ under low-score priority, and $\alpha^*_{uid}(s^*)=1$ under high-score priority. Uniqueness fails only if we are in the neutral priority case where $\bar{w}=0$ and the probability constraint is slack. In this case, $(s_*,s^*)=(s_*(0),s^*(0))=(\underline{s},\overline{s})$. Hence, for the probability constraint not to bind, it must be the case that $c(\overline{s}|\underline{s})<1$. The designer is then indifferent across all allocation rules $\alpha^*_{uid}(s,r)$ for any $r\in[0,1]$. Indeed, for $r'>r$, we have $\alpha^*_{uid}(s,r')-\alpha^*_{uid}(s,r)=(r'-r)\Gamma_{uid}$, so the difference in the designer's payoff is
$(r'-r)\Gamma_{uid}\int_{\underline{s}}^{\overline{s}}w(s)dF(s)=(r'-r)\Gamma_{uid}\bar{w}=0.$
\end{proof}


\begin{proof}[Proof of \propref{PropEucl}]
Let $\alpha$ denote the optimal rule. Using the formulas from \theoremref{ThOptUID} and \theoremref{ThOptUDD}, we get $\alpha'(s)=\mathcal{C}'(0)$ in the \eqref{eq:udd} case. In the \eqref{eq:uid} case, for $s\in[s_*,\hat{s}]$, $\alpha'(s)=\mathcal{C}'\bigl(m(s)-s\bigr)$ increases in $s$ by concavity of $\mathcal{C}$ and since $m(s)$ is decreasing. If instead $s\in[\hat{s},s^*]$, then
$\alpha'(s)=\mathcal{C}'\bigl(s-m^{-1}(s)\bigr)$ decreases in $s$ by concavity of $\mathcal{C}$ and since $m^{-1}(s)$ is decreasing.
\end{proof}


\begin{proof}[Proof of \propref{PropGam}]
Recall that the matching function $m$ is decreasing. This implies that the growth interval increases in $\gamma$ for the inclusion order. It increases strictly for $\gamma<\hat{\gamma}$, and is equal to $[s_*(0),s^*(0)]$ for $\gamma\geq \hat{\gamma}$.

Consider first $\gamma<\gamma'<\hat{\gamma}$, and define the function $\delta(s)=\alpha^*_{\gamma'}(s)-\alpha^*_\gamma(s)$. We denote by $s_*[\gamma]$ and $s^*[\gamma]$ the optimal matching pair under $\gamma$, where we use brackets to distinguish them from the functions $s_*(\nu),s^*(\nu)$.

$\delta(s)$ is equal to 0 for $s\leq s_*[\gamma']$ and $s\geq s^*[\gamma']$. It is equal to $\alpha^*_{\gamma'}(s)$, and therefore increasing and positive, on $\bigl[s_*[\gamma'],s_*[\gamma]\bigr]$. It is equal to  $\alpha^*_{\gamma'}(s)-1$, and therefore increasing and negative, on $\bigl[s^*[\gamma],s^*[\gamma']\bigr]$. 

If the cost function satisfies \eqref{eq:uid}, the derivative of $\delta$ is 
\[\delta'(s)=
\begin{cases}
\left(\frac{1}{\gamma}-\frac{1}{\gamma'}\right)c_s\bigl(m(s)|s\bigr)\leq 0 &\text{if }s\in\bigl[s_*[\gamma],\hat{s}\bigr],\\
\left(\frac{1}{\gamma'}-\frac{1}{\gamma}\right)c_t\bigl(s|m^{-1}(s)\bigr)\leq 0&\text{if }s\in\bigl[\hat{s},s^*[\gamma]\bigr].    
\end{cases}\]
If, instead, the cost function satisfies \eqref{eq:udd}, its derivative on $\bigl[s_*[\gamma],s^*[\gamma]\bigr]$ is 
\[\delta'(s)=\left(\frac{1}{\gamma'}-\frac{1}{\gamma}\right)c_{t^+}\bigl(s|s\bigr)\leq 0.\]
Hence, $\delta$ increases from 0, then decreases weakly, becoming negative, and increases back to 0, which proves point (i) of the proposition.

Next, suppose $\gamma'>\gamma>\hat{\gamma}$. In the low-score priority case, the growth interval under both $\gamma$ and $\gamma'$ is $[s_*(0),\overline{s}]$. The computation of $\delta'$ in this interval is the same as above, implying now that $\delta$ is decreasing on $[s_*(0),\overline{s}]$. Since $\delta\bigl(s_*(0)\bigr)=0$, this proves point (ii). In the high-score priority case, the growth interval under both $\gamma$ and $\gamma'$ is $[\underline{s},s^*(0)]$. The computation of $\delta'$ in this interval is the same as above, implying now that $\delta$ is decreasing on $[\underline{s},s^*(0)]$. Since $\delta\bigl(s^*(0)\bigr)=0$, this proves point (iii).

Finally, if $\gamma\leq \hat{\gamma}<\gamma'$, then $\delta(s)$ is nonnegative below $s_*[\gamma]$ and nonpositive above $s^*[\gamma]$. On $\bigl[s_*[\gamma'],s_*[\gamma]\bigr]$, the same derivative calculation shows that $\delta$ is nonincreasing. Thus $-\delta(s)$ is single-crossing from below.
\end{proof}


\begin{proof}[Proof of \propref{PropWelfFP}] In \cite{perez2022test}, we derive optimal allocation rules without a falsification-proofness constraint for a cost function that satisfies the following \emph{upper triangular inequality}: for all $s \leq m \leq t$, $c(t|m) + c(m|s) \geq c(t|s)$. We also show in \cite{perez2022test} that this triangular inequality is implied by \eqref{eq:uid}. The (unconstrained) optimal allocation rule is 
\begin{equation*}
\alpha^{**}(s,r)=
\begin{cases}
    \Gamma^{**} I(r)&\text{if }s<s_-\\
    \Gamma^{**} I(r)+c(s_+|s_-)-c(s_+|s)&\text{if }s\in[s_-,s_+]\\
    \Gamma^{**} I(r)+c(s_+|s_-)&\text{if }s>s_+
\end{cases},
\end{equation*}
where the probability gap is $\Gamma^{**}=1-c(s_+|s_-)$.

The magnitude of falsification costs, together with score priority, determines the growth interval $[s_-, s_+]$. The relevant cost thresholds are defined as $\hat{c} = c(\overline{s} | \hat{s})$ and $\overline{c} = c\bigl(\overline{s} | s_*(0)\bigr)$. By cost monotonicity, $\overline{c}>\hat{c}$.

The unconstrained optimal rule has three regimes: (1) The \emph{low-cost regime}, if $\overline{c}<1$: then, the growth interval is $[s_-,s_+]=[s_*(0), \overline{s}]$, the probability constraint is slack, and the gap $\Gamma^{**}>0$ is allocated according to priority by $I(r)$. (2) The \emph{intermediate-cost regime}, if $\overline{c}\geq 1 > \hat{c}$: then, $s^+=\overline{s}$, and $s_-$ solves $c(\overline{s} | s_-) = 1$. The probability constraint is binding. (3) The \emph{high-cost regime}, if $\hat{c} \geq 1$: then, $s_-=\hat{s}$, and $s^+$ solves $c(s_+ | \hat{s}) = 1$. The probability constraint is binding.

First, note that, under the low and intermediate-cost regimes, for every ineligible score $s$ that lies in the growth interval of both rules, $\alpha^{*}$ and $\alpha^{**}$,
\begin{equation}
\frac{d\alpha^{**}}{ds}(s) = -c_s\bigl(\overline{s}|s\bigr) \leq -c_s\bigl(m(s)|s\bigr) = \frac{d\alpha^{*}}{ds}(s),\label{eqWelfFP1}
\end{equation}
since, by \eqref{eq:uid}, $c_s(t|s)$ is nondecreasing in $t$.

\medskip

We first treat the \emph{low-score priority} and \emph{neutral priority} cases together:

\step{Low-cost regime.} Both rules have the same growth interval $[s_*(0),s^*(0)]$ with $s^*(0)=\overline{s}$. Furthermore, $\alpha^*(s)=\alpha^{**}(s)=I(r)\Gamma^{**}=I(r)\Gamma$ for $s\leq s_*(0)$. Then, by \eqref{eqWelfFP1}, $\alpha^*(s) > \alpha^{**}(s)$ for all $s\in(s_*(0),\hat{s})$. For  $s\geq \hat{s}$,
\begin{align*}
\alpha^*(s) &= \alpha^*\bigl(m^{-1}(s)\bigr) + c\bigl(s|m^{-1}(s)\bigr) \geq \alpha^{**}\bigl(m^{-1}(s)\bigr) + c\bigl(s|m^{-1}(s)\bigr) \\
&=I(r)\Gamma^{**}+ c\bigl(\overline{s}|s_*(0)\bigr) - c(\overline{s}|m^{-1}(s)\bigr) + c\bigl(s|m^{-1}(s)\bigr) \\
&\geq I(r)\Gamma^{**}+c\bigl(\overline{s}|s_*(0)\bigr) - c(\overline{s}|s\bigr) = \alpha^{**}(s),
\end{align*}
where the first equality holds because \eqref{eqFPC} binds between $m^{-1}(s)$ and $s$, the first inequality holds since $m^{-1}(s)$ is ineligible, and the second inequality is due to the \emph{upper triangular inequality}.

\step{Intermediate-cost regime.} The growth intervals satisfy $s_*\leq s_- \leq \hat{s}< s^*\leq s^+=\overline{s}$, and $\alpha^*(s_*)=0$, while $\alpha^{**}(s_-)=0$. For all $s\in(s_-,\hat{s})$, \eqref{eqWelfFP1} holds, implying $\alpha^*(s)\geq \alpha^{**}(s)$. For  $s\in (\hat{s},s^*)$, using the same ideas as above,
\begin{align*}
\alpha^*(s) &= \alpha^*\bigl(m^{-1}(s)\bigr) + c\bigl(s|m^{-1}(s)\bigr) \geq \alpha^{**}\bigl(m^{-1}(s)\bigr) + c\bigl(s|m^{-1}(s)\bigr) \\
&\geq 1 - c(\overline{s}|m^{-1}(s)\bigr) + c\bigl(s|m^{-1}(s)\bigr) \geq 1 - c(\overline{s}|s\bigr) = \alpha^{**}(s).
\end{align*}
Finally, for $s\geq s^*$, $\alpha^*(s)=1\geq \alpha^{**}(s)$.

\step{High-cost regime.} The growth intervals satisfy $s_*<\hat{s}=s_-<s^*<s_+\leq \overline{s}$. Therefore, $\alpha^*(s)>\alpha^{**}(s)=0$ for $s\in[s_*,\hat{s}]$. Then, for $s\geq \hat{s}$, 
\begin{align*}
\alpha^*(s) &= \alpha^*\bigl(m^{-1}(s)\bigr) + c\bigl(s|m^{-1}(s)\bigr) \geq  c\bigl(s|m^{-1}(s)\bigr) \geq c\bigl(s|\hat{s}\bigr)\\
&\geq  c(s_+|\hat{s}\bigr)- c(s_+|s\bigr)= 1 - c(s_+|s\bigr) = \alpha^{**}(s),
\end{align*}
where the second inequality is by cost monotonicity, the third by the triangular inequality, and the remaining equality is by definition of the growth intervals in the high-cost regime.

\medskip

Next, we treat the \emph{high-score priority} case.

\step{Low-cost regime.} The growth interval is $[s_*(0),s^*(0)]=[\underline{s},s^*(0)]$ for $\alpha^*$, and $[\underline{s},\overline{s}]$ for $\alpha^{**}$. Furthermore,
$$\alpha^*(\underline{s})=1-c\bigl(s^*(0)|\underline{s}\bigr) > 1-c(\overline{s}|\underline{s})=\alpha^{**}(\underline{s}).$$
Hence, $\alpha^*(s)> \alpha^{**}(s)$ for all $s\in[\underline{s},\hat{s}]$ holds by \eqref{eqWelfFP1}. For $s\in \bigl(\hat{s},s^*(0)\bigr)$, using the same ideas as above
\begin{align*}
\alpha^*(s) &= \alpha^*\bigl(m^{-1}(s)\bigr) + c\bigl(s|m^{-1}(s)\bigr) \geq \alpha^{**}\bigl(m^{-1}(s)\bigr) + c\bigl(s|m^{-1}(s)\bigr) \\
&\geq 1 - c(\overline{s}|m^{-1}(s)\bigr) + c\bigl(s|m^{-1}(s)\bigr) \geq 1 - c(\overline{s}|s\bigr) = \alpha^{**}(s).
\end{align*}
For $s\geq s^*(0)$, $\alpha^*(s)=1\geq\alpha^{**}(s)$.

\step{Intermediate-cost regime.} There are two possible cases. If $c(s^*(0)| s_*(0))>1$, the argument is word for word as in the low-score and neutral priority case. Suppose that $c(s^*(0)| s_*(0))<1$. Then 
the growth interval of $\alpha^*$ is $[s_*(0),s^*(0)]$ with $s_*(0)=\underline{s}$, $s^*(0)\leq \overline{s}$, and $\alpha^*\bigl(\underline{s}\bigr)=\Gamma$. The growth interval of $\alpha^{**}$ is such that $s_- >\underline{s}$, and $s^+=\overline{s}$. Therefore, $\alpha^*(s)>\alpha^{**}(s)$ for all $s\leq s_-$. For $s\in(s_-,\hat{s})$, \eqref{eqWelfFP1} holds, therefore $\alpha^*(s)> \alpha^{**}(s)$. For $s\in \bigl(\hat{s},s^*(0)\bigr)$, using the same ideas as above
\begin{align*}
\alpha^*(s) &= \alpha^*\bigl(m^{-1}(s)\bigr) + c\bigl(s|m^{-1}(s)\bigr) \geq \alpha^{**}\bigl(m^{-1}(s)\bigr) + c\bigl(s|m^{-1}(s)\bigr) \\
&\geq 1 - c(\overline{s}|m^{-1}(s)\bigr) + c\bigl(s|m^{-1}(s)\bigr) \geq 1 - c(\overline{s}|s\bigr) = \alpha^{**}(s).
\end{align*}
Finally, for $s\geq s^*(0)$, $\alpha^*(s)=1\geq\alpha^{**}(s)$.

\step{High-cost regime.} Then, the proof is exactly as under low-score or neutral priority.

 \end{proof}


\begin{proof}[Proof of \propref{PropWelfTradeOff}] 
We start by establishing general payoff formulas for a concave Euclidean cost function $\mathcal{C}$ with $d_\mathcal{C}=d$. Let $\bar{d}=\overline{s}-\underline{s}$, and $d^*(0)=s^*(0)-s_*(0)$. The length of the growth interval of $\alpha^*$ is equal to $\ell^*_\mathcal{C}=\min\{d,d^*(0)\}$. For $\alpha^{**}$, it is $$\ell^{**}_\mathcal{C}=\ell^*_\mathcal{C}\Ind_{\bar{w}\leq 0}+\min\{d,\bar{d}\}\Ind_{\bar{w}>0}.$$
In the low and intermediate-cost regimes, all eligible scores falsify, while in the high-cost regime, the scores that falsify are those on the growth interval, which is then a subset of eligible scores. Therefore, we can write the size of the interval of falsifying scores as $f_\mathcal{C}=\min\{\ell^{**}_\mathcal{C},\hat{d}\}=\min\{d,\hat{d}\}$, where we recall that $\hat{d}=\overline{s}-\hat{s}$ is the size of the interval of eligible agents. Finally, let 
$p^*_\mathcal{C}=\mathcal{C}(\ell^*_{\mathcal{C}})$ and $p^{**}_\mathcal{C}=\mathcal{C}(\ell^{**}_{\mathcal{C}})$.

The rule $\alpha^{**}$ is in the low-cost regime if $d\geq\bar{d}\Ind_{\bar{w}>0}+d^*(0)\Ind_{\bar{w}\leq 0}$, in the high-cost regime if $d < \hat{d}$, and in the intermediate-cost regime otherwise. The rule $\alpha^*$ is in its binding regime if $d\leq d^*(0)$ and in its slack regime otherwise. Note that the slack regime corresponds to the low-cost regime of $\alpha^{**}$ under low-score or neutral priority, whereas the transition between the slack and binding regimes of $\alpha^*$ is within the intermediate-cost regime of $\alpha^{**}$ under high-score priority.

We parametrize matching intervals by their length. Let $\psi:[0,d^*(0)]\to[\hat{s},s^*(0)]$ denote the inverse of the continuous, strictly increasing function $t\longmapsto t-m^{-1}(t)$. Thus, the \eqref{eqZAS} interval of length $y$ is $[\psi(y)-y,\psi(y)]$, and $\psi(0)=\hat{s}$. Since $m^{-1}$ is decreasing, for $0\leq y_1<y_2\leq d^*(0)$,
\[
 0\leq\psi(y_2)-\psi(y_1)\leq y_2-y_1.
\]
Consequently, both endpoint maps $y\mapsto\psi(y)$ and $y\mapsto\psi(y)-y$ are Lipschitz, hence absolutely continuous, and $0\leq\psi'(y)\leq1$ for Lebesgue-almost every $y$. At points in $(0,d^*(0))$ where $\psi'$ exists, define
\[
 \eta(y)=\psi'(y)F\bigl(\psi(y)\bigr)
       +\bigl(1-\psi'(y)\bigr)F\bigl(\psi(y)-y\bigr).
\]
At the remaining points, set $\eta(y)=F\bigl(\psi(y)\bigr)$; in particular, $\eta(0)=F(\hat{s})$. This choice on a Lebesgue-null set does not affect integration against $d\mathcal{C}(y)=\mathcal{C}'(y)dy$. For every $y\in[0,d^*(0)]$,
\[
 F\bigl(\psi(y)-y\bigr)\leq\eta(y)\leq F\bigl(\psi(y)\bigr),
 \qquad \lim_{y\downarrow0}\eta(y)=F(\hat{s}),
\]
where the limit follows from the continuity of $F$ and $\psi$. 

We first compute the aggregate payoff of agents under $\alpha^*$. Let $[s_*,s^*]$ denote its growth interval. The rule is constant outside this interval, with top allocation probability $p^*_{\mathcal{C}}+(1-p^*_{\mathcal{C}})I(r)$. Stieltjes integration by parts and the characterization of optimal rules give
\begin{align*}
 A^*(\mathcal{C})
 &=p^*_{\mathcal{C}}+(1-p^*_{\mathcal{C}})I(r)
   -\int_S F(x)d\alpha^*(x)\\
 &=p^*_{\mathcal{C}}+(1-p^*_{\mathcal{C}})I(r)\\
 &\quad-\int_{s_*}^{\hat{s}}F(s)\mathcal{C}'\bigl(m(s)-s\bigr)ds\\
 &\quad-\int_{\hat{s}}^{s^*}F(t)\mathcal{C}'\bigl(t-m^{-1}(t)\bigr)dt.
\end{align*}
The changes of variables $s=\psi(y)-y$ in the first integral and $t=\psi(y)$ in the second are valid because both parameterizations are monotone and absolutely continuous. The first integral becomes
\[
 \int_0^{\ell^*_{\mathcal{C}}}
 F\bigl(\psi(y)-y\bigr)\bigl(1-\psi'(y)\bigr)\mathcal{C}'(y)dy,
\]
and the second becomes
\[
 \int_0^{\ell^*_{\mathcal{C}}}
 F\bigl(\psi(y)\bigr)\psi'(y)\mathcal{C}'(y)dy.
\]
By the definition of $\eta$, we obtain the following general formula for the agents' aggregate payoff:
\[
 A^*(\mathcal{C})=p^*_{\mathcal{C}}
 -\int_0^{\ell^*_{\mathcal{C}}}\eta(y)d\mathcal{C}(y)
 +(1-p^*_{\mathcal{C}})I(r).
\]

Next, we compute the designer's payoff under $\alpha^*$. By \eqref{eq:objdn}--\eqref{eq:objup} and the initial condition \eqref{eq:init},
\begin{align*}
 D^*(\mathcal{C})
 &=\bigl(\bar{w}\bigr)^+ +\int_S\mathcal{W}(x)d\alpha^*(x)\\
 &=\bigl(\bar{w}\bigr)^+
   +\int_{s_*}^{\hat{s}}\mathcal{W}(s)\mathcal{C}'\bigl(m(s)-s\bigr)ds
 +\int_{\hat{s}}^{s^*}\mathcal{W}(t)\mathcal{C}'\bigl(t-m^{-1}(t)\bigr)dt\\
 &=\bigl(\bar{w}\bigr)^+
   +\int_0^{\ell^*_{\mathcal{C}}}\mathcal{W}\bigl(\psi(y)-y\bigr)
      \bigl(1-\psi'(y)\bigr)\mathcal{C}'(y)dy\\
 &\quad+\int_0^{\ell^*_{\mathcal{C}}}\mathcal{W}\bigl(\psi(y)\bigr)
      \psi'(y)\mathcal{C}'(y)dy,
\end{align*}
where the last equality uses the same substitutions. Since the endpoints are matched, $\mathcal{W}(\psi(y)-y)=\mathcal{W}(\psi(y))$. Combining the two integrals yields the designer's payoff under the FP rule:
\[
 D^*(\mathcal{C})=\bigl(\bar{w}\bigr)^+
 +\int_0^{\ell^*_{\mathcal{C}}}\mathcal{W}\bigl(\psi(y)\bigr)d\mathcal{C}(y).
\]

The next calculation computes the aggregate payoff of agents on the growth interval $[s_-,s_+]$ under $\alpha^{**}$. Then
\begin{align*}
\int_{s_-}^{s_+}\alpha^{**}(x)dF(x)&=\mathcal{C}(s_+-s_-)F(s_+)-\int_{s_-}^{s_+}F(x)d\alpha^{**}(x)\\
&=\mathcal{C}(s_+-s_-)F(s_+)-\int_{s_-}^{s_+}F(x)\mathcal{C}'(s_+-x)dx\\
&=\mathcal{C}(s_+-s_-)F(s_+)-\int_{0}^{s_+-s_-}\mathcal{C}'(y)F(s_+-y)dy,
\end{align*}
where the first equality follows from integration by parts, and the second relies on changing the integration variable from $x$ to $y=s_+-x$.

This yields the following formula for the agents' aggregate payoff under $\alpha^{**}$:
$A^{**}(\mathcal{C})=p^{**}_\mathcal{C}-\int_0^{\ell^{**}_\mathcal{C}}F(f_{\mathcal{C}}-y+\hat{s})d\mathcal{C}(y)+(1-p^{**}_{\mathcal{C}})I(r).$

Finally, we compute the designer's payoff under the unconstrained rule $\alpha^{**}$. If $d\leq \hat{d}$, then the designer attains her first-best payoff $V=\bigl(\bar{w}\bigr)^++\mathcal{W}(\hat{s})$. Otherwise, her payoff over eligible agents is $b V$, where $b=p^{**}_\mathcal{C}+I(r)(1-p^{**}_\mathcal{C}) $. Ineligible agents yield a negative payoff
\begin{align*}
 \int_{s_-}^{\hat{s}}\alpha^{**}(x)w(x)dF(x)&=-\mathcal{W}(\hat{s})\alpha^{**}(\hat{s})+ \int_{s_-}^{\hat{s}}\mathcal{W}(x)d\alpha^{**}(x)\\  
 &=-\mathcal{W}(\hat{s})\left(b-\mathcal{C}(\hat{d})\right)+\int_{s_-}^{\hat{s}}\mathcal{W}(x)\mathcal{C}'(\overline{s}-x)dx\\
 &=-\mathcal{W}(\hat{s})\left(b-\mathcal{C}(f_\mathcal{C})\right)+\int_{f_\mathcal{C}}^{\ell^{**}_\mathcal{C}}\mathcal{W}(\overline{s}-y)\mathcal{C}'(y)dy
\end{align*}
This yields the formula:
$D^{**}(\mathcal{C})=\bigl(\bar{w}\bigr)^+ +\mathcal{C}(f_\mathcal{C})\mathcal{W}(\hat{s})+\int_{f_\mathcal{C}}^{\ell^{**}_\mathcal{C}}\mathcal{W}(\overline{s}-y)d\mathcal{C}(y).$

The cost function $\mathcal{C}(x)$ is an increasing and concave function that defines a cumulative distribution function over the interval $[0,d]$. Next, consider a sequence of concave cost functions $\{\mathcal{C}_n\}$ such that $d_{\mathcal{C}_n}=d$ is constant. Since falsifying by more than $d$ is never rational, we consider only the restriction of these cost functions to the interval $[0,d]$. Suppose that $\{\mathcal{C}_n\}$ is an increasing sequence such that $\mathcal{C}_n(x)$ converges to 1 for all $x>0$. It is easy to construct such a sequence by considering a sequence that increases in the concave order. For example, let $\mathcal{C}_0(x)=x/d$ and $\mathcal{C}_{n+1}(x)=g\bigl(\mathcal{C}_n(x)\bigr)$ for any continuously differentiable, strictly increasing, and strictly concave bijective function $g$ from $[0,1]$ to itself. 

Then the sequence $\{\mathcal{C}_n\}$, viewed as cdfs on $[0,d]$, converges in distribution to the Dirac distribution putting all mass at 0. Furthermore, the sequences $p^*_{\mathcal{C}_n}$ and $p^{**}_{\mathcal{C}_n}$ converge to 1. Finally, the sequence $\mathcal{C}_n(f_{\mathcal{C}_n})$ is either constant at 1 or converges to 1. Although $\eta$ need not be continuous away from zero, it is bounded and has right limit $F(\hat{s})$ at zero. Since $\ell^*_{\mathcal{C}_n}=\min\{d,d^*(0)\}>0$ is independent of $n$ and the mass of $d\mathcal{C}_n$ concentrates at zero,
\[
 \lim_{n\to\infty}\int_0^{\ell^*_{\mathcal{C}_n}}\eta(y)d\mathcal{C}_n(y)=F(\hat{s}).
\]
Continuity of $F$ and $\mathcal{W}$ gives the limits of the other payoff integrals. Using these properties, we have the following limits for the payoffs we computed: $\lim_{n\to\infty}A^*(\mathcal{C}_n)=1-F(\hat{s})$; $\lim_{n\to\infty}D^{*}(\mathcal{C}_n)=\bigl(\bar{w}\bigr)^+ +\mathcal{W}(\hat{s})$; $\lim_{n\to\infty}A^{**}(\mathcal{C}_n)=0$ if $d\geq \overline{s}-\hat{s}$, and $\lim_{n\to\infty}A^{**}(\mathcal{C}_n)=1-F(\hat{s}+d)$ otherwise; $\lim_{n\to\infty}D^{**}(\mathcal{C}_n)=\bigl(\bar{w}\bigr)^+ + \mathcal{W}(\hat{s})$.

Putting these together, we obtain that the loss-rate of the designer $L(\mathcal{C}_n)$ always converges to 0, whereas the gain rate of the agents becomes arbitrarily large if $d\geq \overline{s}-\hat{s}$, and converges to $\frac{F(\hat{s}+d)-F(\hat{s})}{1-F(\hat{s}+d)}$ otherwise.
\end{proof}


\begin{proof}[Proof of \lemmaref{lem:lagsuff-fix}] 
We proceed in two steps as in the proof of \lemmaref{LemLagSuff} in the paper. We only provide arguments for the parts of the proof that differ from that of \lemmaref{LemLagSuff}. Recall that we say an allocation rule $\alpha$ is \emph{feasible}, and denote the set of such allocation rules by $\mathbbm{A}$, if it satisfies \eqref{eq:mon}, \eqref{eq:FPCfix}, and \eqref{eq:init}.

\step{Step 1:} \emph{An allocation rule $\alpha\in\mathbbm{A}$  solves \eqref{eqBP} if and only if (A) $\alpha$ satisfies \eqref{eqPC}; and there exists $\nu\geq 0$ such that (B) $\nu=0$ or $\overline{\alpha}-\underline{\alpha}=1$, and (C) $\mathcal{L}(\alpha,\nu)\geq\mathcal{L}(\hat{\alpha},\nu)$ for every  allocation rule $\hat{\alpha}\in\mathbbm{A}$.}

The proof of step 1 is identical to that in \lemmaref{LemLagSuff} of the paper, so we proceed directly to the proof of step 2.

\step{Step 2:} \emph{An allocation rule $\alpha$ and a scalar $\nu\geq 0$ satisfy \eqref{eq:mon}, \eqref{eqPC}, \eqref{eq:init}, and (i)-(v) if and only if $\alpha$ and $\nu$ satisfy  $\alpha\in\mathbbm{A}$, (A), (B) and (C).}

\step{\fbox{$\Rightarrow$}~} The proof of this claim is identical to that in \lemmaref{LemLagSuff}, only replacing \eqref{eqFPC} by \eqref{eq:FPCfix}.

\step{\fbox{$\Leftarrow$}~} The difference with the proof of \lemmaref{LemLagSuff} is that \eqref{eq:FPCfix} does not imply the continuity property (ii). The rest of the proof is identical, so $\alpha\in\mathbbm{A}$ directly implies \eqref{eq:mon}, \eqref{eq:init}, and (iii). (A) directly implies \eqref{eqPC}. (B) and \eqref{eq:init} imply (iv). Additionally, the proof that (C) implies (i) is identical. Finally, having proved (ii), the proof that (C) implies (v) is also identical. Hence, we only need to prove (ii).

We prove continuity by a different argument that relies on \eqref{eq:uid}: we show that (C) implies (ii). The proof is by contradiction. Suppose that $\alpha\bigl(s_*(\nu)^+\bigr)-\alpha\bigl(s_*(\nu)^-\bigr)=\delta>0$. We choose $\tilde{s}\in \bigl(s_*(\nu),\hat{s}\bigr)$ sufficiently close to $s_*(\nu)$ so that $\eta=\delta-c\bigl(\tilde{s}|s_*(\nu)\bigr)>0$. Then consider the function $\hat{\alpha}$ which is constant and equal to $\alpha\bigl(s_*(\nu)^+\bigr)-\eta>0$ on $\bigl(s_*(\nu),\tilde{s}\bigr)$, and equal to $\alpha$ everywhere else. $\hat{\alpha}$ satisfies \eqref{eq:mon} and \eqref{eq:init}. We show that it also satisfies \eqref{eq:FPCfix}. The only case for which an argument is needed is for $s\in\bigl(s_*(\nu),\tilde{s}\bigr),$ and $t\in\bigl[\tilde{s},s^*(\nu)\bigr]$. Then we have:
\begin{align*}
\hat{\alpha}(t)-\hat{\alpha}(s)&=\alpha(t)-\alpha\bigl(s_*(\nu)^+\bigr)+\eta\\
&=\alpha(t)-\alpha\bigl(s_*(\nu)^-\bigr)-c\bigl(\tilde{s}|s_*(\nu)\bigr)\\
&\leq k+c\bigl(t|s_*(\nu)\bigr)-c\bigl(\tilde{s}|s_*(\nu)\bigr)\\
&\leq k+c(t|\tilde{s})\\
&\leq \bar{c}(t|s),
\end{align*}
where the first two equalities are by definition of $\hat{\alpha}$ and $\eta$, the first inequality is because $\alpha$ satisfies \eqref{eq:FPCfix}, the second inequality is by \eqref{eq:uid}, and the last inequality is by cost monotonicity for $c$. Hence, $\hat{\alpha}$ satisfies \eqref{eq:FPCfix}, and is therefore in $\mathbbm{A}$. We show that $\mathcal{L}(\hat{\alpha},\nu)> \mathcal{L}(\alpha,\nu)$, contradicting (C):
\begin{align*}
\int_S\bigl[\mathcal{W}(s)-\nu]d\alpha(s)&=\int_{s_*(\nu)}^{\tilde{s}}\bigl[\mathcal{W}(s)-\nu]d\alpha(s)+\int^{s^*(\nu)}_{\tilde{s}}\bigl[\mathcal{W}(s)-\nu]d\alpha(s)\\
&\leq \bigl[\mathcal{W}(\tilde{s})-\nu]\int_{s_*(\nu)^+}^{\tilde{s}}d\alpha(s)+\int^{s^*(\nu)}_{\tilde{s}}\bigl[\mathcal{W}(s)-\nu]d\alpha(s)\\
&=\bigl[\mathcal{W}(\tilde{s})-\nu]\Bigl\{\alpha(\tilde{s})-\alpha\bigl(s_*(\nu)^+\bigr)\Bigr\}+\int^{s^*(\nu)}_{\tilde{s}}\bigl[\mathcal{W}(s)-\nu]d\alpha(s),
\end{align*}
where the first inequality uses the equality $\mathcal{W}\bigl(s_*(\nu)\bigr)=\nu$ to eliminate the atom at $s_*(\nu)$, and the fact that $\mathcal{W}(\cdot)$ is increasing to the left of $\hat{s}$ by \lemmaref{LemCumSurpl}.

By definition of $\hat{\alpha}$, the first term is strictly dominated by
\begin{align*}
  \bigl[\mathcal{W}(\tilde{s})-\nu]\Bigl\{\alpha(\tilde{s})-\alpha\bigl(s_*(\nu)^+\bigr)+\eta\Bigr\}&=\bigl[\mathcal{W}(\tilde{s})-\nu]\Bigl\{\hat{\alpha}(\tilde{s})-\hat{\alpha}\bigl(s_*(\nu)^+\bigr)\Bigr\}\\
  &=\int_{s_*(\nu)}^{\tilde{s}}\bigl[\mathcal{W}(s)-\nu]d\hat{\alpha}(s),
\end{align*}
while the second term is equal to 
\[\int^{s^*(\nu)}_{\tilde{s}}\bigl[\mathcal{W}(s)-\nu]d\hat{\alpha}(s).\]
Hence, $\int_S\bigl[\mathcal{W}(s)-\nu]d\alpha(s)<\int_S\bigl[\mathcal{W}(s)-\nu]d\hat{\alpha}(s)$, which contradicts (C). To rule out discontinuities at $s^*(\nu)$, we follow a similar construction. 
Therefore, (C) implies (ii).
\end{proof}


\begin{proof}[Proof of \theoremref{ThOptUIDfix}]
To keep notation simple, we only indicate the dependence of $\bar{\alpha}^*_{uid}$ on $\hat{w},r$ when it is useful for the argument. We check that the conditions of \lemmaref{lem:lagsuff-fix} are satisfied. We pick the multiplier $\nu=\mathcal{W}(s_*)\geq 0$. Then $\bar{\alpha}^*_{uid}$ clearly satisfies (i) and it satisfies (ii) by construction. To see that it satisfies (iii) and (iv), note that $\bar{\alpha}^*_{uid}(s^*)-\bar{\alpha}^*_{uid}(s_*)=\bar{c}(s^*|s_*)$ since \eqref{eq:FPCfix} is binding for $(s_*,s^*)$. Hence, by \eqref{eq:boundsfix}, either $\bar{\alpha}^*_{uid}(s^*)-\bar{\alpha}^*_{uid}(s_*)$ is equal to 1 and the probability constraint is binding, or it is strictly less than 1, and then $\nu=0$ and $(s_*,s^*)=(s_*(0),s^*(0))$.

By the optimal transport connection established above, $\bar{\alpha}^*_{uid}$ solves the relaxed program \eqref{pgm:relaxUIDfix}. To show that it satisfies (v), we need to show that it satisfies \eqref{eq:FPCfix} in (ii) for any pair $s,t$ such that $s,t\in[s_*,\hat{s}]$ or $s,t\in[\hat{s},s^*]$. Take, for example, the first case. Then

\begin{align*}
\bar{\alpha}^*_{uid}(t)-\bar{\alpha}^*_{uid}(s)&=-\int_s^tc_s\bigl(m(x)|x)dx\leq -\int_s^tc_s(m(t)|x)dx\tag{by \ref{eq:uid}}\\
&= c\bigl(m(t)|s\bigr)-c\bigl(m(t)|t\bigr)\leq c(t|s)-c(t|t)=c(t|s).\tag{by \ref{eq:uid}}
\end{align*}
The argument is similar in the second case.

For uniqueness, first note that $\bar{c}(s^*(\nu)|s_*(\nu))$ is increasing in $\nu$ so there is a single value of $\nu$ that satisfies \eqref{eq:boundsfix}, and hence satisfies the necessary and sufficient conditions of \lemmaref{lem:lagsuff-fix}. Then for this $\nu$ and the corresponding bounds $(s_*,s^*)$, the solution to the optimal transport problem is uniquely determined up to a constant. This constant is uniquely determined either by the probability constraint if it binds, that is, if $\bar{c}(s^*|s_*)=1$, or by the requirement that $\bar{\alpha}^*_{uid}(s_*)=0$ under low-score priority, and $\bar{\alpha}^*_{uid}(s^*)=1$ under high-score priority. Uniqueness fails only if we are in the neutral priority case where $\bar{w}=0$ and the probability constraint is slack. In this case, $(s_*,s^*)=(s_*(0),s^*(0))=(\underline{s},\overline{s})$. Hence, for the probability constraint not to bind, it must be the case that $\bar{c}(\overline{s}|\underline{s})<1$. The designer is then indifferent across all allocation rules $\bar{\alpha}^*_{uid}(s,r)$ for any $r\in[0,1]$. Indeed, for $r'>r$, we have $\bar{\alpha}^*_{uid}(s,r')-\bar{\alpha}^*_{uid}(s,r)=(r'-r)\bar{\Gamma}_{uid}$, so the difference in the designer's payoff is
$(r'-r)\bar{\Gamma}_{uid}\int_{\underline{s}}^{\overline{s}}w(s)dF(s)=(r'-r)\bar{\Gamma}_{uid}\bar{w}=0.$
\end{proof}


\begin{proof}[Proof of \propref{PropOutOpt}]
Let $\tilde{\alpha}^*(\hat{w})$ denote the correspondence mapping $\hat{w}$ to the set of solutions of  \eqref{eqBP}. By \lemmaref{lem:mon}, we can write \eqref{eqBP} as an optimization problem over the set of nondecreasing functions from $S$ to $[0,1]$ satisfying \eqref{eqFPC} and \eqref{eqPC}. This space is compact, by Helly's theorem, and convex. Furthermore, the objective function is linear and therefore continuous in $\alpha$. Hence, Berge's maximum theorem implies that $\tilde{\alpha}^*(\hat{w})$ is an upper hemicontinuous correspondence. By \theoremref{ThOptUID} and \theoremref{ThOptUDD}, the correspondence is singleton-valued for $\hat{w}\neq \bar{w}$, and for $\hat{w}=\bar{w}$ if  $c(\overline{s}|\underline{s})<1$, so the continuity results with respect to $\hat{w}$ follow.    

The space of feasible and non-decreasing allocation rules is also a lattice with respect to the partial order $\alpha\succeq \beta \Leftrightarrow ~\forall s~~\alpha(s)\geq \beta(s)$, with the corresponding strict ordering $\alpha\succ \beta$ if $\alpha\succeq \beta$ and $\alpha(s)>\beta(s)$ for some $s$. Indeed, it is easy to see that, for two such allocation rules $\alpha$ and $\beta$, their meet $\alpha\wedge\beta$ and their join $\alpha\vee\beta$ are also nondecreasing and feasible. Furthermore, the objective function is supermodular in $\alpha$ and has strictly increasing differences in $(-\hat{w},\alpha)$. Hence, by Milgrom and Shannon's monotone selection theorem \citep{MilShan94}, $\alpha^*(\cdot,\hat{w},r)$ is strictly decreasing in $\hat{w}$ for the $\succeq$ order (recalling that the allocation rule is independent of $r$ except when $\hat{w}=\bar{w}$, where $r$ pins down the selection. 

Furthermore, it is easy to see  that $\alpha^*(s,\bar{w},r)$ is strictly increasing in $r$ for every $s$, both in the \eqref{eq:uid} and \eqref{eq:udd} cases, since $I(\bar{w},r)=r$. Together with the continuity of the correspondence at $\hat{w}=\bar{w}$, this implies the results on the left and right limits of $\alpha(\cdot,\hat{w},r)$ as $\hat{w}\to\bar{w}$.
\end{proof}


\begin{proof}[Proof of \corref{CorMass}]
This follows almost directly from \propref{PropOutOpt}. To complete the argument, we only need to notice that, since the solution $\alpha^*(s,\hat{w},r)$ is continuous in $s$, the result that $\alpha^*(\cdot,\hat{w},r)\succ \alpha^*(\cdot,\hat{w}',r)$ for $\hat{w}<\hat{w}'$, implies $\alpha^*(s,\hat{w},r)> \alpha^*(s,\hat{w}',r)$ for all $s$ on a subinterval of $S$, so $A^*(\hat{w},r)>A^*(\hat{w}',r)$.
\end{proof}


\begin{proof}[Proof of \theoremref{ThWithinSol}] The proof relies on the following Lemma which collects classical results in optimization theory \citep[see, for example,][chapter 8]{luen69}. Necessity of (i) holds because \eqref{eqPin} is linear in $\alpha$.
  
\begin{Lem}\label{LemAux}
The following statements are equivalent: \emph{(i)} $\alpha$ solves \eqref{eqPin}; \emph{(ii)} There exists an outside option $\hat{w}(\rho)$ such that $\alpha$ solves \eqref{eqBP} and $\int_S\alpha(s)dF(s)=\rho$; \emph{(iii)} There exists $\hat{w}$ such that $(\alpha,\hat{w})$ is a saddle-point for the Lagrangian of the within problem $\int_S\alpha(s)\{w(s)-\hat{w}\}dF(s)+\hat{w}\rho$. 

Furthermore, the value function of the constrained problem is concave in $\rho$, and its derivative $\overline{W}'(\rho)$ exists almost everywhere, and is equal to $\hat{w}(\rho)$.
\end{Lem}

The expected worth function $w(s)$ is bounded by assumption. Let $w^-=w(\underline{s})$ and $w^+=w(\overline{s})$ be its bounds. Then it is easy to see $\alpha^*(s,w^-,r)=A^*(w^-,r)=1$ and $\alpha^*(s,w^+,r)=A^*(w^+,r)=0$. By the continuity and strict monotonicity results of \corref{CorMass}, it follows that there exists a unique value of $\hat{w}\in [w^-,w^+]$, and, if $\hat{w}=\bar{w}$, a unique value of $r\in [0,1]$, such that $A^*(\hat{w},r)=\rho$. By \lemmaref{LemAux}, $\alpha^*(\cdot,\hat{w},r)$ is then the unique solution to the within problem \eqref{eqPin}. The continuity and monotonicity results of \corref{CorMass} also imply continuity and monotonicity of $\hat{w}(\rho)$ and $r(\rho)$. By \lemmaref{LemAux}, $\overline{W}(\rho)$ is concave on $[0,1]$, and since $\hat{w}(\rho)$ is unique, it is differentiable everywhere, and $\overline{W}'(\rho)=\hat{w}(\rho)$. In particular, $\overline{W}(\rho)$ is strictly concave at every $\rho$ such that $\hat{w}(\rho)$ is strictly decreasing, that is whenever $\hat{w}(\rho)\neq \bar{w}$ or $c(\overline{s}|\underline{s})\geq 1$.
\end{proof}


\begin{proof}[Proof of \theoremref{ThAcrossSol}]The objective function of the across problem is continuous and concave in $\bm{\rho}$ by \theoremref{ThWithinSol}, and the feasible set is nonempty, compact and convex. Therefore, it admits a solution characterized by the Kuhn-Tucker conditions (i)-(iii), recalling that, by \theoremref{ThWithinSol}, $W_i'(\rho_i)=\hat{w}_i(\rho_i)/\mu$, and the outside option value $\hat{w}_i(\rho_i)$ is the one that solves the within problem as defined by (iv). The condition for uniqueness holds because the objective function is then strictly concave by \theoremref{ThWithinSol}.
\end{proof}


\begin{proof}[Proof or \propref{prop:algo}]
The sequence $(Q^k,R^k)$ is increasing and bounded above by $(I,1)$ in the $(\subseteq,\leq)$ order on $2^I\times\{0,1\}$, so the algorithm stops in finitely many steps. Let $k$ be the step at which it stops. Let $\lambda_R=\hat{w}^k_i$ and $\lambda_i=0$ for all $i\notin Q^k$. This is consistent since $\hat{w}^k_i$ must be equal across all  $i\notin Q^k$. Let $\lambda_\ell=\lambda_R-\hat{w}^k_\ell$ for $\ell\in Q^k$.
Then it is easy to verify the vector of multipliers $\bm{\lambda}$, $\bm{\rho}^k$, $\hat{w}^k$ and  $\bm{r}^k$ satisfy all the conditions of \theoremref{ThAcrossSol}. Therefore $\bm{\rho}^k$ is a solution to the across problem.
\end{proof}


\begin{proof}[Proof of \propref{PropAcrossWelf}]
Increasing $\gamma_i$ lowers falsification costs, which tightens \eqref{eqFPC} and therefore shrinks the set of feasible allocation rules in the original problem, therefore weakly decreases its value function $\overline{W}$. Suppose $\tilde{F}_i$ first-order stochastically dominates $F_i$, and let $F^x_i=x\tilde{F}_i+(1-x)F_i$. Then $F^x_i$ increases with $x$ in the FOSD order. 

Consider the within problem for group $i$ under the score distribution $F^x_i$. To clarify the dependence on $x$, we denote its value function by $W_i(\rho_i|x)$ in this proof. By \lemmaref{LemAux}, 
$$W_i(\rho_i|x)=\min_{\hat{w}\in[w^-,w^+]}\max_{\alpha}\int_{S_i}\alpha(s)\{w_i(s)-\hat{w}\}dF^x_i(s)+\hat{w}\rho_i/\mu_i,$$
and $\bigl(\alpha^*_i(\cdot,\hat{w}_i(\rho_i),r_i(\rho_i)|x),\hat{w}_i(\rho_i)\bigr)$ is the unique solution to this saddle-point problem. In what follows, let $\alpha^*_i(s)$ denote the function $\alpha^*_i(\cdot,\hat{w}_i(\rho_i),r_i(\rho_i)|x)$.

Let $\mathcal{L}(\alpha,\hat{w},x)=\int_{S_i}\alpha(s)\{w_i(s)-\hat{w}\}dF^x_i(s)+\hat{w}\rho_i/\mu_i$ be the objective function. It is continuously differentiable in $x$ since it is linear. Furthermore, the saddle-point problem admits a solution for every $x\in[0,1]$ by \theoremref{ThWithinSol}. The interval $[w^-,w^+]$ and the space of nondecreasing continuous functions in which $\alpha$ is taken is also compact by Helly's selection theorem. Therefore, we can apply the envelope theorem for saddle-points of \citet[][Theorem 5]{mil_seg02}, and our uniqueness result to obtain that $W_i(\rho_i|x)$ is differentiable in $x$, and
\begin{align*}
\frac{\partial W_i(\rho_i|x)}{\partial x}&=\frac{\partial\mathcal{L}\Bigl(\alpha_i^*\bigl(\cdot,\hat{w}_i(\rho_i),r_i(\rho_i)\bigr),\hat{w}_i(\rho_i),x\Bigr)}{\partial x}\\
&=\int_{S_i}\alpha^*_i(s)\{w_i(s)-\hat{w}_i(\rho_i)\}d\tilde{F}_i(s)-\int_{S_i}\alpha^*_i(s)\{w_i(s)-\hat{w}_i(\rho_i)\}dF_i(s).
\end{align*}

Using the differential form, we have:
\[\frac{\partial W_i(\rho_i|x)}{\partial x}=\int_{S_i}\alpha_i^{*'}(s)\left(\tilde{\mathcal{W}}_i(s,\hat{w}_i(\rho_i))-\mathcal{W}_i(s,\hat{w}_i(\rho_i))\right)ds.\]

Next, we show this must be nonnegative. Indeed, note first 
\begin{align*}
\tilde{\mathcal{W}}^+_i(s,\hat{w}_i(\rho_i))-\mathcal{W}^+_i(s,\hat{w}_i(\rho_i))&=\bigl(1-\tilde{F}_i(s)\bigr)\int_s^{\overline{s}_i}\bigl\{w_i(y)-\hat{w}_i(\rho_i)\bigr\}d\frac{\tilde{F}_i(y)}{1-\tilde{F}_i(s)}\\
&~~~~-\bigl(1-F_i(s)\bigr)\int_s^{\overline{s}_i}\bigl\{w_i(y)-\hat{w}_i(\rho_i)\bigr\}d\frac{F_i(y)}{1-F_i(s)}.
\end{align*}
By first-order stochastic dominance, $1-\tilde{F}_i(s)\geq 1-F_i(s)$, and the stochastic dominance ordering of the conditional distributions on $[s,\overline{s}_i]$ is preserved since $\frac{\tilde{F}_i(y)}{1-\tilde{F}_i(s)}\leq \frac{F_i(y)}{1-F_i(s)}$. Since $w_i(\cdot)$ is an increasing function, this implies 
\[\tilde{\mathcal{W}}^+_{i}(s,\hat{w}_i(\rho_i))-\mathcal{W}^+_i(s,\hat{w}_i(\rho_i))\geq 0.\]

If $\bar{w}_{F_i}\leq \bar{w}_{\tilde{F}_i}\leq \hat{w}_i(\rho_i)$, then the group has low-score or neutral priority under both distributions. Using the differential version of the objective function \eqref{eq:objdn}, the difference in welfare is given by

\[\int_{\underline{s}_i}^{\overline{s}_i}\alpha^{*'}_i(s)\bigl\{\tilde{\mathcal{W}}^+_i(s,\hat{w}_i(\rho_i))-\mathcal{W}^+_i(s,\hat{w}_i(\rho_i))\bigr\}ds.\]
Since $\alpha_i^{*'}(s)\geq 0$, and the difference in cumulative surplus is positive, then $\frac{\partial W_i(\rho_i|x)}{\partial x}\geq 0$.

Suppose instead, $\bar{w}_{F_i}\leq \hat{w}_i(\rho_i)\leq \bar{w}_{\tilde{F}_i}$, so the shift in score distributions switches the priority of the group. Then, using \eqref{eq:objdn} for $F_i$, \eqref{eq:objup} for $\tilde{F}_i$, and the relationship $\mathcal{W}^-(s,\hat{w})=\mathcal{W}^+(s,\hat{w})-(\bar{w}-\hat{w})$, we can write the welfare change as
\[\bigl(\bar{w}_{\tilde{F}_i}-\hat{w}_i(\rho_i)\bigr)\left(1-\int_{\underline{s}_i}^{\overline{s}_i}\alpha_i^{*'}(s)ds\right)\int_{\underline{s}_i}^{\overline{s}_i}\alpha_i^{*'}(s)\bigl\{\tilde{\mathcal{W}}^+_i(s,\hat{w}_i(\rho_i))-\mathcal{W}^+_i(s,\hat{w}_i(\rho_i))\bigr\}ds,\]
where the second term is positive for the same reasons as in the previous case, and the first term is equal to $\bigl(\bar{w}_{\tilde{F}_i}-\hat{w}_i(\rho_i)\bigr)\alpha^{*'}_i(\underline{s}_i)\geq 0$. Hence, again, $\frac{\partial W_i(\rho_i|x)}{\partial x}\geq 0$.

Suppose finally $\hat{w}_i(\rho_i)\leq \bar{w}_{F_i}\leq  \bar{w}_{\tilde{F}_i}$ so the priority is to high scores under both distributions. Then, using \eqref{eq:objup} and the relationship $\mathcal{W}^-(s,\hat{w})=\mathcal{W}^+(s,\hat{w})-(\bar{w}-\hat{w})$, we can write the welfare change as
\[\bigl(\bar{w}_{\tilde{F}_i}-\bar{w}_{F_i}\bigr)\left(1-\int_{\underline{s}_i}^{\overline{s}_i}\alpha_i^{*'}(s)ds\right)\int_{\underline{s}_i}^{\overline{s}_i}\alpha_i^{*'}(s)\bigl\{\tilde{\mathcal{W}}^+_i(s,\hat{w}_i(\rho_i))-\mathcal{W}^+_i(s,\hat{w}_i(\rho_i))\bigr\}ds,\]
and both terms are positive, and then $\frac{\partial W_i(\rho_i|x)}{\partial x}\geq 0$.

Now, consider the across problem. Applying the (classical) envelope theorem to this problem, and letting $\bm{\rho}^*$ denote its unique solution, we obtain
\[\frac{\partial \overline{W}(x)}{\partial x}=\mu_i \frac{\partial W_i(\rho^*_i|x)}{\partial x}\geq 0.\]
\end{proof}

\section{Proofs of \secref{sec:general fals}}

As a preliminary, we show that the cost function satisfies the standard triangular inequality
\begin{equation}\label{eq:tri}
\forall s,m,t,~~~~~c(t|m)+c(m|s)\geq c(t|s)\tag{TRI}
\end{equation}
\begin{Lem}[Triangular inequality]\label{lem-tri}
If $c$ satisfies \eqref{eq:dtri},\eqref{eq:uid}, and downward monotonicity, then it satisfies \eqref{eq:tri}.
\end{Lem}
\begin{proof}
\eqref{eq:uid} proves the claim for $s<m<t$, and \eqref{eq:dtri} for $t<m<s$, while cost monotonicity proves it for all other configurations.
\end{proof}


First, we show that we can, without loss of generality, restrict attention to falsification-proof allocation rules.

\begin{Prop}[Falsification-proofness principle]\label{prop:FPprinc}
Assume \eqref{eq:tri}. Let $\alpha$ be an allocation rule, and $\phi$ a best-response of the agents (i.e. $\phi$ satisfies \eqref{eq:IC}). Then there exists an allocation rule $\tilde{\alpha}$ that satisfies: (i) falsification-proofness \eqref{eqFPC}; (ii) $\phi$ is a best response to $\tilde{\alpha}$; (iii) for every $s$, the equilibrium allocation probability is unchanged, $\int\alpha(t)d\phi(t|s)=\int \tilde{\alpha}(t)d\phi(t|s)$.
\end{Prop}
\begin{proof}
Let $u(s)=\max_t \alpha(t)-c(t|s)$ be the value function of the agents under $\alpha$. Then we define the new allocation rule $\tilde{\alpha}(s)=u(s)$. It satisfies $\tilde{\alpha}(s)=\alpha(s)$ if $s\in T(\alpha,s)$, and $\tilde{\alpha}(s)>\alpha(s)$ otherwise.  

\step{Claim 1:} \emph{If $s\notin T(\alpha,s)$, then for every $s'\in S$, $s\notin T(\alpha,s')$.}

\medskip

Let $t\in T(\alpha,s)$. Since $s\notin T(\alpha,s)$, $\alpha(t)-c(t|s)> \alpha(s)$. Suppose, by contradiction, that $s\in T(\alpha,s')$ for some $s'\neq s$. Then, $\alpha(s)-c(s|s')\geq \alpha(t)-c(t|s')$. Combining these two inequalities, we obtain $c(t|s)+c(s|s')<c(t|s')$, which contradicts \eqref{eq:tri}. 

In particular, claim 1 implies that for any score $t$ that is a target under $\alpha$, so $t\in \cup_{s\in S}T(\alpha,s)$, we have $\tilde{\alpha}(t)=\alpha(t)$.

\step{Claim 2:} \emph{$T(\alpha,s)\cup\{s\}\subseteq T(\tilde{\alpha},s)$.}

\medskip

Let $t\in T(\alpha,s)$, so $\alpha(t)=\tilde{\alpha}(t)$ by Claim 1. Consider any $t'\neq s$. If $t'$ is a target under $\alpha$, then $\tilde{\alpha}(t')=\alpha(t')$ by Claim 1, and $\alpha(t)-c(t|s)\geq \alpha(t')-c(t'|s)$ (since $t\in T(\alpha,s)$) implies $\tilde{\alpha}(t)-c(t|s)\geq \tilde{\alpha}(t')-c(t'|s)$. Then suppose that $t'$ is not a target. Then, there exists $t''$ such that $\tilde{\alpha}(t')=\alpha(t'')-c(t''|t')$. Then, $t\in T(\alpha,s)$ and \eqref{eq:tri} imply
\[\tilde{\alpha}(t)-c(t|s)=\alpha(t)-c(t|s)\geq \alpha(t'')-c(t''|s)=\tilde{\alpha}(t')+c(t''|t')-c(t''|s)\geq \tilde{\alpha}(t')-c(t'|s).\]

Next, we show that $s\in T(\tilde{\alpha},s)$. Suppose $s\notin T(\alpha,s)$ since we can otherwise use the same argument as above. Consider any $t'\neq s$. If $t'$ is a target under $\alpha$, then $\tilde{\alpha}(t')=\alpha(t')$ and $\tilde{\alpha}(s)=u(s)\geq \alpha(t')-c(t'|s)=\tilde{\alpha}(t')-c(t'|s)$. Suppose instead that $t'$ is not a target under $\alpha$. Then, there exists $t''$ such that $\tilde{\alpha}(t')=\alpha(t'')-c(t''|t')$. Then, \eqref{eq:tri} implies
\[\tilde{\alpha}(s)=u(s)\geq \alpha(t'')-c(t''|s)=\tilde{\alpha}(t')+c(t''|t')-c(t''|s)\geq \tilde{\alpha}(t')-c(t'|s).\]

\step{Conclusion:} Claim 2 implies that $\phi$ is  a best-response to $\tilde{\alpha}$ since, for every $s$, $T(\alpha,s)\subseteq T(\tilde{\alpha},s)$. It also implies that $\tilde{\alpha}$ satisfies the falsification-proofness constraint since, for every $s$, $s\in T(\tilde{\alpha},s)$. Finally, (iii) holds since $\alpha(t)=\tilde{\alpha}(t)$ for every $t\in T(\alpha,s)$, and $\supp \phi(\cdot|s)\subseteq T(\alpha,s)$ by \eqref{eq:IC}.
\end{proof}


We define $s^{**}_\zeta=\max\,\bigl\{t\in[\hat{\sigma}(\zeta),\overline{s}]\,:\,c\bigl(t|\hat{\sigma}(\zeta)\bigr)\leq 1\bigr\}$. It is the cutoff above which the unconstrained optimal rule $\alpha^{**}$ under the shifted worth function allocates with maximum probability. Then, $s^{**}_\zeta$ is decreasing in $\zeta$ and $s^{**}_\zeta<\overline{s}$ if and only if $c\bigl(\overline{s}|\hat{\sigma}(\zeta)\bigr)>1$. We show that we can restrict attention to allocation rules that are flat above $s^{**}_\zeta$.

\begin{Lem}[Top-Flatness]\label{lem:topflat}
Assume \eqref{eq:tri}. Let $\alpha$ be an allocation rule, and $\phi$ a best-response to $\alpha$. There exists an allocation rule $\tilde{\alpha}$ that satisfies \eqref{eqFPC} and is flat above $s^{**}_\zeta$, and a best-response $\tilde{\phi}$ to $\tilde{\alpha}$ such that: (i) $\iint \tilde{\alpha}(t)w_\zeta(s)d\tilde{\phi}(t|s)dF(s)\geq \iint \alpha(t)w_\zeta(s)d\phi(t|s)dF(s)$ and $\iint c(t|s)d\tilde{\phi}(t|s)dF(s)\leq \iint c(t|s)d\phi(t|s)dF(s)$; (ii) $\tilde{\phi}(s)=\delta_s$ for all $s\geq s^{**}_\zeta$ and all $s\leq \hat{\sigma}(\zeta)$.
\end{Lem}
\begin{proof}
We start by applying the falsification-proofness principle of \propref{prop:FPprinc} to $\alpha$, so we might as well assume that $\alpha$ satisfies \eqref{eqFPC}. We can also assume that ineligible scores do not falsify under $\phi$, as it is a best-response to $\alpha$ by \eqref{eqFPC}; it lowers $\iint c(t|s)d\phi(t|s)dF(s)$ and increases $\iint \alpha(t)w_\zeta(s)d\phi(t|s)dF(s)$. 

If $s^{**}_\zeta=\overline{s}$, there is nothing to prove. Suppose, therefore, that $s^{**}_\zeta<\overline{s}$. Then, we let 
\[\hat{\alpha}(s)=\begin{cases}1&\text{if } s\geq s^{**}_\zeta\\ 1-c(s^{**}_\zeta|s)&\text{if } s<s^{**}_\zeta \text{ and }c(s^{**}_\zeta|s)\leq 1\\ 0&\text{otherwise}\end{cases}.\] 
And we let $\tilde{\alpha}(s)=\max\bigl\{\alpha(s),\hat{\alpha}(s)\bigr\}$. By definition of $s^{**}_\zeta$, $c(s^{**}_{\zeta}|\hat{\sigma}(\zeta))=1$, hence $\tilde{\alpha}(s)=\alpha(s)$ for all $s\leq \hat{\sigma}(\zeta)$. Scores above $s^{**}_\zeta$ have no incentive to falsify and receive a higher allocation probability under $\tilde{\alpha}$ than under $\alpha$. 

We partition $[\underline{s},s^{**}_\zeta)$ into two sets: $\hat{S}^+=\bigl\{s\in[\underline{s},s^{**}_\zeta)\,:\,\alpha(s)\geq \hat{\alpha}(s)\bigr\}$, and $\hat{S}^-=\bigl\{s\in[\underline{s},s^{**}_\zeta)\,:\,\alpha(s)< \hat{\alpha}(s)\bigr\}$. Hence, $\hat{S}^-$ is the set where $\tilde{\alpha}(s)=\hat{\alpha}(s)$. Furthermore, all scores in $\hat{S}^-$ are eligible for the shifted worth function since they satisfy $s\geq\hat{\sigma}(\zeta)$.

Consider the falsification strategy:
\[
\tilde{\phi}(s)=\begin{cases}
\phi(s)&\text{if }s\in \hat{S}^+\\
\pi(s) \delta_s + \bigl(1-\pi(s)\bigr)\delta_{s^{**}_\zeta}&\text{if }s\in \hat{S}^-\\
\delta_s&\text{if }s\geq s^{**}_\zeta
\end{cases}.
\]
For $s\geq s^{**}_\zeta$, not falsifying is optimal under $\tilde{\alpha}$. This yields less falsification from these scores and a higher allocation probability.

If $s\in \hat{S}^+$, the only new potential targets for $s$ under $\tilde{\alpha}$ are $s^{**}_\zeta$ and scores in $\hat{S}^-$. We show that there is no gain from falsifying to any such score. By definition, we have $\tilde{\alpha}(s)=\alpha(s)\geq 1-c(s^{**}_\zeta|s)$, so not falsifying dominates falsifying to $s^{**}_{\zeta}$. But then, \eqref{eq:tri} implies that, for any $\tilde{s}\in \hat{S}^-$,
\[\tilde{\alpha}(s)=\alpha(s)\geq 1-c(s^{**}_\zeta|s)\geq 1-c(s^{**}_\zeta|\tilde{s})-c(\tilde{s}|s)=\tilde{\alpha}(\tilde{s})-c(\tilde{s}|s)\geq \alpha(\tilde{s})-c(\tilde{s}|s).\]
Hence, not falsifying also dominates falsifying as $\tilde{s}$ both under $\alpha$ and under $\tilde{\alpha}$. Therefore, $\tilde{\phi}(s)=\phi(s)$ is an optimal strategy for $s$ under $\tilde{\alpha}$.

If $s\in \hat{S}^-$, then $s$ must be eligible, as noted above. We choose $\pi(s)=1$ if $\tilde{\alpha}(s)\geq \int\alpha(t)d\phi(t|s)$. This yields a higher allocation probability for $s$ and a lower cost of falsification. Otherwise, we choose $\pi(s)$ so that
\[\pi(s)\tilde{\alpha}(s)+1-\pi(s)=\int_{T(\alpha,s)}\alpha(t)d\phi(t|s),\]
leaving the final expected allocation probability unchanged. By \eqref{eqFPC}, $ \alpha(t)=\alpha(s)+c(t|s)$ for all $t\in T(\alpha,s)$. Hence,
\[\pi(s)=\frac{1-\int_{T(\alpha,s)}\alpha(t)d\phi(t|s)}{c(s^{**}_\zeta|s)}=\frac{1-\alpha(s)-\int_{T(\alpha,s)}c(t|s)d\phi(t|s)}{c(s^{**}_\zeta|s)}.\]
The cost of falsification for $s$ under $\tilde{\phi}$ is
\begin{align*}   
\int c(t|s)d\tilde{\phi}(t|s)&=\bigl(1-\pi(s)\bigr)c(s^{**}_\zeta|s)\\
&=c(s^{**}_\zeta|s)-1+\alpha(s)+\int_{T(\alpha,s)}c(t|s)d\phi(t|s)\\
&<\int_{T(\alpha,s)}c(t|s)d\phi(t|s),
\end{align*}
since $s\in \hat{S}^-$.

Overall, we have obtained a falsification strategy $\tilde{\phi}$ that is optimal under $\tilde{\alpha}$; induces a higher allocation probability for eligible scores and the same allocation probability for ineligible scores; and has a lower cost of falsification than $\phi$. If $\tilde{\alpha}$ does not satisfy \eqref{eqFPC}, we can apply to it the transformation in \propref{prop:FPprinc} which clearly preserves flatness above $s^{**}_\zeta$. 
\end{proof}


\begin{proof}[Proof of \theoremref{thm:fp-optimal}]
By  \lemmaref{lem:topflat}, it is without loss of optimality for \eqref{pgm:wwp} to restrict attention to the set $\mathcal{A}$ of allocation rules that satisfy \eqref{eqFPC}, and the top-flatness property, paired with an honest falsification strategy for scores outside of $\bigl(\hat{\sigma}(\zeta),s^{**}_\zeta\bigr)$. Introducing the aggregate cost of falsification $\Phi$ as an auxiliary optimization variable, we can rewrite \eqref{pgm:wwp} as
\begin{align*}
\max_{\Phi\geq 0}\max_{\alpha\in \mathcal{A}}\max_{\phi}~&\iint \alpha(t)w_\zeta(s)d\phi(t|s)dF(s)-\zeta \Phi\\
\text{s.t.}~&\eqref{eq:IC}, \iint c(t|s)d\phi(t|s)dF(s)=\Phi 
\end{align*}
This allows us to isolate the following optimization problem 
\begin{align}\label{PgmFixFals}
Z(\Phi)=\max_{\alpha\in \mathcal{A}}\max_{\phi}~&\Pi(\alpha,\phi)\tag{FixFals}\\
\text{s.t.}~&\eqref{eq:IC}, \iint c(t|s)d\phi(t|s)dF(s)=\Phi \notag 
\end{align}
where we defined $\Pi(\alpha,\phi)=\iint \alpha(t)w_\zeta(s)d\phi(t|s)dF(s)$
to be the allocative surplus with shifted worth function. The problem consists in choosing an allocation rule and an optimal falsification strategy of the agent so as to maximize this (modified) allocative surplus at a fixed aggregate falsification cost $\Phi$. Note that $\alpha$ satisfies \eqref{eqFPC} so any amount of falsification can be shut down without affecting agent optimality.

The original problem then becomes $\max_{\Phi\geq 0}Z(\Phi)-\zeta\Phi$. To prove the theorem, we show that $Z$ has bounded upward differences with bounding factor $w_\zeta\bigl(s^{**}_\zeta\bigr)$, so that the solution to this problem must be $\Phi=0$ whenever $\zeta\geq w_\zeta\bigl(s^{**}_\zeta\bigr)$ (uniquely if the inequality is strict), implying that $\alpha^*_\zeta$ then solves the weighted welfare problem. This yields the condition $ w\bigl(s^{**}_\zeta\bigr)\leq 0$, which is equivalent to $s^{**}_\zeta\leq \hat{s}$. By definition of $s^{**}_\zeta$, this is equivalent to \eqref{eq:honest}.

\step{Bounded upward differences:} Let $\Phi'>\Phi\geq 0$ and let $(\alpha',\phi')$ solve \eqref{PgmFixFals} under $\Phi'$, and $(\alpha,\phi)$ under $\Phi$. Then 
\[\iint c(t|s)d\phi'(t|s)dF(s)=\Phi'>\Phi=\iint c(t|s)d\phi(t|s)dF(s).\]
As noted above, because $\alpha'$ satisfies \eqref{eqFPC}, we can shut down any amount of falsification in $\phi'$ without affecting agent optimality. Therefore, there exists an optimal falsification strategy $\tilde{\phi}$ of the agents under $\alpha'$ such that $\iint c(t|s)d\tilde{\phi}(t|s)dF(s)=\Phi$. Then we have
\begin{align*}
Z(\Phi')-Z(\Phi)&=\Pi(\alpha',\phi')-\Pi(\alpha',\tilde{\phi})+\Pi(\alpha',\tilde{\phi})-\Pi(\alpha,\phi)\\
&\leq \Pi(\alpha',\phi')-\Pi(\alpha',\tilde{\phi})\\
&=\int_S \int_{T(\alpha',s)} \alpha'(t)w_\zeta(s)\bigl[d\phi'(t|s)-d\tilde{\phi}(t|s)\bigr]dF(s)\\
&=\int_S \int_{T(\alpha',s)}\bigl(\alpha'(s)+c(t|s)\bigr)w_\zeta(s)\bigl[d\phi'(t|s)-d\tilde{\phi}(t|s)\bigr]dF(s)\\
&=\int_S w_\zeta(s)\int_{T(\alpha',s)} c(t|s)\bigl[d\phi'(t|s)-d\tilde{\phi}(t|s)\bigr]dF(s)\\
&\leq w_\zeta(s^{**}_\zeta)\iint c(t|s)\bigl[d\phi'(t|s)-d\tilde{\phi}(t|s)\bigr]dF(s)\\
&=w_\zeta(s^{**}_\zeta)\bigl(\Phi'-\Phi\bigr).
\end{align*}
Since $(\alpha,\phi)$ is optimal for \eqref{PgmFixFals} at $\Phi$, $\Pi(\alpha',\tilde{\phi})\leq \Pi(\alpha,\phi)$, implying the first inequality.  The second equality is because both $\phi'(s)$ and $\tilde{\phi}(s)$ are supported on $T(\alpha',s)$. The third equality is due to \eqref{eqFPC} which implies that $\alpha'(t)-c(t|s)=\alpha'(s)$ for all $t\in T(\alpha',s)$. The fourth equality is because $\phi'(s)$ and $\tilde{\phi}(s)$ are probability distributions fully supported on $T(\alpha',s)$, so $\int_{T(\alpha',s)}d\phi'(t|s)=\int_{T(\alpha',s)}d\tilde{\phi}(t|s)=1$. The second inequality is because $\phi'(s)=\tilde{\phi}(s)=\delta_s$ for $s\geq s^{**}_\zeta$ by the top-flatness property. 
\end{proof}


\begin{proof}[Proof of \propref{PropShift}]
Let $\tilde{\alpha}^*(\zeta)$ denote the correspondence mapping $\zeta$ to the set of solutions of  \eqref{eqBP} with shifted worth. By \lemmaref{lem:mon}, we can write \eqref{eqBP} as an optimization problem over the set of nondecreasing functions from $S$ to $[0,1]$ satisfying \eqref{eqFPC} and \eqref{eqPC}. This space is compact, by Helly's theorem, and convex. Furthermore, the objective function is linear and therefore continuous in $\alpha$. Hence, Berge's maximum theorem implies that $\tilde{\alpha}^*(\zeta)$ is an upper hemicontinuous correspondence. 


The space of feasible and non-decreasing allocation rules is also a lattice with respect to the partial order $\alpha\succeq \beta \Leftrightarrow ~\forall s~~\alpha(s)\geq \beta(s)$, with the corresponding strict ordering $\alpha\succ \beta$ if $\alpha\succeq \beta$ and $\alpha(s)>\beta(s)$ for some $s$. Indeed, it is easy to see that, for two such allocation rules $\alpha$ and $\beta$, their meet $\alpha\wedge\beta$ and their join $\alpha\vee\beta$ are also nondecreasing and feasible. Furthermore, the objective function is supermodular in $\alpha$ and has strictly increasing differences in $(\zeta,\alpha)$. Hence, by Milgrom and Shannon's monotone selection theorem \citep{MilShan94}, $\alpha^*_\zeta(\cdot,r)$ is strictly increasing in $\zeta$ for the $\succeq$ order (recalling that the allocation rule is independent of $r$ except when $\zeta=-\bar{w}$, where $r$ pins down the selection). 

\end{proof}

\section{Comparative statics and shape}\label{app-sec-comp-stat} 

In this section, we provide additional comparative statics results for the optimal falsification-proof allocation rule of the paper.

\subsection{Shape under shifted linear costs} \label{subsec-shiftLin} 

\begin{Prop}[Baseline rules under shifted linear costs]\label{PropShftLin}
In the shifted linear cost model, the baseline rule is concave in the \eqref{eq:udd} case. In the \eqref{eq:uid} case, it is convex below the eligibility threshold if $\frac{\kappa^{''}(s)}{\kappa'(s)}\leq \frac{2}{\overline{s}-s}$, and always concave above the eligibility threshold.
\end{Prop}

\begin{proof}
Let $\alpha$ denote the baseline rule. For the shifted linear cost family, we have $c_{t}(t|s) = \kappa(s)$, $c_{s}(t|s) = -\kappa(s)+(t- s)\kappa'(s)$ and $c_{ts}(t|s)=\kappa'(s)$. If $\kappa'(s) \leq 0$, we are in the \eqref{eq:udd} case. On the growth interval, $\alpha'(s) = c_{t}(s|s) = \kappa(s)$ and $\alpha''(s) = \kappa'(s) \leq 0$ so the baseline rule is concave along the growth interval.

If $\kappa'(s) \geq 0$, we are in the \eqref{eq:uid} case. For ineligible scores on the growth interval,
\[\alpha'(s) = -c_{s}(m(s)|s) = \kappa(s) -(m(s)- s)\kappa'(s),\]
and
\[\alpha{''}(s) = (2-m'(s))\kappa'(s) - (m(s)-s)\kappa^{''}(s).\]
Note that, for the linear shifted cost to satisfy our basic assumptions on costs, we must have $c_s(t|s)=-\kappa(s) +(t- s)\kappa'(s)<0$ for every $t$, that is, $\frac{\kappa'(s)}{\kappa(s)}\leq \frac{1}{\overline{s}-s}$. Since $m'(s)< 0$, and $\kappa'(s)\geq 0$, we have $\alpha''(s)\geq 2\kappa'(s)- (m(s)-s)\kappa^{''}(s)$. Then $\alpha''(s)\geq 0$, whenever
\[\frac{\kappa''(s)}{\kappa'(s)}\leq \frac{2}{\overline{s}-s}.\]
For eligible scores, $\alpha'(s) = c_{t}\bigl(s|m^{-1}(s)\bigr) = \kappa(m^{-1}(s))$ and \[\alpha''(s)=\frac{1}{m'\bigl(m^{-1}(s)\bigr)}\kappa'(m^{-1}(s))\leq 0\] since $m$ is decreasing. 
\end{proof}

\subsection{Returns to scale in falsification} \label{subsec-RTS} 

We investigate how the shape of the falsification cost function influences optimal allocations. Specifically, we focus on the class $\mathcal{E}$ of Euclidean cost functions, $c(t | s) = \mathcal{C}\bigl(|t - s|\bigr)$, where $\mathcal{C}$ is either concave or convex, and examine the effect of increasing the convexity of the cost function. Intuitively, greater convexity captures lower economies of scale in the magnitude of falsification. 


To compare cost functions, we need some normalization. We normalize costs so that the maximum amount of falsification an agent is willing to undertake to get the good is identical for all cost functions and less than $s^*(0) - s_*(0)$. That is, we consider two cost functions $\mathcal{C}, \hat{\mathcal{C}}$ such that $\mathcal{C}^{-1}(1) = \hat{\mathcal{C}}^{-1}(1) < s^*(0) - s_*(0)$. We say that $\hat{\mathcal{C}}$ is more convex than $\mathcal{C}$, and denote $\hat{\mathcal{C}} \succeq_{vex} \mathcal{C}$, if either $\hat{\mathcal{C}}$ is convex and $\mathcal{C}$ is concave, or both are concave and $\mathcal{C}$ is more concave than $\hat{\mathcal{C}}$ in the usual sense, or both are convex and $\hat{\mathcal{C}}$ is more convex than $\mathcal{C}$ in the usual sense.\footnote{
That is, there exists an increasing and concave function $g: [0,1] \to [0,1]$ such that $\mathcal{C} = g \circ \hat{\mathcal{C}}$ when both are concave, or an increasing and convex function $h: [0,1] \to [0,1]$ such that $\hat{\mathcal{C}} = h \circ \mathcal{C}$ if both are convex.} We denote the corresponding optimal falsification-proof allocation rules by $\alpha^*, \hat{\alpha}^*$, and their growth intervals by $I^*, \hat{I}^*$.

\begin{Prop}[Effect of lowering economies of scale]\label{PropScale}
If $\hat{\mathcal{C}}$ is more convex than $\mathcal{C}$, then:
\begin{enumerate}[(i)]
\item $I^*\subseteq \hat{I}^*\subseteq [s_*(0),s^*(0)]$. Furthermore, $I^*= \hat{I}^*\subset [s_*(0),s^*(0)]$ if both cost functions are concave. 
\item If $I^*\subset [s_*(0),s^*(0)]$, there exists a threshold $\tilde{s}\in I^*$ such that $\hat{\alpha}^*(s)\leq \alpha^*(s)$ for $s>\tilde{s}$, and $\hat{\alpha}^*(s)\geq \alpha^*(s)$ for $s<\tilde{s}$.
\item If $I^*=\hat{I}^*=[s_*(0),s^*(0)]$, then both cost functions are convex and $\hat{\alpha}^*(s)\leq\alpha^*(s)$ for all $s$ if the group has low-score priority, but $\hat{\alpha}^*(s)\geq\alpha^*(s)$ for all $s$ if the group has high-score priority.
\end{enumerate}
\end{Prop}

\begin{proof}
When $\mathcal{C}$ is concave, the growth interval is determined by the equation $m(s_*)-s_*=L$, and does not vary with the cost function, and it is a subset of $[s_*(0),s^*(0)]$ since we assumed $L<s^*(0)-s_*(0)$. For convex cost functions, the growth interval is given by the equation $m(s_*)-s_*=\min\{s^*(0)-s_*(0),1/\mathcal{C}'(0)\}$ by \propref{PropEucl}. Furthermore, if both cost functions are convex, the convex ordering and our normalization imply $\hat{\mathcal{C}}'(0)\leq \mathcal{C}'(0)$, hence $I^*\subseteq \hat{I}^*$. 

Next, let $\delta(s)=\hat{\alpha}^*(s)-\alpha^*(s)$. Suppose first that both cost functions are concave, and let $I^*=[s_*,s^*]$ be their common growth interval. In particular, $\delta(s_*)=\delta(s^*)=0$. Furthermore, $\delta$ is differentiable and 
\[\delta'(s)=
\begin{cases}
\bigl\{1-g'\circ\hat{\mathcal{C}}\bigl(m(s)-s\bigr)\bigr\}\hat{\mathcal{C}}'\bigl(m(s)-s\bigr) & \text{if }s\in[s_*,\hat{s}],\\
\bigl\{1-g'\circ\hat{\mathcal{C}}\bigl(s-m^{-1}(s)\bigr)\bigr\}\hat{\mathcal{C}}'\bigl(s-m^{-1}(s)\bigr) & \text{if }s\in[\hat{s},s^*]
\end{cases},\]
where $g$ is an increasing and concave bijection of $[0,1]$ such that $\mathcal{C}=g\circ \hat{\mathcal{C}}$. As such, $g'(0)\geq 1 \geq g'(1)$ and $g'$ is a non-increasing function. Since
$\mathcal{C}'\geq 0$, this implies $\delta'$ is single crossing from the positives to the negatives on  $[s_*,\hat{s}]$ and from the negatives to the positives on $[\hat{s},s^*]$. Therefore, there exists a single threshold $\tilde{s}\in[s_*,s^*]$ such that $\delta(s)\geq 0$ for $s\leq \tilde {s}$ and $\delta(s)\leq 0$ for $s\geq \tilde{s}$.

If the two cost functions are convex, then for $\hat{\mathcal{C}}$ to be more convex than $\mathcal{C}$, it must be that $\hat{\mathcal{C}}'(0)\leq \mathcal{C}'(0)$, which implies (iii). 

Let $\tilde{\mathcal{C}}$ be the unique linear cost function that belongs to our normalized class of functions. Since $\tilde{\mathcal{C}}$ is both concave and convex, point (ii) is satisfied when comparing $\tilde{\mathcal{C}}$ to a concave cost function $\mathcal{C}$, and also when comparing a convex cost function $\hat{\mathcal{C}}$ to $\tilde{\mathcal{C}}$. Since $\hat{\alpha}^*-\alpha^*=\hat{\alpha}^*-\tilde{\alpha}^*+\tilde{\alpha}^*-\alpha^*$, it is also satisfied when comparing $\hat{\mathcal{C}}$ to $\mathcal{C}$.

If $I^*=\hat{I}^*=[s_*(0),s^*(0)]$, then both functions must be convex by (i). If the group has low-score priority, then $\alpha^*\bigl(s_*(0)\bigr)=\hat{\alpha}^*\bigl(s_*(0)\bigr)=0$, and both allocation rules are linear with respective slopes $\mathcal{C}'(0)\geq \hat{\mathcal{C}}'(0)$, implying $\hat{\alpha}^*(s)\leq \alpha^*(s)$ for all $s$. If instead the group has high-score priority, the slopes compare in the same way, but the allocation rules are tied at $s^*(0)$ instead of $s_*(0)$, implying $\hat{\alpha}^*(s)\geq \alpha^*(s)$ for all $s$.
\end{proof}

Lower economies of scale, like higher gaming ability, benefit low-score agents and hurt high-score agents. However, if the diseconomies of scale become too strong, as in case (iii), the effect is uniform across all scores, resembling the impact of high gaming abilities as described in \propref{PropGam}.

\subsection{Score distribution} \label{subsec-score} 

We examine what changes in the score distribution result in a uniformly higher optimal allocation probability. To simplify notation, we write $s$ for the surplus, so $s = w(s) - \hat{w}$
which effectively transforms the score distribution $F$. Consequently, the eligibility threshold is fixed at 0. We consider two atomless score distributions, $\hat{F}$ and $\tilde{F}$, whose common support $[\underline{s}, \overline{s}]$ includes a neighborhood of 0. The function $\Delta(s) = \tilde{F}(s) - \hat{F}(s)$ denotes the change in the score distribution.

All distributional effects on the allocation rule are transmitted through the matching functions $\hat{m}(s)$ and $\tilde{m}(s)$. We begin by showing that the allocation rules satisfy $\tilde{\alpha}(s) \geq \hat{\alpha}(s)$ for every $s$, and for every cost function that meets the conditions in \eqref{eq:uid} or \eqref{eq:udd}, if and only if $\tilde{m}(s) \leq \hat{m}(s)$ for every $s \leq 0$. We then provide a necessary and sufficient condition on the distributions for the matching function to decrease.

Note that the matching function was defined on the interval $[s_*(0), 0]$, but the lower bound $s_*(0)$ may now depend on the specific score distribution used. To ease the exposition, we extend each matching function $\hat{m}(s)$ and $\tilde{m}(s)$ to the left by setting $\hat{m}(s) = \hat{m}\bigl(\hat{s}_*(0)\bigr)$ for $s \leq \hat{s}_*(0)$, and $\tilde{m}(s) = \tilde{m}\bigl(\tilde{s}_*(0)\bigr)$ for $s \leq \tilde{s}_*(0)$.

\begin{Prop}[Effect of score distribution]\label{PropDist}
The following statements are equivalent:
\begin{enumerate}[(a)]
\item The allocation rules satisfy $\tilde{\alpha}(s)\geq \hat{\alpha}(s)$ for every $s$, and every cost function that satisfies \eqref{eq:uid} or \eqref{eq:udd}.
\item The matching functions satisfy  $\tilde{m}(s)\leq \hat{m}(s)$ for every $s\leq 0$.
\item \label{pdistc} For every $0>s\geq \max\{\hat{s}_*(0),\tilde{s}_*(0)\}$, \label{ccond}
$$ \int_s^{\tilde{m}(s)}xd\tilde{F}(x)\geq \int_s^{\tilde{m}(s)}xd\hat{F}(x).$$
\item \label{pdistd} For every $0>s\geq \max\{\hat{s}_*(0),\tilde{s}_*(0)\}$,
$$ \int_s^{0}\big\{\Delta(s)-\Delta(x)\bigr\}dx+ \int_0^{\tilde{m}(s)}\big\{\Delta\bigl(\tilde{m}(s)\bigr)-\Delta(x)\bigr\}dx\geq 0.$$
\end{enumerate}
\end{Prop} 
\begin{proof}~\\
\step{\textbullet\ \underline{$(b)\Rightarrow (a)$}.} Suppose (b) holds. 
\medskip

\textopenbullet\ \emph{We start by showing (i) $\tilde{s}_*(0)\leq \hat{s}_*(0)$ and (ii) $\tilde{s}^*(0)\leq \hat{s}^*(0)$.}
\medskip

First suppose $\tilde{s}^*(0)=\overline{s}$. Then (ii) must hold, and (i) also because, otherwise, we would have the following contradiction
$$\overline{s}=\tilde{s}^*(0)=\tilde{m}\bigl(\tilde{s}_*(0)\bigr)\leq \hat{m}\bigl(\tilde{s}_*(0)\bigr)<\hat{m}\bigl(\hat{s}_*(0)\bigr)=\hat{s}^*(0),$$
where the first inequality is by (b), and the second inequality because $\hat{m}$ is decreasing on $[\hat{s}_*(0),0]$.

Next, suppose $\hat{s}_*(0)=\underline{s}$. Then (i) must hold, and (ii) also because, otherwise, we would have the following contradiction
$$\tilde{s}_*(0)=\tilde{m}^{-1}\bigl(\tilde{s}^*(0)\bigr)<\tilde{m}^{-1}\bigl(\hat{s}^*(0)\bigr)\leq \hat{m}^{-1}\bigl(\hat{s}^*(0)\bigr)=\hat{s}_*(0)=\underline{s},$$
where the first inequality is because $\tilde{m}^{-1}$ is decreasing on $[0,\tilde{s}^*(0)]$, and the second inequality is by (b).

If neither of these cases hold, by \lemmaref{LemCumSurpl}, (iii), we must have $\tilde{s}_*(0)=\underline{s}$ and $\hat{s}^*(0)=\overline{s}$, which imply (i) and (ii).

An implication of (i) and (ii) is (iii): if $\hat{F}$ has high-score priority, then so does $\tilde{F}$, and if $\tilde{F}$ has low-score priority, then so does $\hat{F}$.

\medskip

\textopenbullet\ \emph{Next, consider a cost function that satisfies \eqref{eq:udd}.} 
\medskip

(b) implies $\tilde{m}(\hat{s}_*)\leq \hat{m}(\hat{s}_*)=\hat{s}^*$, therefore
$$\int_{\hat{s}_*}^{\tilde{m}(\hat{s}_*)}c_{t^+}(x|x)dx\leq \int_{\hat{s}_*}^{\hat{s}^*}c_{t^+}(x|x)dx=1.$$
Then, using \eqref{eq:bounds} and point (i) we just proved, we must have $\tilde{s}_*\leq \hat{s}_*$.

Then, for every $s\in[\hat{s}_*,\tilde{s}^*]$, 
$$\tilde{\alpha}(s)-\hat{\alpha}(s)=\tilde{\Gamma}_{udd}\Ind_{\mathcal{E}}+\int_{\tilde{s}_*}^{\hat{s}_*}c_{t^+}(x|x)dx\geq 0,$$
where $\mathcal{E}$ is the event in which only $\tilde{F}$ has high-score priority (the event in which only $\hat{F}$ has high-score priority is impossible by (iii)). This also implies $\tilde{s}^*\leq \hat{s}^*$, so, for any $s\geq \tilde{s}^*$, we also have $1=\tilde{\alpha}(s)\geq \hat{\alpha}(s)$. Finally, for $s\leq\hat{s}_*$, we have $\tilde{\alpha}(s)\geq \hat{\alpha}(s)=0$.

\medskip

\textopenbullet\ \emph{Finally, consider a cost function that satisfies \eqref{eq:uid}.}

\medskip
(b) implies $\tilde{m}(\hat{s}_*)\leq \hat{m}(\hat{s}_*)=\hat{s}^*$, therefore
$$c\bigl(\tilde{m}(\hat{s}_*)|\hat{s}_*\bigr)\leq c\bigl(\hat{s}^*|\hat{s}_*\bigr)\leq 1.$$
Then, using \eqref{eq:bounds} and point (i) we just proved, we must have $\tilde{s}_*\leq \hat{s}_*$.

(b) also implies $\hat{m}^{-1}(\tilde{s}^*)\geq \tilde{m}^{-1}(\tilde{s}^*)=\tilde{s}_*$, therefore 
$$c\bigl(\tilde{s}^*|\hat{m}^{-1}(\tilde{s}^*)\bigr) \leq c(\tilde{s}^*|\tilde{s}_*)\leq 1.$$
Then, using \eqref{eq:bounds} and point (ii) we just proved, we must have $\tilde{s}^*\leq \hat{s}^*$.

Then, for every $s\in[\hat{s}_*,0]$, 
\begin{align*}
\tilde{\alpha}(s)-\hat{\alpha}(s)&=\tilde{\Gamma}_{uid}\Ind_{\mathcal{E}}-\int_{\tilde{s}_*}^{\hat{s}_*}c_s\bigl(\tilde{m}(x)|x\bigr)dx-\int_{\hat{s}_*}^s\bigl\{c_s\bigl(\tilde{m}(x)|x\bigr)-c_s\bigl(\hat{m}(x)|x\bigr)\bigr\}dx\\
&\geq 0,
\end{align*}
where $\tilde{\Gamma}_{uid}\geq 0$ by definition, the second term is nonnegative by cost monotonicity, and the third term is nonnegative by \eqref{eq:uid} and (b).

And for every $s\in [0,\tilde{s}^*]$, 
\begin{align*}
\tilde{\alpha}(s)-\hat{\alpha}(s)&=\hat{\Gamma}_{uid}\Ind_{\mathcal{E}'}+\int_{\tilde{s}^*}^{\hat{s}^*}c_t\bigl(x|\hat{m}^{-1}(x)\bigr)dx\\
&~~~~+\int_{s}^{\tilde{s}^*}\bigl\{c_t\bigl(x|\hat{m}^{-1}(x)\bigr)-c_t\bigl(x|\tilde{m}^{-1}(x)\bigr)\bigr\}dx\\
&\geq 0,
\end{align*}
where $\hat{\Gamma}_{uid}\geq 0$ by definition, $\mathcal{E}'$ is the event in which only $\hat{F}$ has low-score priority, the second term is nonnegative by cost monotonicity, and the third term is nonnegative by \eqref{eq:uid} and (b).

\step{\textbullet\ \underline{$(a)\Rightarrow (b)$}.} Suppose (a) holds, and consider the family of linear cost functions $c(t|s)=\beta |t-s|$, for $\beta>0$. By choosing $\beta$ sufficiently low, we can ensure neither of the allocation rules saturates the probability constraint. In this case, $\hat{\alpha}_\beta(\underline{s})>0$ if and only if $\hat{F}$ has high-score priority, but then (a) implies $\tilde{F}$ must have high-score priority as well. Similarly $\tilde{\alpha}_\beta(\overline{s})<1$ if and only if $\tilde{F}$ has low-score priority, and then (a) implies $\hat{F}$ has low-score priority as well.  

Then $\tilde{\alpha}_\beta(s)=\beta(s-\tilde{s}_*)$ on $[\tilde{s}_*,\tilde{s}^*]$, and $\hat{\alpha}_\beta(s)=\beta(s-\hat{s}_*)$ on $[\hat{s}_*,\hat{s}^*]$. By varying $\beta$ from 0 to infinity, we have $\tilde{s}_*$ span $[\tilde{s}_*(0),0)$, and  $\hat{s}_*$ span $[\hat{s}_*(0),0)$. For $\beta$ sufficiently large, we have both $\tilde{s}_*>\tilde{s}_*(0)$ and $\hat{s}_*>\hat{s}_*(0)$. Pick such a value of $\beta$, then by (a), we have
$$-\beta \tilde{s}_*=\tilde{\alpha}_\beta(0)\geq \hat{\alpha}_\beta(0)= -\beta\hat{s}_*,$$
so $\tilde{s}_*\leq \hat{s}_*$. Furthermore, for such a value of $\beta$, we must have
$$\tilde{m}(\tilde{s}_*)=\frac{1}{\beta}+\tilde{s}_*\leq \frac{1}{\beta}+\hat{s}_*=\hat{m}(\hat{s}_*)\leq \hat{m}(\tilde{s}_*).$$
Varying $\beta$ so $\tilde{s}_*$ spans $[s_*(0),0)$, this shows (b).

 \step{\textbullet\ \underline{$(b)\Leftrightarrow (c)\Leftrightarrow (d)$}.} Since, for all $s<0$, every $x$ between $\tilde{m}(s)$ and $\hat{m}(s)$ is nonnegative, $\tilde{m}(s)\leq \hat{m}(s)$ is equivalent to $\int_{\tilde{m}(s)}^{\hat{m}(s)}xd\hat{F}(x)\geq 0$. By definition of the matching functions,
$$\int_{s}^{\tilde{m}(s)}xd\tilde{F}(x)=\int_{s}^{\hat{m}(s)}xd\hat{F}(x)=0,$$
therefore
$$\int_{\tilde{m}(s)}^{\hat{m}(s)}xd\hat{F}(x)-\int_{s}^{\tilde{m}(s)}xd\tilde{F}(x)=\int_{s}^{\tilde{m}(s)}xd\tilde{F}(x)-\int_{s}^{\tilde{m}(s)}xd\hat{F}(x).$$

This shows the equivalence between (b) and (c). The inequality in (d) results from applying integration by parts to (c).
\end{proof}

Hence, a simple first-order stochastic dominance shift is not sufficient to increase the allocation probability for all scores. Since it is challenging to interpret conditions (\ref{pdistc}) and (\ref{pdistd}), we provide a more easily interpretable sufficient condition on $\Delta$. We say that $\Delta$ \emph{divests} an interval $I \subseteq S$ if every score in $I$ (formally, every measurable subset of $I$) is less likely under $\tilde{F}$ than under $\hat{F}$. In other words, for every $[s, s'] \subseteq I$,
$$
\Delta(s') - \Delta(s) = \bigl\{\tilde{F}(s') - \tilde{F}(s)\bigr\} - \bigl\{\hat{F}(s') - \hat{F}(s)\bigr\} \leq 0,
$$
or equivalently, if $\Delta$ is nonincreasing on $I$. Conversely, if $\Delta$ is nondecreasing on $I$, we say it \emph{invests} $I$.

\begin{Prop}\label{PropSuffDist}
Suppose there exists $a\in[\underline{s},0)$ and $b\in(0,\overline{s}]$ such that
\begin{enumerate}
\item $\Delta(a)=\Delta(b)=0$, $\Delta(s)\geq 0$ for all $s\leq a$, and all $s\geq b$;
\item $\Delta$ divests $[a,0]$ and invests $[0,b]$;
\item $\int_{\underline{s}}^0\Delta(x)dx\leq 0$ and $\int_0^{\overline{s}}\Delta(x)dx\leq 0$. 
\end{enumerate}
Then the allocation rules satisfy $\tilde{\alpha}(s)\geq \hat{\alpha}(s)$ for every $s$.
\end{Prop}

In particular, a change in the distribution that shifts mass from ineligible scores to eligible ones satisfies the conditions of \propref{PropSuffDist}, thereby uniformly increasing the allocation probability.

\begin{proof}[Proof of \propref{PropSuffDist}]
We show the conditions of the proposition imply that, for every $z<0<y$, $\int_z^0\bigl\{\Delta(z)-\Delta(x)\bigr\}dx\geq 0$, and
$\int_0^y\bigl\{\Delta(y)-\Delta(x)\bigr\}dx\geq 0$, which implies condition (d) of \propref{PropDist}.

If $z<a$, then  
$$\int_z^0\bigl\{\Delta(z)-\Delta(x)\bigr\}dx=-z\Delta(z)-\int_z^0\Delta(x)dx\geq -z\Delta(z)-\int_{\underline{s}}^0\Delta(x)dx\geq 0,$$
where the first inequality is from condition 1, and the second inequality is from condition 3 and condition 1, as $\Delta(z)\geq 0$. If $z\geq a$, then condition 2 implies $\Delta(z)\geq \Delta(x)$ for every $x\in[z,0]$. The proof is symmetric for the integral from 0 to $y$.
\end{proof}


\section{A continuum as a single-agent}\label{AppMech} 


In this appendix, we explain why treating the continuum as a single agent is without loss of generality. The continuum of agents is interpreted as a limit case where the size of the population becomes arbitrarily large. We have already discussed why there is no loss of generality in considering allocation rules that only depend on the observed score profile and group identity. In the finite population case, an agent $j$ in group $i$ then receives the good with ex post allocation probability $\alpha_{i,j}(s_j,s_{-j})$. As is often the case in mechanism design, the problem can be reformulated as one of choosing interim allocation probabilities $\alpha_{i,j}(s_j)=\Exp_{s_{-j}}\alpha_{i,j}(s_j,s_{-j})$. Furthermore, given the symmetry of our setup, we can assume symmetry across agents of the same group, so we can write $\alpha_i(s)$ for the interim allocation probability for an agent with score $s$ in group $i$. Then the \emph{interim problem} of optimizing over symmetric interim allocation rules in any finite population is exactly the program we solve in the continuum.\footnote{The same approach is used in \cite{mylovanov2017optimal}.} However, to find a solution to the initial program, we need to ensure that the interim allocation rules that solve the interim program are feasible in the sense that they can be obtained from an ex-post allocation rule. In the finite population case, the exact condition for this to be possible can be derived from \citet{che2013generalized} which generalizes the condition of \citet{border1991implementation} to setups with multiple goods and quotas. In the limit case of the continuum, however, the interim rules can be used directly as ex post allocation rules that only depend on each agent's score, so feasibility is automatically satisfied.


{
\singlespacing
\bibliographystyle{econ}
\bibliography{
FalsproofBiblio}

@article{condorelli2013market,
  author = {Condorelli, Daniele},
  title = {Market and non-market mechanisms for the optimal allocation of scarce resources},
  journal = {Games and Economic Behavior},
  year = {2013},
  volume = {82},
  pages = {582--591},
  doi = {10.1016/j.geb.2013.08.008},
  url = {https://doi.org/10.1016/j.geb.2013.08.008}
}

@article{akbarpour2020redistributive,
  author = {Akbarpour, Mohammad and Dworczak, Piotr and Kominers, Scott Duke},
  title = {Redistributive Allocation Mechanisms},
  journal = {Journal of Political Economy},
  year = {2024},
  volume = {132},
  number = {6},
  pages = {1831--1875},
  doi = {10.1086/728111},
  url = {https://doi.org/10.1086/728111}
}

@unpublished{bjerrenielsen2023playing,
  author = {Andreas Bjerre-Nielsen and Lykke Sterll Christensen and Mikkel H{\o}st Gandil and Hans Henrik Sievertsen},
  title = {Playing the system: address manipulation and access to schools},
  year = {2023},
  note = {Working paper, arXiv:2305.18949},
  url = {https://arxiv.org/abs/2305.18949},
  eprint = {2305.18949},
  archiveprefix = {arXiv},
  primaryclass = {econ.GN}
}

@techreport{IPP2022,
  author = {Askenazy, Philippe and Breda, Thomas and Moreau, Flavien and Pecheu, Vladimir},
  title = {Les entreprises sous-d\'eclarent-elles leur effectif \`a 49 salari\'es pour contourner la loi?},
  institution = {Institut des Politiques Publiques},
  type = {Note IPP},
  year = {2022},
  number = {82},
  month = {March}
}

@techreport{lee2018manipulation,
  author = {Bolton, Lee},
  title = {Manipulation of the Waitlist Priority of the Organ Allocation System through the Escalation of Medical Therapies},
  institution = {{OPTN/UNOS Ethics Committee}},
  type = {Public comment proposal},
  year = {2018},
  url = {https://www.hrsa.gov/sites/default/files/hrsa/optn/ethics_publiccomment_20180122.pdf}
}

@article{mcmichael2022stealing,
  author = {McMichael, Benjamin J.},
  title = {Stealing Organs?},
  journal = {Indiana Law Journal},
  year = {2022},
  volume = {97},
  number = {1},
  pages = {135},
  url = {https://www.repository.law.indiana.edu/ilj/vol97/iss1/3/}
}

@article{schummer2021influencing,
  author = {Schummer, James},
  title = {Influencing waiting lists},
  journal = {Journal of Economic Theory},
  year = {2021},
  volume = {195},
  pages = {105263},
  doi = {10.1016/j.jet.2021.105263},
  url = {https://doi.org/10.1016/j.jet.2021.105263}
}

@article{galbiati2012tax,
  author = {Galbiati, Roberto and Zanella, Giulio},
  title = {The tax evasion social multiplier: Evidence from {Italy}},
  journal = {Journal of Public Economics},
  year = {2012},
  volume = {96},
  number = {5--6},
  pages = {485--494},
  doi = {10.1016/j.jpubeco.2012.01.003},
  url = {https://doi.org/10.1016/j.jpubeco.2012.01.003}
}

@article{ajzenman2018power,
  author = {Ajzenman, Nicol{\'a}s},
  title = {The power of example: Corruption spurs corruption},
  journal = {American Economic Journal: Applied Economics},
  year = {2021},
  volume = {13},
  number = {2},
  pages = {230--257},
  doi = {10.1257/app.20180612},
  url = {https://doi.org/10.1257/app.20180612}
}

@inproceedings{milli2019social,
  author = {Milli, Smitha and Miller, John and Dragan, Anca D. and Hardt, Moritz},
  title = {The social cost of strategic classification},
  booktitle = {Proceedings of the Conference on Fairness, Accountability, and Transparency},
  year = {2019},
  pages = {230--239},
  publisher = {Association for Computing Machinery},
  doi = {10.1145/3287560.3287576},
  url = {https://doi.org/10.1145/3287560.3287576}
}

@article{mil_seg02,
	author = {Milgrom, P. and Segal, I.},
	journal = {Econometrica},
	pages = {583--601},
	title = {Envelope Theorems for Arbitrary Choice Sets},
	volume = {70},
	year = {2002}}

@book{luen69,
	address = {New York},
	author = {Luenberger, David G.},
	publisher = {John Wiley and Sons},
	title = {Optimization by Vector Space Methods},
	year = {1969}}

@article{che2013generalized,
	author = {Che, Yeon-Koo and Kim, Jinwoo and Mierendorff, Konrad},
	journal = {Econometrica},
	number = {6},
	pages = {2487--2520},
	publisher = {Wiley Online Library},
	title = {Generalized Reduced-Form Auctions: A Network-Flow Approach},
	volume = {81},
	year = {2013}}

@article{border1991implementation,
	author = {Border, Kim C},
	journal = {Econometrica: Journal of the Econometric Society},
	pages = {1175--1187},
	publisher = {JSTOR},
	title = {Implementation of reduced form auctions: A geometric approach},
	year = {1991}}

@article{frankel2019muddled,
  author = {Frankel, Alex and Kartik, Navin},
  title = {Muddled information},
  journal = {Journal of Political Economy},
  year = {2019},
  volume = {127},
  number = {4},
  pages = {1739--1776},
  doi = {10.1086/701604},
  url = {https://doi.org/10.1086/701604}
}

@article{perez2022test,
  author = {Perez-Richet, Eduardo and Skreta, Vasiliki},
  title = {Test design under falsification},
  journal = {Econometrica},
  year = {2022},
  volume = {90},
  number = {3},
  pages = {1109--1142},
  doi = {10.3982/ECTA16346},
  url = {https://doi.org/10.3982/ECTA16346}
}

@unpublished{perez2023score,
  author = {Perez-Richet, Eduardo and Skreta, Vasiliki},
  title = {Score-based mechanisms},
  year = {2024},
  note = {Working paper, arXiv:2403.08031},
  url = {https://arxiv.org/abs/2403.08031},
  eprint = {2403.08031},
  archiveprefix = {arXiv}
}

@book{galichon2018optimal,
  author = {Galichon, Alfred},
  title = {Optimal transport methods in economics},
  year = {2018},
  publisher = {Princeton University Press},
  isbn = {9780691183466}
}

@article{mye81,
  author = {Myerson, Roger B.},
  title = {Optimal Auction Design},
  journal = {Mathematics of Operations Research},
  year = {1981},
  volume = {6},
  number = {1},
  pages = {58--73},
  doi = {10.1287/moor.6.1.58},
  url = {https://doi.org/10.1287/moor.6.1.58}
}

@article{bdl2019,
  author = {Ben-Porath, Elchanan and Dekel, Eddie and Lipman, Barton L.},
  title = {Mechanisms with Evidence: Commitment and Robustness},
  journal = {Econometrica},
  year = {2019},
  volume = {87},
  number = {2},
  pages = {529--566},
  doi = {10.3982/ECTA12432},
  url = {https://doi.org/10.3982/ECTA12432}
}

@article{ben2014optimal,
  author = {Ben-Porath, Elchanan and Dekel, Eddie and Lipman, Barton L.},
  title = {Optimal allocation with costly verification},
  journal = {American Economic Review},
  year = {2014},
  volume = {104},
  number = {12},
  pages = {3779--3813},
  doi = {10.1257/aer.104.12.3779},
  url = {https://doi.org/10.1257/aer.104.12.3779}
}

@article{mylovanov2017optimal,
  author = {Mylovanov, Tymofiy and Zapechelnyuk, Andriy},
  title = {Optimal allocation with ex post verification and limited penalties},
  journal = {American Economic Review},
  year = {2017},
  volume = {107},
  number = {9},
  pages = {2666--2694},
  doi = {10.1257/aer.20140494},
  url = {https://doi.org/10.1257/aer.20140494}
}

@article{erlanson2019note,
  author = {Erlanson, Albin and Kleiner, Andreas},
  title = {A note on optimal allocation with costly verification},
  journal = {Journal of Mathematical Economics},
  year = {2019},
  volume = {84},
  pages = {56--62},
  doi = {10.1016/j.jmateco.2019.06.002},
  url = {https://doi.org/10.1016/j.jmateco.2019.06.002}
}

@article{chua2019optimal,
  author = {Geoffrey A. Chua and Gaoji Hu and Fang Liu},
  title = {Optimal multi-unit allocation with costly verification},
  journal = {Social Choice and Welfare},
  year = {2023},
  volume = {61},
  number = {3},
  pages = {455--488},
  month = {October},
  doi = {10.1007/s00355-023-01463-5},
  url = {https://doi.org/10.1007/s00355-023-01463-5}
}

@inproceedings{epitropou2019optimal,
  author = {Epitropou, Markos and Vohra, Rakesh},
  title = {Optimal On-Line Allocation Rules with Verification},
  editor = {Fotakis, Dimitris and Markakis, Evangelos},
  booktitle = {Algorithmic Game Theory: 12th International Symposium, {SAGT} 2019},
  series = {Lecture Notes in Computer Science},
  year = {2019},
  volume = {11801},
  pages = {3--17},
  publisher = {Springer},
  doi = {10.1007/978-3-030-30473-7_1},
  url = {https://doi.org/10.1007/978-3-030-30473-7_1}
}

@article{li2020mechanism,
  author = {Li, Yunan},
  title = {Mechanism design with costly verification and limited punishments},
  journal = {Journal of Economic Theory},
  year = {2020},
  volume = {186},
  pages = {105000},
  doi = {10.1016/j.jet.2020.105000},
  url = {https://doi.org/10.1016/j.jet.2020.105000}
}

@inproceedings{hartrough2008,
  author = {Hartline, Jason D. and Roughgarden, Tim},
  title = {Optimal Mechanism Design and Money Burning},
  booktitle = {Proceedings of the Fortieth Annual {ACM} Symposium on Theory of Computing},
  year = {2008},
  pages = {75--84},
  publisher = {Association for Computing Machinery},
  month = {May},
  doi = {10.1145/1374376.1374390},
  url = {https://doi.org/10.1145/1374376.1374390}
}

@article{dworczak2022equity,
  author = {Dworczak, Piotr},
  title = {How to Allocate Money?},
  journal = {American Economic Journal: Microeconomics},
  year = {2026},
  volume = {18},
  number = {1},
  pages = {312--339},
  doi = {10.1257/mic.20230263},
  url = {https://doi.org/10.1257/mic.20230263}
}

@unpublished{patel2023costly,
  author = {Patel, Rohit and Urgun, Can},
  title = {Costly Verification and Money Burning},
  year = {2025},
  month = {March},
  note = {Working paper},
  url = {https://curgun.scholar.princeton.edu/document/66}
}

@unpublished{kattwinkel2019allocation,
  author = {Kattwinkel, Deniz},
  title = {Allocation with Correlated Information: Too good to be true},
  year = {2019},
  note = {Working paper},
  url = {https://economics.yale.edu/sites/default/files/2023-03/kattwinkel.pdf}
}

@article{kattknoe2022,
  author = {Kattwinkel, Deniz and Knoepfle, Jan},
  title = {Costless Information and Costly Verification: A Case for Transparency},
  journal = {Journal of Political Economy},
  year = {2023},
  volume = {131},
  number = {2},
  pages = {504--548},
  doi = {10.1086/721618},
  url = {https://doi.org/10.1086/721618}
}

@unpublished{augiasperez2023,
  author = {Augias, Victor and Perez-Richet, Eduardo},
  title = {Non-Market Screening with Investment},
  year = {2024},
  note = {Working paper},
  url = {https://sites.google.com/view/victoraugias/}
}

@article{frankel2020improving,
  author = {Frankel, Alex and Kartik, Navin},
  title = {Improving information from manipulable data},
  journal = {Journal of the European Economic Association},
  year = {2022},
  volume = {20},
  number = {1},
  pages = {79--115},
  month = {February},
  doi = {10.1093/jeea/jvab017},
  url = {https://doi.org/10.1093/jeea/jvab017}
}

@article{ball2020scoring,
  author = {Ball, Ian},
  title = {Scoring Strategic Agents},
  journal = {American Economic Journal: Microeconomics},
  year = {2025},
  volume = {17},
  number = {1},
  pages = {97--129},
  month = {February},
  doi = {10.1257/mic.20230275},
  url = {https://doi.org/10.1257/mic.20230275}
}

@article{lan_pla16,
  author = {Augustin Landier and Guillaume Plantin},
  title = {Taxing the Rich},
  journal = {The Review of Economic Studies},
  year = {2017},
  volume = {84},
  number = {3},
  pages = {1186--1209},
  doi = {10.1093/restud/rdw033},
  url = {https://doi.org/10.1093/restud/rdw033}
}

@article{kephart2016revelation,
  author = {Kephart, Andrew and Conitzer, Vincent},
  title = {The Revelation Principle for Mechanism Design with Signaling Costs},
  journal = {ACM Transactions on Economics and Computation},
  year = {2021},
  volume = {9},
  number = {1},
  pages = {6:1--6:35},
  doi = {10.1145/3434408},
  url = {https://doi.org/10.1145/3434408}
}

@article{deneckere2017screening,
  author = {Deneckere, Raymond and Severinov, Sergei},
  title = {Signalling, screening and costly misrepresentation},
  journal = {Canadian Journal of Economics},
  year = {2022},
  volume = {55},
  number = {3},
  pages = {1334--1370},
  doi = {10.1111/caje.12614},
  url = {https://doi.org/10.1111/caje.12614}
}

@unpublished{severinov2019screening,
  author = {Severinov, Sergei and Tam, Terry Yin-Chi},
  title = {Screening Under Fixed Cost of Misrepresentation},
  year = {2019},
  note = {Working paper},
  url = {https://sites.google.com/site/terrytamyinchi/research}
}

@article{tan2023price,
  author = {Tan, Teck Yong},
  title = {Price Discrimination with Manipulable Observables},
  journal = {The RAND Journal of Economics},
  year = {2026},
  volume = {57},
  number = {3},
  pages = {705--723},
  doi = {10.1111/1756-2171.70058},
  url = {https://doi.org/10.1111/1756-2171.70058}
}

@unpublished{li2024screening,
  author = {Li, Yingkai and Qiu, Xiaoyun},
  title = {Screening Signal-Manipulating Agents via Contests},
  year = {2023},
  note = {Working paper, arXiv:2302.09168v1},
  url = {https://arxiv.org/abs/2302.09168v1},
  eprint = {2302.09168v1},
  archiveprefix = {arXiv},
  primaryclass = {econ.TH}
}

@article{lac_wei89,
  author = {Lacker, Jeffrey M. and Weinberg, John A.},
  title = {Optimal Contracts under Costly State Falsification},
  journal = {Journal of Political Economy},
  year = {1989},
  volume = {97},
  number = {6},
  pages = {1345--1363},
  doi = {10.1086/261657},
  url = {https://doi.org/10.1086/261657}
}

@unpublished{carlier2012optimal,
  author = {Carlier, Guillaume},
  title = {Optimal transportation and economic applications},
  year = {2010},
  month = {June},
  note = {Lecture notes for the {IMA} programme ``New Mathematical Models in Economics and Finance''},
  url = {https://www.ceremade.dauphine.fr/~carlier/IMA-transport-Lecture-Notes.pdf}
}

@article{daskalakis2017strong,
  author = {Daskalakis, Constantinos and Deckelbaum, Alan and Tzamos, Christos},
  title = {Strong duality for a multiple-good monopolist},
  journal = {Econometrica},
  year = {2017},
  volume = {85},
  number = {3},
  pages = {735--767},
  doi = {10.3982/ECTA12618},
  url = {https://doi.org/10.3982/ECTA12618}
}

@unpublished{kolesnikov2022beckmann,
  author = {Kolesnikov, Alexander V. and Sandomirskiy, Fedor and Tsyvinski, Aleh and Zimin, Alexander P.},
  title = {{Beckmann}'s approach to multi-item multi-bidder auctions},
  year = {2022},
  note = {Working paper, arXiv:2203.06837},
  url = {https://arxiv.org/abs/2203.06837},
  eprint = {2203.06837},
  archiveprefix = {arXiv}
}

@unpublished{arieli2022persuasion,
  author = {Arieli, Itai and Babichenko, Yakov and Sandomirskiy, Fedor},
  title = {Feasible Conditional Belief Distributions},
  year = {2024},
  month = {November},
  note = {Working paper, arXiv:2307.07672v2},
  doi = {10.48550/arXiv.2307.07672},
  url = {https://arxiv.org/abs/2307.07672v2},
  eprint = {2307.07672v2},
  archiveprefix = {arXiv},
  primaryclass = {econ.TH}
}

@article{kolotilin2022persuasion,
  author = {Kolotilin, Anton and Corrao, Roberto and Wolitzky, Alexander},
  title = {Persuasion and Matching: Optimal Productive Transport},
  journal = {Journal of Political Economy},
  year = {2025},
  volume = {133},
  number = {4},
  pages = {1334--1381},
  doi = {10.1086/734095},
  url = {https://doi.org/10.1086/734095}
}

@article{lin2022credible,
  author = {Lin, Xiao and Liu, Ce},
  title = {Credible Persuasion},
  journal = {Journal of Political Economy},
  year = {2024},
  volume = {132},
  number = {7},
  pages = {2228--2273},
  doi = {10.1086/728745},
  url = {https://doi.org/10.1086/728745}
}

@article{boerma2021sorting,
  author = {Boerma, Job and Tsyvinski, Aleh and Zimin, Alexander P.},
  title = {Sorting with Teams},
  journal = {Journal of Political Economy},
  year = {2025},
  volume = {133},
  number = {2},
  pages = {421--454},
  doi = {10.1086/732891},
  url = {https://doi.org/10.1086/732891}
}

@article{amador2006commitment,
  author = {Amador, Manuel and Werning, Ivan and Angeletos, George-Marios},
  title = {Commitment vs. Flexibility},
  journal = {Econometrica},
  year = {2006},
  volume = {74},
  number = {2},
  pages = {365--396},
  doi = {10.1111/j.1468-0262.2006.00666.x},
  url = {https://doi.org/10.1111/j.1468-0262.2006.00666.x}
}

@article{amador2020regulating,
  author = {Amador, Manuel and Bagwell, Kyle},
  title = {Regulating a monopolist with uncertain costs without transfers},
  journal = {Theoretical Economics},
  year = {2022},
  volume = {17},
  number = {4},
  pages = {1719--1760},
  doi = {10.3982/TE4691},
  url = {https://doi.org/10.3982/TE4691}
}

@article{MilShan94,
  author = {Milgrom, Paul and Shannon, Chris},
  title = {Monotone Comparative Statics},
  journal = {Econometrica},
  year = {1994},
  volume = {62},
  number = {1},
  pages = {157--180},
  doi = {10.2307/2951479},
  url = {https://doi.org/10.2307/2951479}
}
}


\end{document}